\documentclass[acmtog,nonacm]{acmart}

\usepackage{booktabs} 
\usepackage{setspace}
\usepackage{graphicx}
\usepackage{subfigure}
\usepackage{ulem}
\usepackage{enumitem}
\usepackage{xcolor}
\usepackage{adjustbox}
\usepackage{color}
\usepackage{soul}

\usepackage{algorithm2e} 

\SetAlFnt{\small}
\SetAlCapFnt{\small}
\SetAlCapNameFnt{\small}
\SetAlCapHSkip{0pt}

\acmJournal{TOG}

\newcommand{\moduleOneFull}{{\emph{Local Disentanglement}}}
\newcommand{\moduleTwoFull}{{\emph{Global Fusion}}}

\newcommand{\moduleOne}{{\emph{LD}}}
\newcommand{\moduleTwo}{{\emph{GF}}}

\newcommand{\sysName}{{DeepFaceEditing}}

\newcommand{\ie}{i.e.}

\begin{document}
\title{\sysName: Deep Face Generation and Editing with Disentangled Geometry and Appearance Control}

\normalem 

\author{Shu-Yu Chen}

\authornotemark[2]
\affiliation{%
 \institution{Institute of Computing Technology, CAS and University of Chinese Academy of Sciences}}

\email{chenshuyu@ict.ac.cn}
 
\author{Feng-Lin Liu} 
\authornotemark[2]
\affiliation{
 \institution{Institute of Computing Technology, CAS and University of Chinese Academy of Sciences}
}
 
\author{Yu-Kun Lai}
\affiliation{%
\institution{School of Computer Science and Informatics, Cardiff University}
}
\email{LaiY4@cardiff.ac.uk}

\author{Paul L. Rosin}
\affiliation{%
\institution{School of Computer Science and Informatics, Cardiff University}
}
\email{RosinPL@cardiff.ac.uk}

\author{Chunpeng Li}
\affiliation{%
\institution{Institute of Computing Technology, CAS and University of Chinese Academy of Sciences }
}
\email{cpli@ict.ac.cn}

\author{Hongbo Fu}
\affiliation{%
\institution{School of Creative Media, City University of Hong Kong}
}
\email{hongbofu@cityu.edu.hk}
 
\author{Lin Gao}
\authornotemark[1]
\affiliation{
\institution{Institute of Computing Technology, CAS and University of Chinese Academy of Sciences }
}
\email{gaolin@ict.ac.cn}


\authorsaddresses{
$\dag$ Authors contributed equally.\\
\ $\ast$ Corresponding author.\\
Webpage: \url{http://geometrylearning.com/DeepFaceeEditing/} \\This is the author's version of the work. It is posted here for your personal use. Not for redistribution. }

\begin{abstract}
Recent facial image synthesis methods have been mainly based on conditional generative models.
Sketch-based conditions can effectively describe the geometry of faces, including the contours of facial components, hair structures, as well as salient edges (e.g., wrinkles) on face surfaces but lack effective control of appearance, which is influenced by color, material, lighting condition, etc.
To have more control of generated results, one possible approach is to apply existing disentangling works to disentangle face images into geometry and appearance representations. However, existing disentangling methods are not optimized for human face editing, and cannot achieve fine control of facial details such as wrinkles. To address this issue, we propose \sysName, a structured disentanglement framework specifically designed for face images to support face generation and editing with disentangled control of geometry and appearance. 
We adopt a local-to-global approach to incorporate the face domain knowledge: local component images are decomposed into geometry and appearance representations, which are fused consistently using a global fusion module to improve generation quality. We exploit sketches to assist in extracting a better geometry representation, which also supports intuitive geometry editing via sketching. The resulting method can either extract the geometry and appearance representations from face images, or directly extract the geometry representation from face sketches. Such representations allow users to easily edit and synthesize face images, with decoupled control of their geometry and appearance.
Both qualitative and quantitative evaluations show the superior detail and appearance control abilities of our method compared to state-of-the-art methods.

\end{abstract}



\begin{teaserfigure}
  \centering
  \includegraphics[width=1\linewidth]{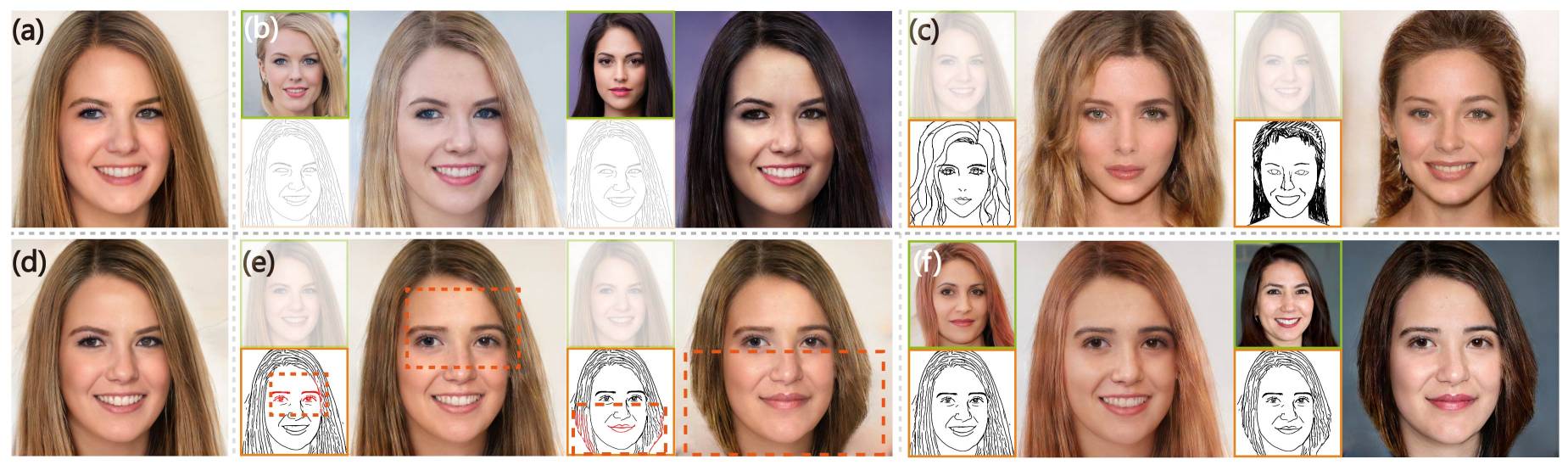}
  \caption{Our \sysName~method allows users to intuitively edit a face image to manipulate its geometry
  and appearance with detailed control. Given a portrait image (a), our method disentangles its geometry and appearance, and the resulting representations can faithfully reconstruct the input image (d). We show a range of flexible face editing tasks that can be achieved with our unified framework: (b) changing the appearance according to the given reference images while retaining the geometry, (c) replacing the geometry of the face with a sketch while keeping the appearance, (e) editing the geometry using sketches, and (f) editing both the geometry and appearance. The inputs used to control the appearance and geometry are shown as small images with green and orange borders, respectively.
}
  \label{fig:teaser}
\end{teaserfigure}

\maketitle

\section{Introduction}

Generating realistic face images is an active research topic in image synthesis.  Recent techniques such as StyleGAN~\cite{Karras_2019_CVPR,Karras2019stylegan2} allow random generation of realistic face images. However, many of them lack direct control of facial geometry or appearance. Adding conditions to the generation process is necessary to generate specific face images of interest, since face images are results of multiple factors, including geometry, appearance, head pose, viewpoint, etc.

Sketch-based conditions have been commonly used in existing image-to-image translation techniques {\cite{Isola_2017_CVPR,Wang_2018_CVPR}}, since sketching provides an easy and expressive means to depict desired geometry, including shape contours and salient surface edges. Multiple sketch-to-image techniques {\cite{Chen_2018_CVPR,Sangkloy_2017_CVPR,DeepFacePencil,10.1145/3386569.3392386}} have been proposed to generate realistic face images from edge maps or even rough sketches. However, since it is difficult to infer the appearance information from sketches alone, the above techniques lack enough control of appearance during the generation process. On the other hand, there exist various techniques {\cite{kim2020deformable,Kolkin_2019_CVPR,Yoo_2019_ICCV}} which allow the generation of realistic faces with their appearance as similar to given reference images as possible. However, such techniques focus on global appearance transfer and often do not well preserve geometric details of source images. Recent image disentangling works~\cite{park2020swapping,kim2020deformable} offer promising frameworks to have decoupled control of multiple attributes such as geometry and appearance. However, disentangling techniques for processing general image types do not provide optimal results for face generation. These methods treat images holistically and do not provide detailed, localized control, which is essential for face image editing.

In this work, we present \sysName, a novel face generation and editing framework that allows disentangled control of geometry and appearance (Fig.~\ref{fig:teaser}). The key enabler is a structured disentanglement framework specifically designed for face images. Observing that faces have a fixed structure of facial components, we adopt a local-to-global framework, consisting of two modules: \moduleOneFull~(\moduleOne) and \moduleTwoFull~(\moduleTwo). 
With paired images and sketches, the \moduleOne~module disentangles facial components into two independent representations: geometry and appearance. The former
encodes the geometry information including shapes of face and facial components as well as salient geometric details such as wrinkles. This can be specified explicitly as a sketch or implicitly as a geometry reference image. In contrast, the appearance representation encodes the information related to color, material, lighting conditions, etc. 
The \moduleTwo~module is trained to fuse local representations to obtain globally consistent face images. Unlike many other forms of disentanglement, our disentanglement makes the resulting representations easy to manipulate, and in particular, the geometry is represented using sketches, which can be edited intuitively for both overall shapes and details.

To ensure reliable disentanglement, our geometry space is designed to be a shared space of both sketches and images. In this way, we can not only make use of the geometry information in a geometry reference image but also develop a sketch-based interface for intuitive geometry manipulation. Figure \ref{fig:teaser} shows representative editing examples achieved with our interface, which supports decoupled control of appearance and geometry. We will also show the applications of our technique to face generation and face image style transfer.

We perform extensive qualitative and quantitative experiments, which show that disentangled control of geometry and appearance allows flexible face image generation and editing, and our approach outperforms state-of-the-art methods for various applications. 

The main contributions of our work are summarized as follows:
\begin{itemize}

\item We present a structured disentanglement framework for face image representation and synthesis, which ensures geometry and appearance are well decoupled and can be manipulated separately.
\item To achieve this, our network architecture involves local disentanglement (LD) modules for individual facial components to make learning more efficient, and a global fusion (GF) module for effective fusion of feature maps generated by LD modules. The LD modules are designed to embed both images and sketches of local regions to a shared space to ensure appropriate disentanglement, and allow sketches to be used as an explicit representation for the geometry, which is essential for detailed editing.
\item We present a novel interactive system that supports real-time editing of portrait images, 
which allows detailed editing of the geometry through an intuitive sketch-based interface, as well as modifying the appearance by providing an appearance reference image.

\end{itemize}

\section{Related work}

Our work is closely related to several topics, including neural face image synthesis, neural face image editing, neural image disentanglement, and style transfer.

\subsection{Neural Face Image Synthesis}
In recent years, conditional generative models like Generative Adversarial Networks (GANs)~\cite{goodfellow2014generative} have been widely used in face generation. They are capable of generating realistic face images from a random sampling of the Gaussian distribution~\cite{Karras_2019_CVPR}. 
These methods are not specifically designed to control the detail or appearance of faces explicitly. Several solutions \cite{Conditional_GAN_Face, pSp} have been proposed to control the image synthesis with specific conditions by using conditional GANs~\cite{Conditional_GAN}.
Also inspired by conditional GANs~\cite{Conditional_GAN},
Lee et al.~\shortcite{CelebAMask-HQ} and Gu et al.~\shortcite{gu2019mask} take a semantic mask as input and synthesize a new face based on a reference portrait. Instead of directly feeding a semantic layout as input, Park et al.~\shortcite{SPADE} propose spatially-adaptive normalization to progressively inject semantic information and achieve better visual quality for scene images.

Such methods support controllable image editing by editing semantic masks, but lack precise control of details within label maps. Another user-friendly interactive mode of geometry control is sketching. 
Compared with label maps, sketches can be easily drawn by users and artists, and thus have often been adopted for depicting the shape of desired content~\cite{Sketch2Photo}.
Based on Pix2Pix~\shortcite{Isola_2017_CVPR} for general image-to-image translation, SketchyGAN~\cite{Chen_2018_CVPR} and LinesToFacePhoto~\cite{Li:2019:LFP:3343031.3350854} synthesize realistic images by converting sketches to distance maps.
On the basis of pix2pixHD~\shortcite{Wang_2018_CVPR}, an extension of Pix2Pix for generating high-resolution images, Li et al.~\shortcite{DeepFacePencil} further introduce DeepFacePencil, a novel sketch-based face image synthesis framework that is robust to hand-drawn sketches.
To produce high-quality faces from rough or incomplete sketches, Chen et al.~\shortcite{10.1145/3386569.3392386} present DeepFaceDrawing, which takes a local-to-global approach and leverages manifold projection to enhance the generation quality and robustness from freehand sketches.
Besides 2D image generation, Han et al.~\shortcite{DeepSketch2Face} develop a sketching system for 3D face and caricature modeling. The above works use sketches to depict target geometry, and have little or no control of appearance during editing.
However, the appearance plays a central role for realistic image generation.
Sangkloy et al. \shortcite{Sangkloy_2017_CVPR} propose to use additional color strokes to indicate preferred colors for objects. While they achieve impressive results, the synthesized images still exhibit various artifacts like color leaking. For generating high-quality face images, it is essential to represent the complex appearance, which is difficult to achieve by only using color strokes. It also needs extra effort to paint the color strokes in addition to the geometry sketches, and the consistency between different color strokes should also be maintained for visual realism. It is thus not straightforward to extend the method by Sangkloy et al. \shortcite{Sangkloy_2017_CVPR} to generate realistic face images from freehand drawings and through interactive editing.

Sketches can be used to depict not only shape contours (similar to label maps) but also fine-level details like wrinkles on faces. However, as discussed previously, sketches themselves cannot support appearance control. To address this issue,
Zhang et al.~\shortcite{zhang2020cross} and Lee et al.~\shortcite{Lee_2020_CVPR1} propose exemplar-based image translation methods to generate photo-realistic images by learning dense cross-domain correspondence. These methods could be applied to different datasets including  face edge maps but the face edge maps used in their work are not freehand sketches.
Liu et al. \shortcite{liu2020selfsupervised} further present an exemplar-based image synthesis method with freehand sketches. The line-sketch generator is adopted to produce multiple sketches to improve the model’s robustness and thus handle freehand sketches. 
However, these methods are not specifically designed for photo-realistic face image synthesis considering the face structures. Robust high-quality face synthesis methods with detailed control using freehand sketches and appearance references are still missing.

\subsection{Neural Face Editing}
Face image editing has been a long-standing topic with the ability of providing flexible human-computer interaction, whose results have been significantly improved since the emergence of deep neural networks. Here, we mainly review the neural-network-based approaches. These approaches can be further grouped into three categories according to the intermediary they provide to the user for editing, namely direct editing, semantic-mask-guided editing and sketch-based editing.
For direct editing, Brock et al.~\shortcite{Neural_Photo_Editing} propose an introspective adversarial network (IAN), which turns rough painting strokes into user-desired photo-realistic images by directly manipulating images.
For semantic-mask-guided editing, pioneering works such as ~\cite{gu2019mask,CelebAMask-HQ,Zhu_2020_CVPR} incorporate semantic masks with conditional GANs \cite{Conditional_GAN} for interactive editing of face images, which have been discussed in previous sub-sections.
For the sketch-based methods, FaceShop~\cite{portenier2018faceshop} introduces a face image editing system to synthesize realistic images according to user inputs: masks, simple sketches, and colored strokes.
DeepFillv2~\cite{Yu_2019_ICCV} introduces gated convolution to learn dynamic features and presents a patch-based GAN discriminator to generate high quality inpaintings in a free-form manner. It is also applied to face editing under the guidance of input sketches.
Based on U-Net~\cite{ronneberger2015u}, Jo et al.~\shortcite{Jo_2019_ICCV} present an extra style loss to generate more robust and higher quality results, requiring minimal effort from users.
While these methods can generate high quality edited faces with guiding colored strokes and simple sketches within local areas, our method further supports editing of faces in both the local and global manners. In addition, since these methods treat face editing as an image completion problem with sketch guidance, they can only generate images based on the appearance of input images (i.e. not support using appearance from other images). 
Meanwhile, as demonstrated in \cite{Jo_2019_ICCV}, non-photorealistic artifacts could be generated with purely sketch and color strokes.
Recently, Yang et al.~\shortcite{yang2020deep} present another face editing method based only on sketches. With a sketch refinement strategy, their method is robust for human-drawn sketches but lacks the control of face appearance.

\subsection{Neural Image Disentanglement}

Learning a disentangled representation aims at modeling the factorization of data variation.
Previous works have introduced the disentanglement into image-to-image translation to facilitate multi-domain translation or allow the manipulation of certain image attributes while retaining others. For example, Liu et al.\shortcite{NEURIPS2018_84438b7a} disentangle images into domain-invariant representations to generate realistic results across multiple domains by changing the domain labels.
The work of \cite{NEURIPS2018_dc6a7071} decouples images into a shared part and an exclusive part to achieve multi-modal image translation. Furthermore, several works (e.g., \cite{NEURIPS2019_5a142a55,lee2019drit}) have extended the disentangled representation to provide domain-invariant and domain-specific representations to perform multi-domain and multi-modal translations simultaneously. 
However, these methods have focused on multi-domain translation and their disentanglement mainly aims at holistic attributes whereas our method focuses on disentangled representations that respect facial component structure and support detailed editing.

Another group of methods apply disentangled representations of latent code to control predefined attributes of faces.
For instance, the method by Zhang et al.~\shortcite{Zhang_Huang_Li_Zhao_Zhang_2019} is trained on labeled input data pairs by swapping designated parts of embeddings to control specific attributes. Deng et al. \shortcite{Deng_2020_CVPR} imitate
the 3D rendering process and introduce contrastive learning to learn a disentangled latent space. Many other works (e.g., \cite{harkonen2020ganspace,Shen_2020_CVPR,Tewari_2020_CVPR,abdal2020styleflow}) have tried to analyze and disentangle the latent code of some pretrained GAN space \cite{Karras_2019_CVPR} also with labeled data of specific attributes. 
Although these works successfully disentangle the latent space, they could only control a limited number of predefined attributes such as gender, expression,  and age, due to the use of labeled data in the training stage.
In contrast, our method provides users with more freedom to edit faces by changing sketches and to swap the face appearance by changing their appearance code.

Several research works have also proposed to disentangle specific elements and manipulate images.
Nguyen-Phuoc et al.~\shortcite{HoloGAN} and Schwarz et al.~\shortcite{schwarz2021_GRAF} utilize 3D representations and disentangle images into pose, shape, and appearance. However, these methods can only generate images randomly and do not support editing with detailed control, except for changing image viewpoints.
Wang et al. ~\shortcite{8803416} explicitly disentangle face deformations and appearance with two parallel networks to achieve expression editing.
While their work is limited to only changing facial expressions, the method in \cite{park2020swapping} also encodes images into geometry and appearance components with the idea of swapping  and co-occurrent patch discriminating statistics. 
Compared with this work, our method leverages a sketch as a constraint instead of the patch discriminator for precise facial detail editing. 
MichiGAN~\cite{tan2020michigan} is specifically designed for photo-realistic hair image generation of portraits conditioned on decoupled attributes. 
In our work, we aim to disentangle geometry and appearance of face images for  intuitive and detailed control of face generation. This benefits further applications such as freehand face editing and sketching.

\subsection{Style Transfer}
Style transfer focuses on changing the style of an image while preserving its key content. One category of style transfer methods (e.g.,~\cite{Cycle_GAN,Multmodel_I2I_Zhu,MUNIT,FUNIT}) requires a collection of target images and treats this problem as image-to-image translation. 
Another category of methods achieves style transfer with an arbitrary style image.
Starting from the pioneering work ~\cite{Gatys_style_transfer}, the follow-up works (e.g.,~\cite{Markov_style_transfer_Li, Mechrez_Style_transfer_contextual_loss, Kolkin_2019_CVPR}) take various content and style representations to generate attractive stylized images by iterative optimization. DST~\cite{kim2020deformable} further develops a novel geometry-aware style transfer method by deforming a content image to match the geometry of a style image. Another path of improvement is to eliminate the time-consuming optimization process. These works (e.g.,~\cite{deep_image_analogy, Huang_2017_ICCV, Univeral_style_transfer}) find dense correspondence or manipulate features in pre-trained networks to synthesize high-quality stylized images in real time. Although the above works generate interesting artistic results, these are not particularly suitable for photo-realistic image style transfer.
To handle this problem, WCT$^2$~\cite{Yoo_2019_ICCV} employs wavelet operations and progressive stylization to synthesize high resolution stylized photo-realistic images within a few seconds. 
In our work, we disentangle the geometry and appearance of face images to represent content and style information, respectively. Our method not only supports photo-realistic synthesis of stylized faces but also allows detailed editing with sketches.

\begin{figure*}[h]
    \centering
    \includegraphics[width=0.99 \linewidth]{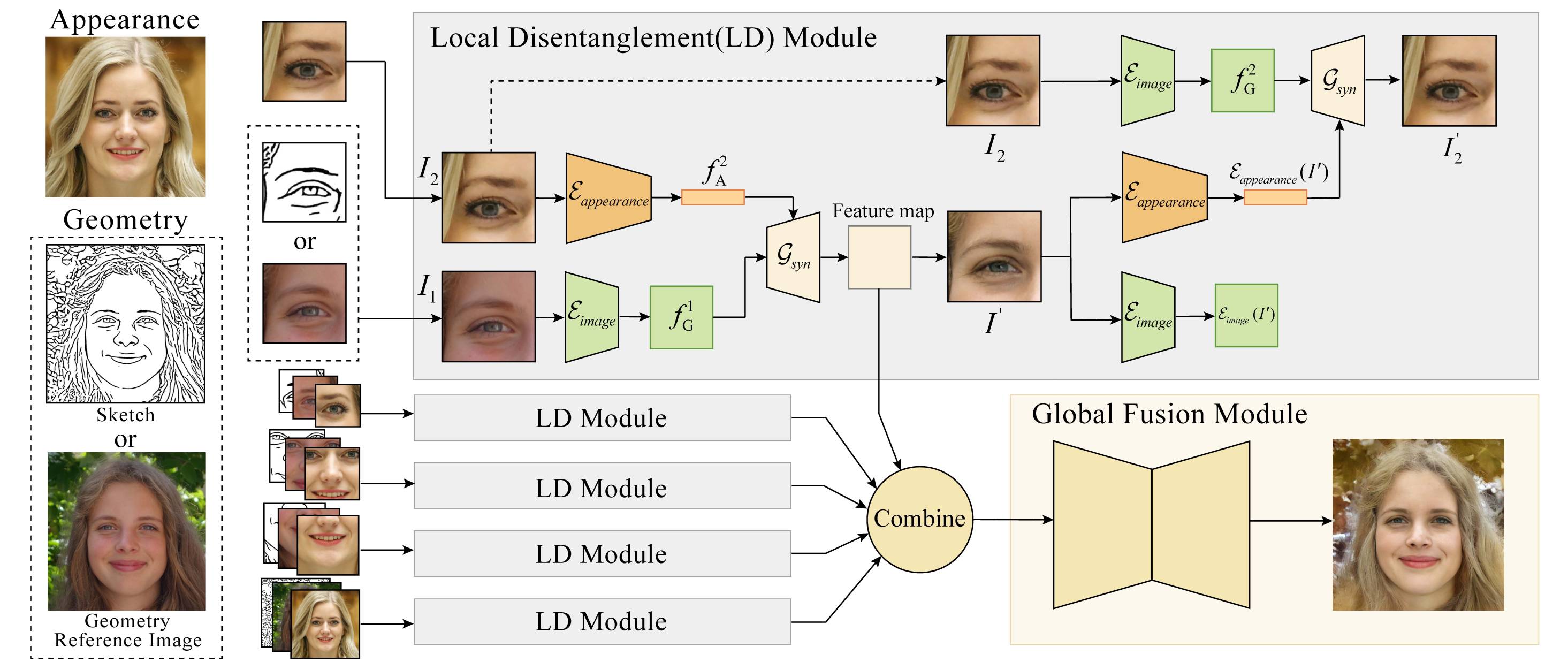}
    \caption{An overview of our framework. Our pipeline leverages a local-to-global strategy. The upper-right shows an illustration of a Local Disentanglement (LD) module (for ``right eye''), which uses $\mathcal{E}_{image}$ (when using a geometry reference image) or
    $\mathcal{E}_{sketch}$ (when using a sketch for controlling the geometry)
    to extract the geometry features and $\mathcal{E}_{appearance}$ to extract the appearance features. We use a swapping strategy to generate a new image $I^\prime$ with the geometry of $I_1$ and appearance of $I_2$. Taking the generated image $I^\prime$ as the appearance reference, along with the geometry of $I_2$, should lead to the reconstructed image $I_2^\prime$ that ideally matches $I_2$, allowing us to utilize a cycle loss to regularize training. Note that in our LD modules, such swapping and cycle constraints are applied to local regions corresponding to ``left eye'', ``right eye'', ``nose'', ``mouth'', and ``background''. The required geometry information can be explicitly given by a sketch or implicitly given by a reference image.
    Original images courtesy of Mats Adamczak and Senterpartiet (Sp).
    } 
    \label{fig:pipline}
\end{figure*}

\section{Methodology}
In this section, we formalize the structure of our proposed face image generation architecture in detail. 
Inspired by Huang et al.~\shortcite{Face_Rotation_ICCV} and DeepFaceDrawing~\cite{10.1145/3386569.3392386}, we leverage a local-to-global framework to generate high-quality face images. Specifically, we decompose a face image into five components (``left-eye'', ``right-eye'', ``nose'', ``mouth'', and ``background'') and process them using individual network modules. After component-level generation, we fuse the image patches into globally consistent results. Therefore, our architecture, called {\sysName}, comprises 5 \moduleOneFull ~\ (Sec.~\ref{sec:LDM}) ~\ modules responsible for disentangling the geometry and appearance for each component, and a \moduleTwoFull ~\ (Sec.~\ref{sec:GFM})~\ module responsible for fusing component features and generating high-quality results with global consistency. During training, we adopt a swapping scheme~\cite{park2020swapping} with a cycle-consistency constraint to enhance the robustness and generalization ability of our framework. 

\subsection{\moduleOneFull}\label{sec:LDM}

By incorporating this module, we aim to extract both the geometry and appearance features for each face component and generate local image patches from these features. To achieve this, we design a Geometry Encoder and an Appearance Encoder for obtaining geometry features and appearance features, respectively.
Furthermore, we design an Image Synthesis Generator to combine the geometry features and appearance features from a pair of image and sketch or two images (one providing the geometry features while the other providing appearance features) to obtain the translated component-level image patches.

\paragraph{Geometry Encoder}
Sketches depict the contours of real images, and are inherently suitable for geometry information extraction. Thus, for sketch inputs, we can directly extract pure geometry information using an auto-encoder network. The main challenge here is to extract geometry features from real images. Given a real facial component image as input, an intuitive approach extracting its geometry features is to first use a pre-trained image-to-sketch translation network to translate the real image into the sketch domain and then send the generated sketch into the geometry encoder for sketches. While this approach is promising, we show that there exists a unified way to extract the geometry information from sketches and real images.

\begin{figure}
    \centering
    \includegraphics[width=0.80 \linewidth]{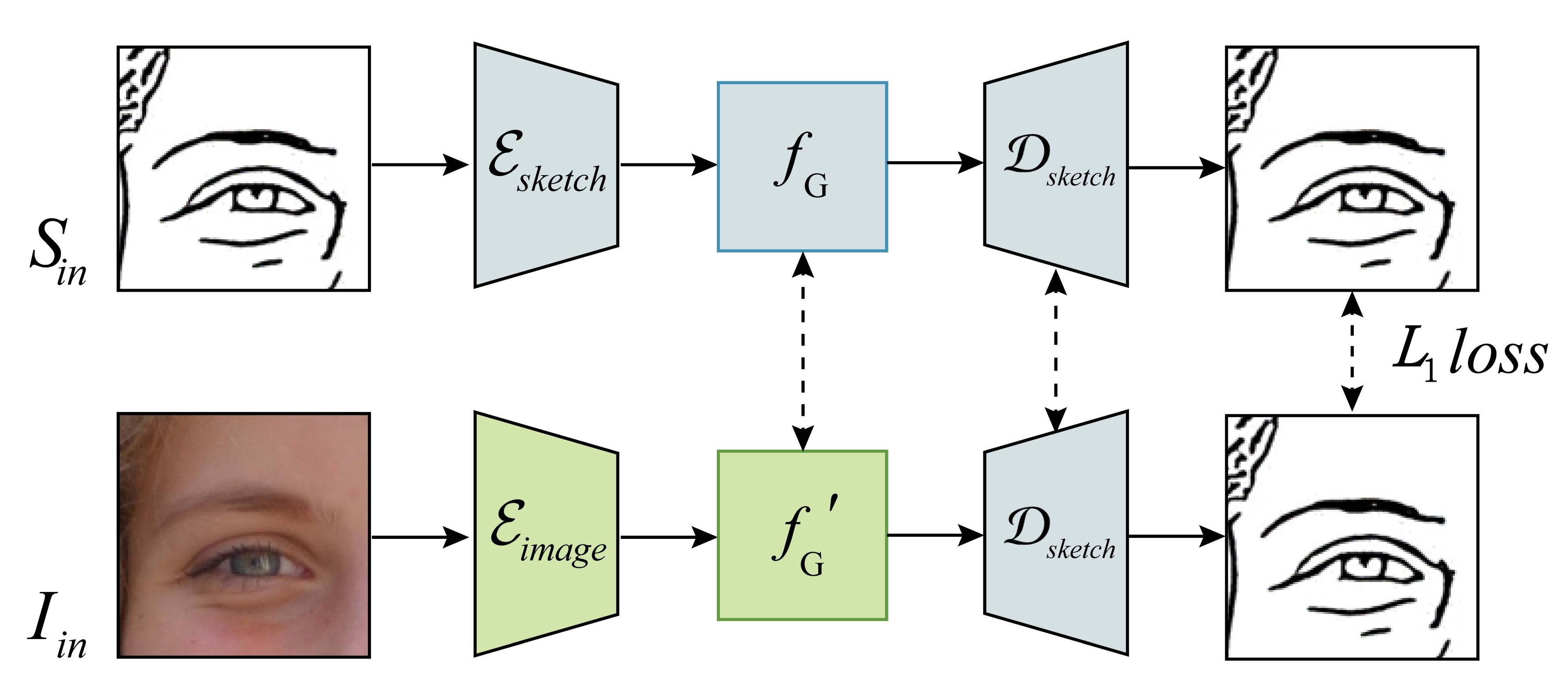}
    \caption{Illustration of Geometry Encoder. With pre-trained $\mathcal{E}_{sketch}$ and $\mathcal{D}_{sketch}$, which take a sketch $S_{in}$ as input, we train Geometry Encoder $\mathcal{E}_{image}$, which takes a real image $I_{in}$ {as input} and generates a geometry feature map $f_G^\prime$ that matches $f_G$. During training, we fix the parameters of $\mathcal{E}_{sketch}$ and $\mathcal{D}_{sketch}$, and add $L1$ losses to enforce consistency between the geometry latent feature maps $f_G$ and $f_G^\prime$, outputs of layers of $\mathcal{D}_{sketch}$ and output sketches of $\mathcal{D}_{sketch}$, as indicated by the dashed arrows. Original images courtesy of Mats Adamczak.
    } 
    \label{fig:Sketch Structure Encoder}
\end{figure}

Specifically, we achieve this by training two auto-encoders, one for sketches and the other for images, and aligning the latent distribution of the image space to that of the sketch space, ensuring that only the geometry information is encoded. We first train a network consisting of an encoder $\mathcal{E}_{sketch}$ and a decoder $\mathcal{D}_{sketch}$ to generate an intermediate feature of sketches, as shown Fig.~\ref{fig:Sketch Structure Encoder} (top). To retain essential spatial information, the latent space in the bottleneck layer is not in the form of a vector but instead a low-resolution feature map of dimension $h_G \times w_G \times c_G$, where $h_G$, $w_G$ and $c_G$ are the height, width and channel number for the geometry latent feature map.
Both the input and output of this network are sketches. Note that the sketches here can be either edge maps extracted from images or hand-drawn sketches. For hand-drawn sketches, especially incomplete ones during the sketching process, we apply sketch manifold projection~\cite{10.1145/3386569.3392386} during preprocessing to improve robustness. 

Let $I_{in}$ denote a real image and $S_{in}$ denote its corresponding sketch. We extract the geometry feature of $S_{in}$ through the pre-trained $\mathcal{E}_{sketch}$, formulated as $f_{G} = \mathcal{E}_{sketch}(S_{in})$.
Then we train an encoder $\mathcal{E}_{image}$ to map the corresponding image $I_{in}$ into the latent geometry space of sketches, denoted as $f^\prime_G = \mathcal{E}_{image}(I_{in})$. To encourage $f_{G}$ and $f^\prime_{G}$ to follow the same distribution, we impose constraints on each layer of $\mathcal{D}_{sketch}$ when we feed $f_{G}$ and $f^\prime_{G}$ into the pre-trained decoder $\mathcal{D}_{sketch}$. We also add an $L1$ loss between the outputs $\mathcal{D}_{sketch}(f_G)$ and $\mathcal{D}_{sketch}(f_{G}^\prime)$. 

\paragraph{{Appearance Encoder}}

Appearance is another important aspect of a facial image.
The mapping between facial geometry and real face images is clearly one-to-many, but by specifying the appearance, such ambiguities can be resolved.
We employ another encoder $\mathcal{E}_{appearance}$ to extract the appearance feature. The appearance encoder leverages global average pooling (i.e., for each feature channel, taking the average over all the spatial locations in the feature map), to eliminate spatial information and extract an appearance feature which is independent from its geometry feature. As appearance features are extracted for individual local regions, dropping spatial information does not cause significant loss of useful information. Results of the experiments on disentangled interpolation on face appearance and geometry (see Fig.~\ref{fig:disentangled}) demonstrate that  $\mathcal{E}_{appearance}$ can learn a continuous and smooth facial appearance space efficiently.

\paragraph{Image Synthesis Generator}

Given independent geometry and appearance representations, an image synthesis generator takes them as inputs and generates reconstruction or swapping results. 
To control the appearance of generated face images, we adopt Adaptive Instance Normalization (AdaIN)~\cite{Huang_2017_ICCV} in our facial image synthesis generator.
Specifically, this generator comprises 4 residual blocks and 4 up-sampling layers, and the appearance features are injected into them.
Then, we obtain an embedding feature which has the same resolution as the input image but with 64 channels. 
Finally, an image consistent with the input geometry and appearance features is predicted via a single convolution layer.

\begin{figure}
    \centering
    \includegraphics[width=0.99 
    \linewidth]{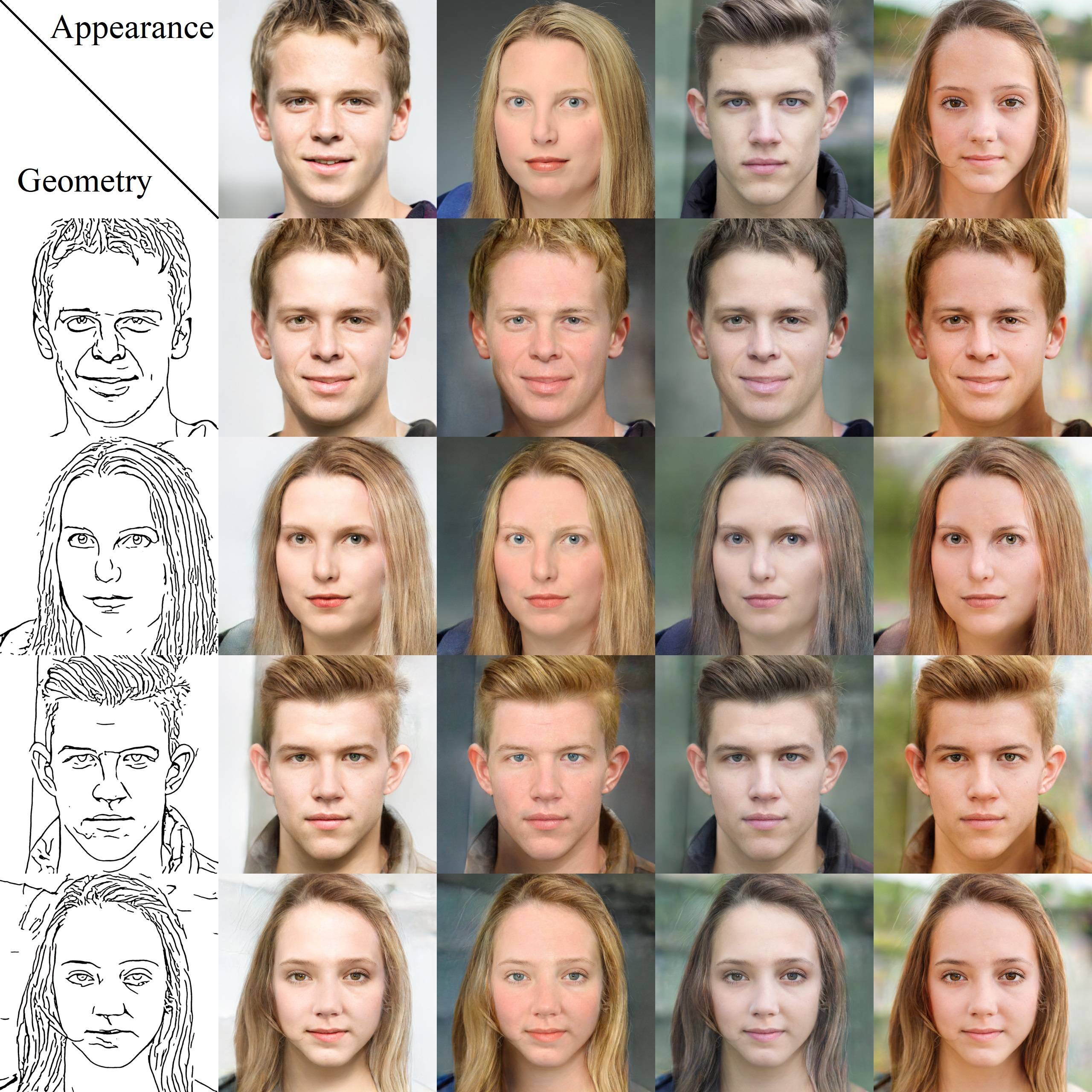}
    \caption{
    Face generation with different geometry and appearance inputs. 
    The geometry inputs (first column) use edge maps from the same set of images as those used for appearance (first row).
    Each face image is generated with the sketch in the same row for geometry and appearance reference image in the same column for appearance. 
    Therefore, the images on the main diagonal look the same as the corresponding appearance images. This indicates the accurate reconstruction ability of our method.
    Original images courtesy of Greg Mooney, BLSZ\_BD and christopher simon.
    }
    \label{fig:Swapping}
\end{figure}

Unlike those sketch-to-image tasks, our framework takes real facial images as inputs and disentangles appearance features with geometry features under the assistance of sketches. 
Furthermore, a network that could provide controllable sketch-based face editing would be more appealing to users.
Therefore, to achieve this, we randomly feed both the features generated by $\mathcal{E}_{sketch}$ and $\mathcal{E}_{image}$ into the image synthesis generator during the training stage. 
During inference, either geometry reference images or sketches can provide the geometry information and generate photo-realistic facial images with specific appearance. With appearance and geometry features disentangled, we can generate face images possessing the geometry and appearance from different sources, as shown in Fig.~\ref{fig:Swapping}.

\subsection{\moduleTwoFull}\label{sec:GFM}

This module generates the final face images from local image features encoded by the \moduleOneFull ~\ modules.
To translate local image features into a complete and natural-looking face image, one possible way is to directly combine the local image patches generated by applying one or more convolution layers to the component-level feature maps 
(like the process of generating $I^\prime$ in Fig.~\ref{fig:pipline}).
However, this straightforward approach is prone to exhibiting artifacts on the boundaries of different components. 
So instead of directly combining component-level image patches, our method combines the generated intermediate feature maps of the \moduleOneFull ~\  modules before sending them into an image generation network. In this way, our network aggregates more information flow and is capable of generating high-quality images.

More concretely, our \moduleTwoFull ~\ module comprises three units: an encoder, residual blocks and a decoder, like DeepFaceDrawing~\shortcite{10.1145/3386569.3392386} and pix2pixHD~\cite{Wang_2018_CVPR}. Given the feature map of the background component, we replace certain patches of it with the corresponding generated features of other components, in the order of ``mouth'', ``nose'', ``left-eye'', and ``right-eye'' to reduce the impact of overlapping between components. 
Then we feed the combined feature map into the \moduleTwoFull \ module to generate a new face with desired appearance and geometry.

\subsection{Training Process}\label{sec:train}

In this part, we introduce the training process of the aforementioned modules at length. We train the entire framework in a step-by-step manner. Specifically, we first train the \moduleOneFull ~\ modules, and then train the \moduleTwoFull ~\ module with parameters of the \moduleOneFull ~\ modules fixed.

\paragraph{{Training Dataset}}
To train our network, we need a large-scale dataset of sketch-image pairs.
At the same time, the sketches in the sketch-image pairs are required to be salient, natural and similar to hand-drawn sketches. 
Traditional edge extraction methods, such as HED~\cite{xie15hed} and Canny~\cite{canny1986a}, often fail to produce ideal edge maps.
Therefore, we follow DeepFaceDrawing~\cite{10.1145/3386569.3392386} and use the Photocopy filter in Photoshop followed by sketch simplification to build this dataset.
We use FFHQ~\cite{Karras_2019_CVPR} as our training data. In this way, we generate
32.2k sketch-image pairs as our dataset and randomly select 29.9k pairs for training and 2300 pairs for testing. The resolution of both images and sketches is set to 512 $\times$ 512.

\paragraph{{Disentanglement Training}}

The training process of each \moduleOneFull ~\ module consists of three steps. 
First, as described in Sec. \ref{sec:LDM}, $\mathcal{E}_{sketch}$ and $\mathcal{D}_{sketch}$ are trained to learn a geometry latent space for sketches using L1 reconstruction loss.
Once $\mathcal{E}_{sketch}$ is trained, we can represent the geometry feature as $f_{G} = \mathcal{E}_{sketch}(S_{in})$. 
Then, we train the network $\mathcal{E}_{image}$, which takes a real image $I_{in}$ as input and predicts a geometry feature $f_{G}^\prime = \mathcal{E}_{image}(I_{in})$ following the same distribution as the learned geometry space. 
The loss function $L_{image}$ is defined as follows:
\begin{equation}
    \begin{aligned}
        {L_{image}}(\mathcal{E}_{image}) =\sum_{i=0}^N ||\mathcal{D}_{sketch}^{(i)}(f_{G}^\prime) - \mathcal{D}_{sketch}^{(i)}(f_{G})||_1, 
    \end{aligned}
\end{equation}
where $N=7$ is the number of layers of decoder $\mathcal{D}_{sketch}$. The index 0 corresponds to the input feature map, the index $N$ corresponds to the output image, and other indices are intermediate feature maps.
Note that when optimizing the parameters of $\mathcal{E}_{image}$, we fix the weights of $\mathcal{E}_{sketch}$ and $\mathcal{D}_{sketch}$.
Finally, we train the appearance encoder $\mathcal{E}_{appearance}$ and the image synthesis generator $\mathcal{G}_{syn}$ with weights of $\mathcal{E}_{sketch}$ and $\mathcal{E}_{image}$ fixed.
During the last step of training, we randomly feed the geometry feature $f_{G}$ of sketches or $f_{G}^\prime$ of real images into $\mathcal{G}_{syn}$.
In the following parts, we denote both $f_{G}$ and $f_{G}^\prime$ as $G$, without distinguishing their sources. We further introduce a swapping strategy and adopt cycle-consistency loss to disentangle a real facial image into appearance and geometry.
To ensure photorealism of generated images, we also adopt the muti-scale discriminator~\cite{Wang_2018_CVPR} and adversarial loss. 

More concretely, as shown in Fig.~\ref{fig:pipline}, given two component images $I_1$ ($I_1$ could be either a real image or sketch) and $I_2$ ($I_2$ should be a real image) in the training set, we can extract geometry features $f_G^1$ and $f_G^2$ from $I_1$ and $I_2$ by passing them through the pre-trained $\mathcal{E}_{sketch}$ or $\mathcal{E}_{image}$. The appearance feature $f_A^2$ from $I_2$ is extracted by using $\mathcal{E}_{appearance}$.
By swapping the geometry feature of $I_2$ with that of $I_1$, we generate image $I^\prime$ using the geometry feature $f_G^1 = \mathcal{E}_{image}(I_1)$ of image $I_1$ and the appearance feature $f_A^2 = \mathcal{E}_{appearance}(I_2)$ of image $I_2$, as 
$I^\prime = \mathcal{G}_{syn}(f_G^1,f_A^2)$. 
With the geometry of $I_1$ and the appearance of $I^\prime$, we can establish a cyclic reconstruction of image $I_2$, as $I_2^\prime = \mathcal{G}_{syn}(f_G^2,\mathcal{E}_{appearance}(I^\prime))$. 
We also introduce a self-reconstruction loss: when both $\mathcal{E}_{image}$ and $\mathcal{E}_{appearance}$ take the same image (such as $I_1$) as input, we can reconstruct it using its geometry and appearance features. The self-reconstruction of $I_1$ can be formulated as $I_1^\prime=\mathcal{G}_{syn}(f_G^1,\mathcal{E}_{appearance}(I_1))$. 
Overall, we adopt the following loss terms to train our \moduleOneFull ~\ modules:
\begin{itemize}[leftmargin=*]
\item[-] {Self-reconstruction loss:}
When the geometry and appearance come from the same image, i.e. $I_1=I_2=I$, the self-consistency of our framework requires that we can reconstruct $I$ after passing it through our framework.
The self-reconstruction loss contains three terms: 1) Perceptual loss~\cite{DBLP:journals/corr/JohnsonAL16} ${L}_{VGG}$, which measures the visual similarity between the generated images and input images by a pre-trained VGG-19 model; 2) Feature matching loss~\cite{Wang_2018_CVPR} ${L}_{FM}$ of discriminators, which aims to stabilize the training process; 3) Lab color loss ${L}_{Lab}$~\cite{tan2020michigan}, 
which calculates the chromatic distance in the $a$ and $b$ channels for controlling the color tone by converting images to the CIE-LAB color space.
The self-reconstruction loss can be formulated as follows:
\begin{equation}
        {L}_{Recon} = \alpha_1{L}_{Lab}(I_1,I_1^\prime) + \alpha_2{L}_{FM}(I_1,I_1^\prime) + \alpha_3{L}_{VGG}(I_1,I_1^\prime),\\
\end{equation}
In our experiments, we empirically set $\alpha_1,\alpha_2,\alpha_3=1,10,10$.

\begin{figure*}[h]
    \centering
    \setlength{\fboxrule}{0.5pt}
    \setlength{\fboxsep}{-0.01cm}
    \begin{tabular}{cc}
    
     \framebox{\includegraphics[width=0.13\linewidth]{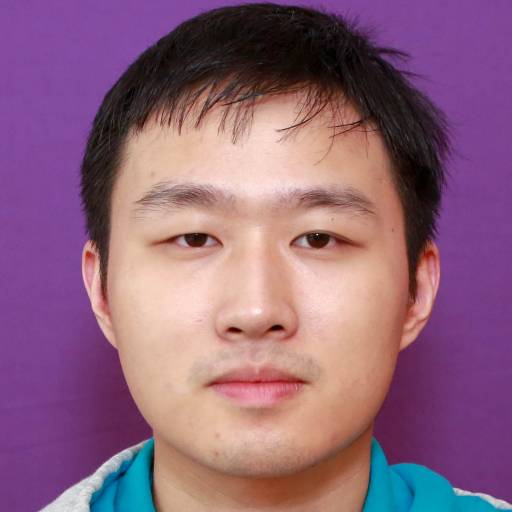}}
     \framebox{\includegraphics[width=0.13\linewidth]{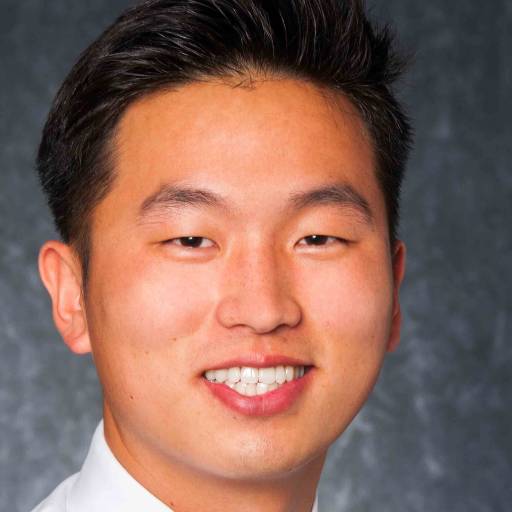}} &
     \framebox{\includegraphics[width=0.13\linewidth]{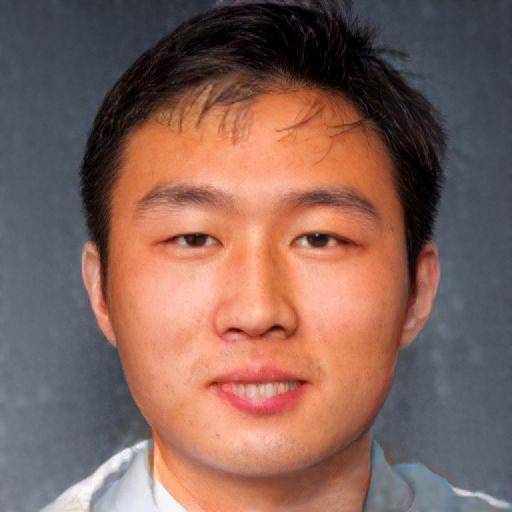}}
     \framebox{\includegraphics[width=0.13\linewidth]{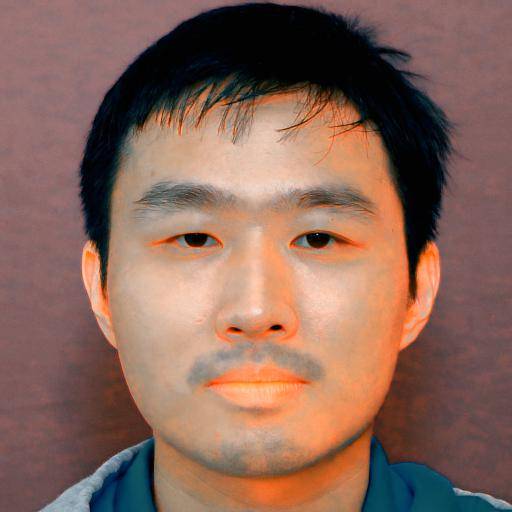}}
     \framebox{\includegraphics[width=0.13\linewidth]{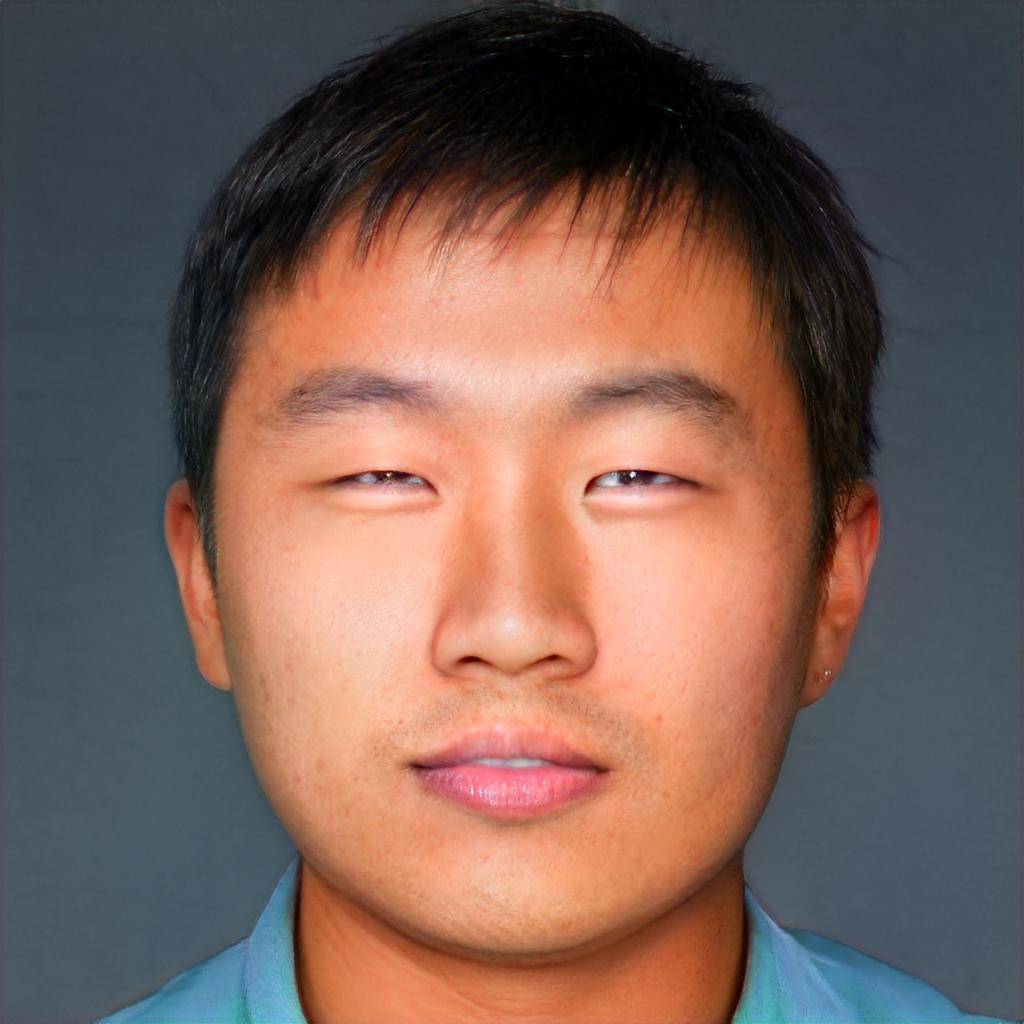}}
     \framebox{\includegraphics[width=0.13\linewidth]{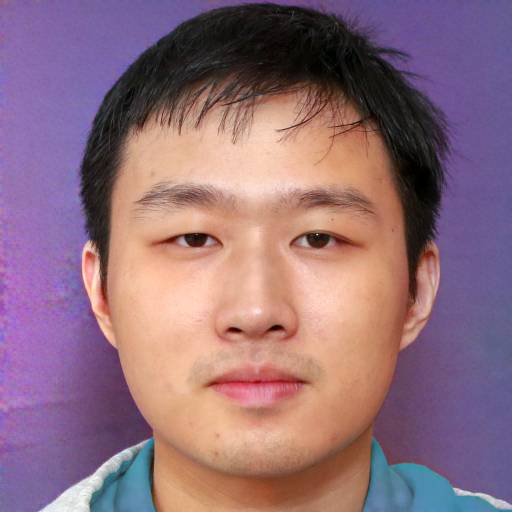}}
     \framebox{\includegraphics[width=0.13\linewidth]{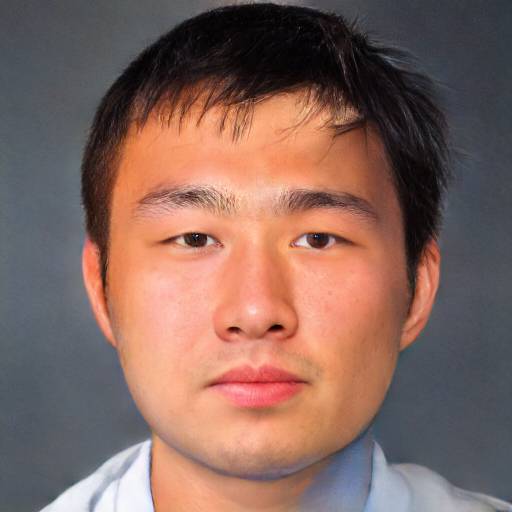}} \\

    
     \framebox{\includegraphics[width=0.13\linewidth]{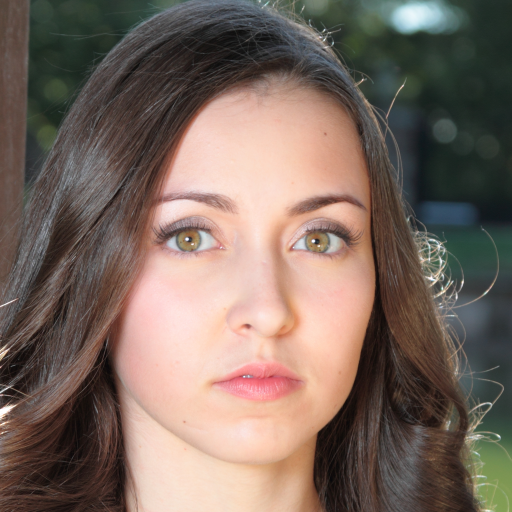}}
     \framebox{\includegraphics[width=0.13\linewidth]{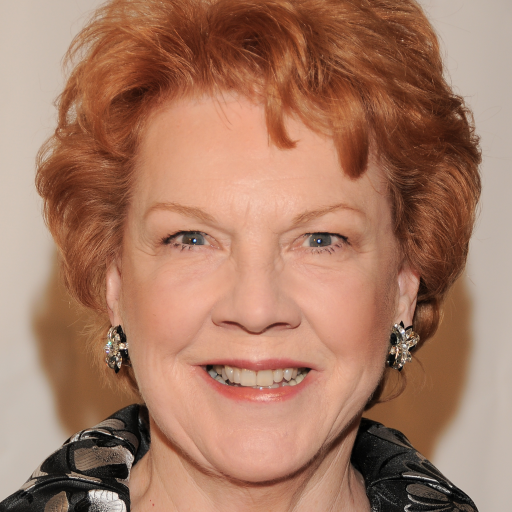}} &
     \framebox{\includegraphics[width=0.13\linewidth]{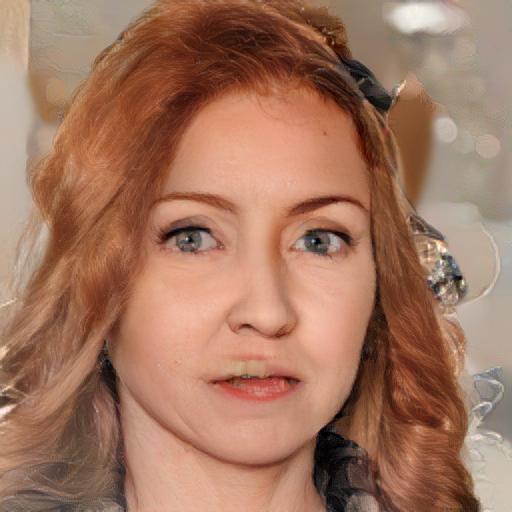}}
     \framebox{\includegraphics[width=0.13\linewidth]{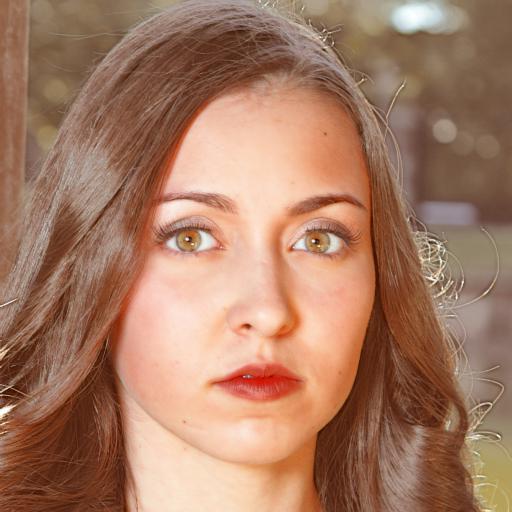}}
     \framebox{\includegraphics[width=0.13\linewidth]{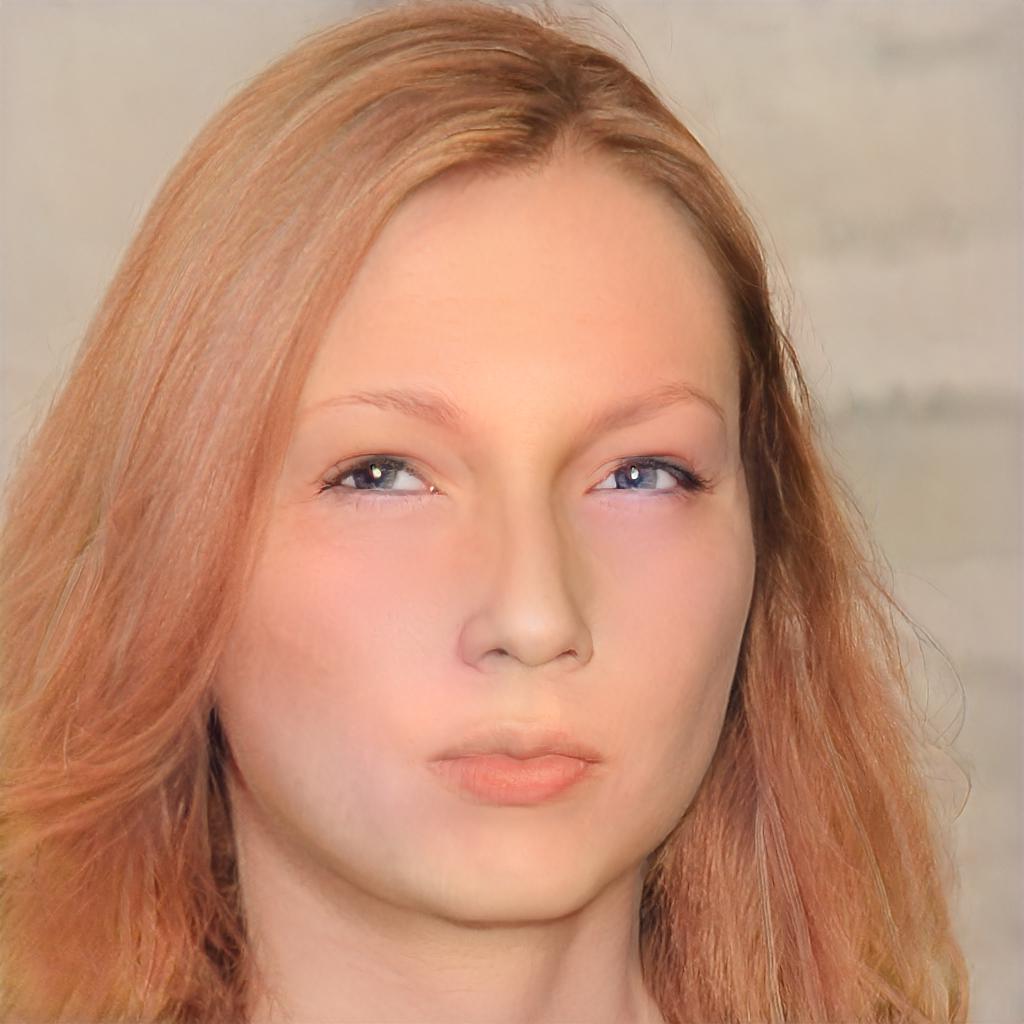}}
     \framebox{\includegraphics[width=0.13\linewidth]{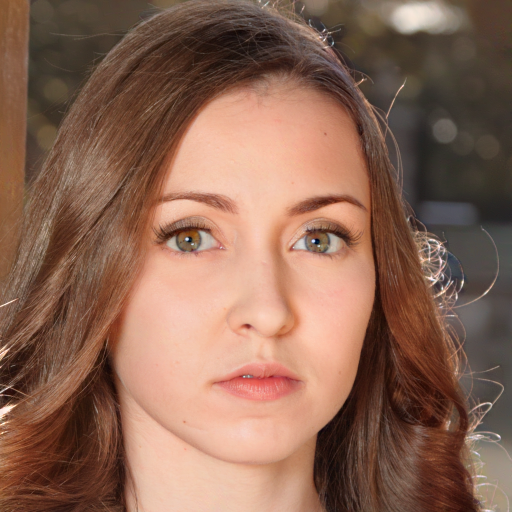}}
     \framebox{\includegraphics[width=0.13\linewidth]{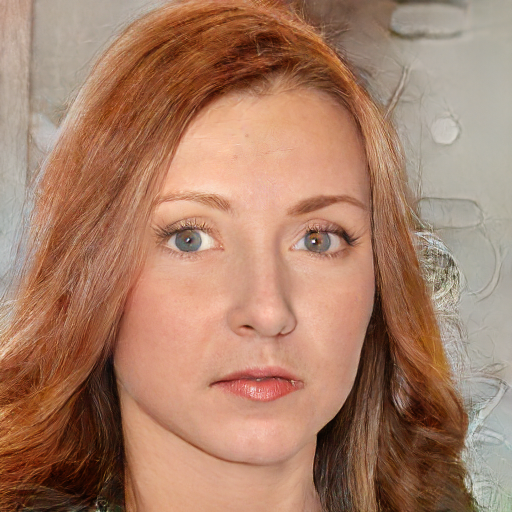}}\\
    
     \framebox{\includegraphics[width=0.13\linewidth]{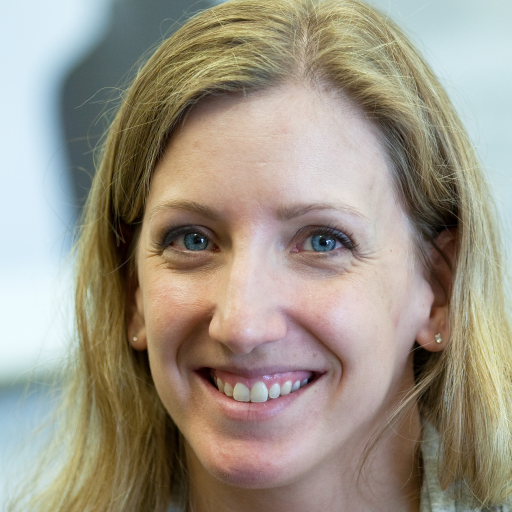}}
     \framebox{\includegraphics[width=0.13\linewidth]{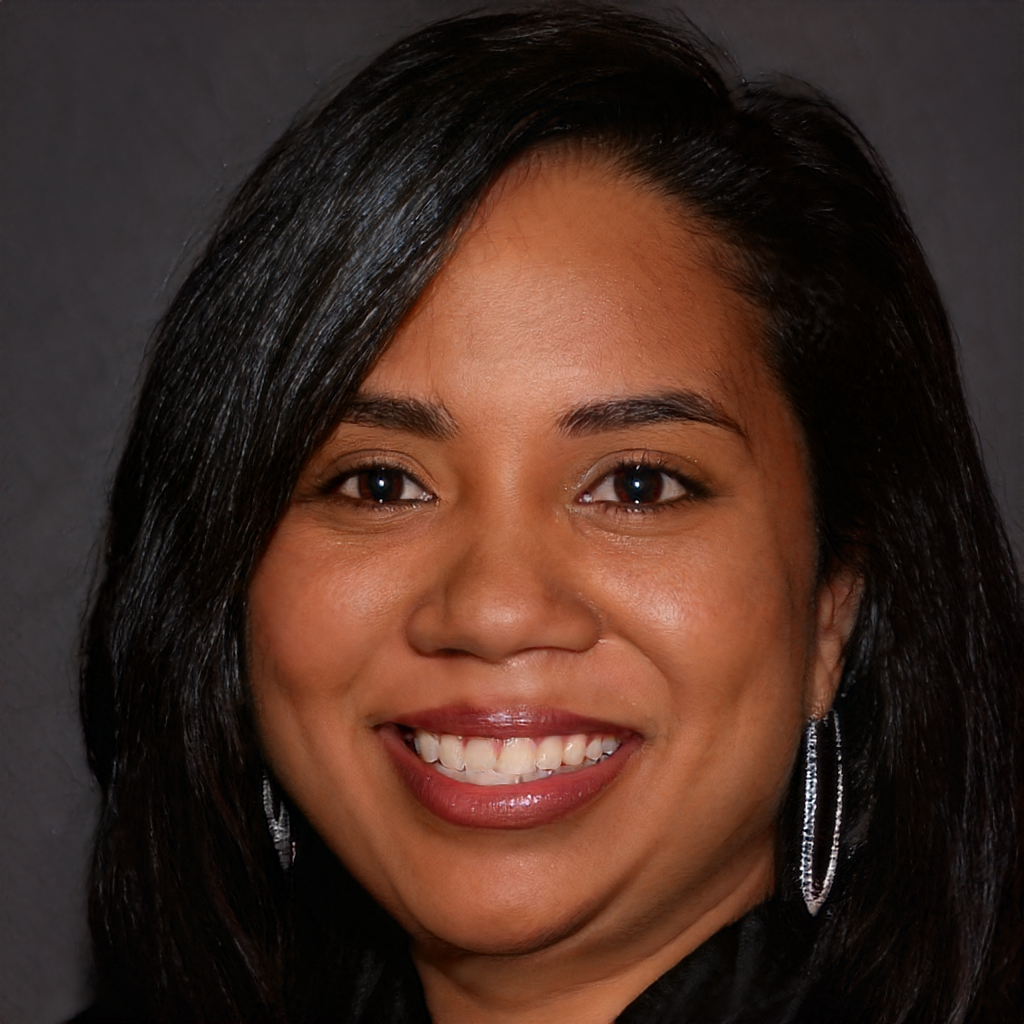}} &
     \framebox{\includegraphics[width=0.13\linewidth]{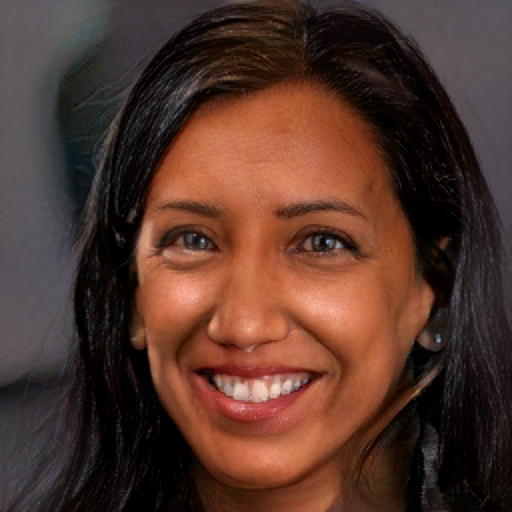}}
     \framebox{\includegraphics[width=0.13\linewidth]{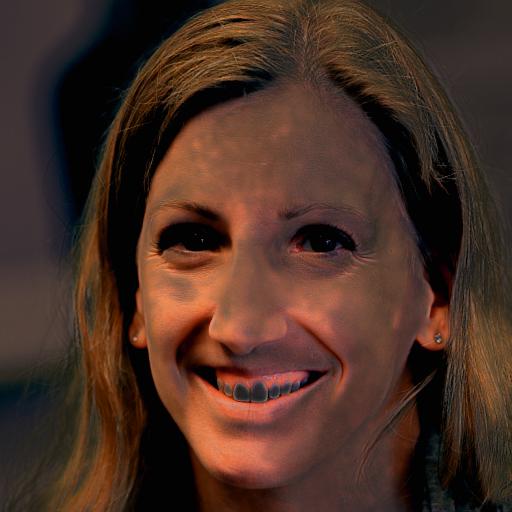}}
     \framebox{\includegraphics[width=0.13\linewidth]{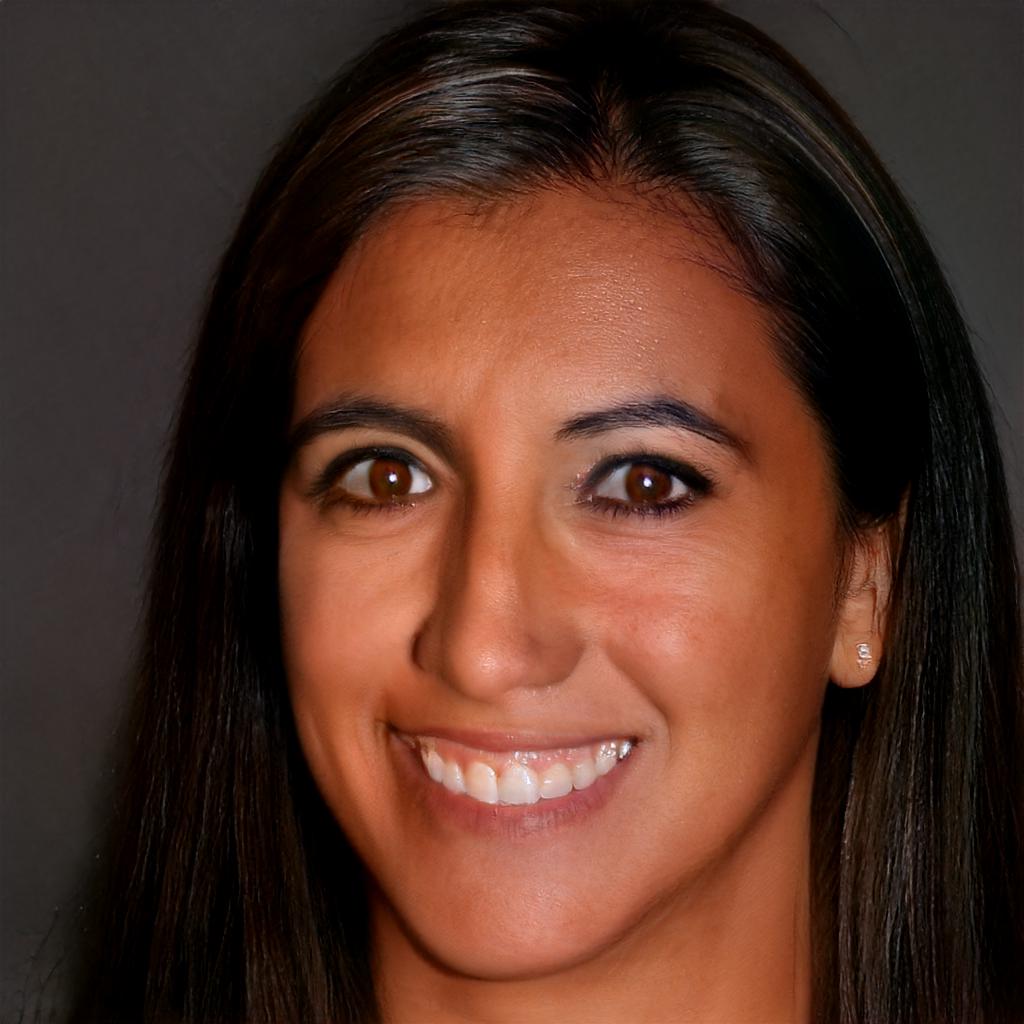}}
     \framebox{\includegraphics[width=0.13\linewidth]{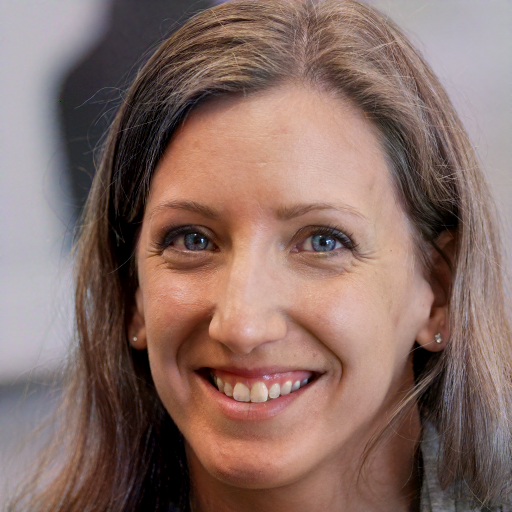}}
     \framebox{\includegraphics[width=0.13\linewidth]{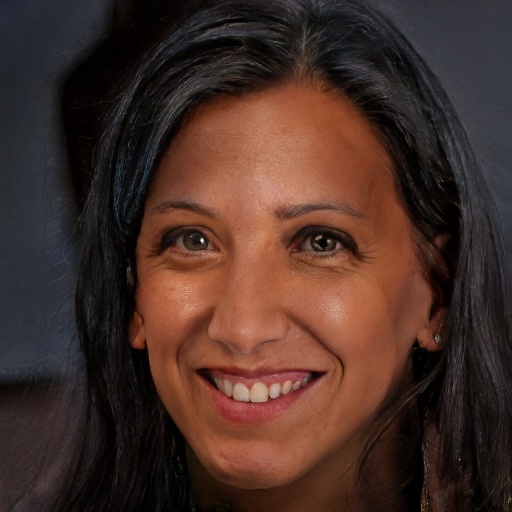}}\\

    \vspace{-5.5mm}\\
    \subfigure[Geometry]{\framebox{\includegraphics[width=0.13\linewidth]{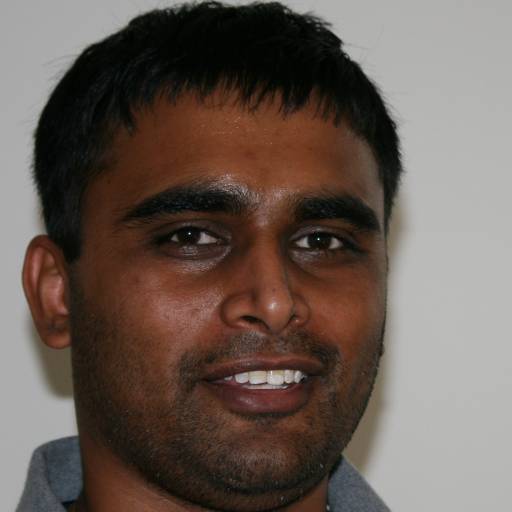}}}
    \subfigure[Appearance]{\framebox{\includegraphics[width=0.13\linewidth]{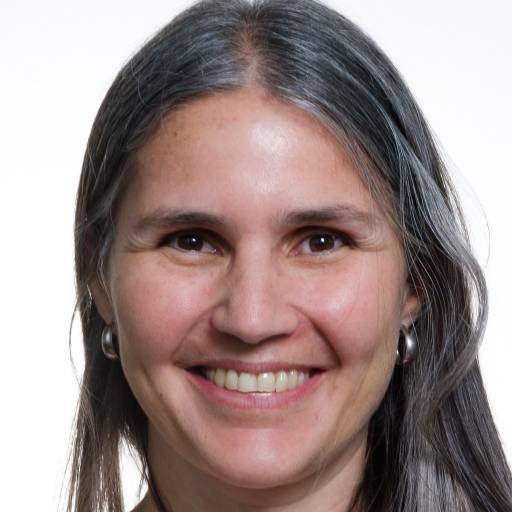}}} &
    \subfigure[Strotss]{\framebox{\includegraphics[width=0.13\linewidth]{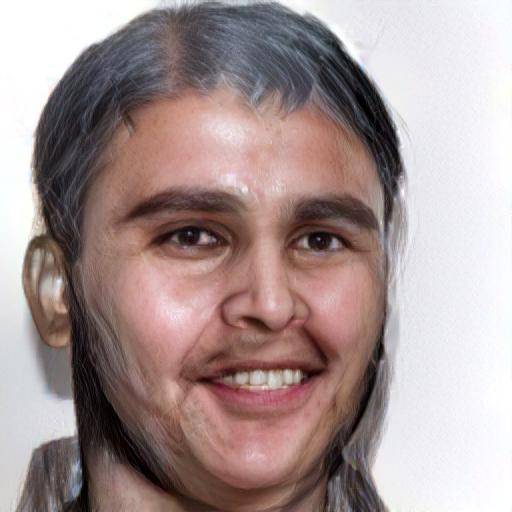}}}
    \subfigure[WCT$^2$]{\framebox{\includegraphics[width=0.13\linewidth]{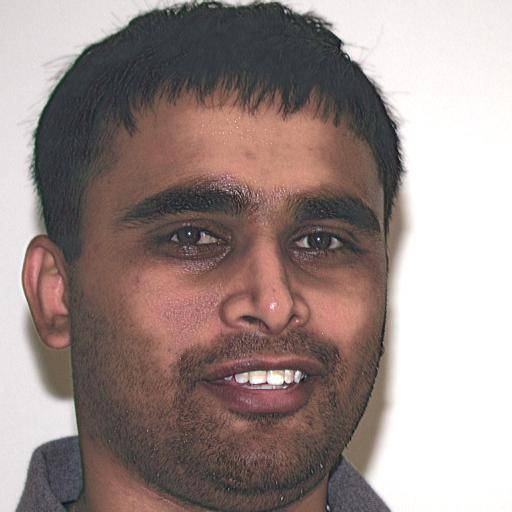}}}
    \subfigure[StyleGAN2]{\framebox{\includegraphics[width=0.13\linewidth]{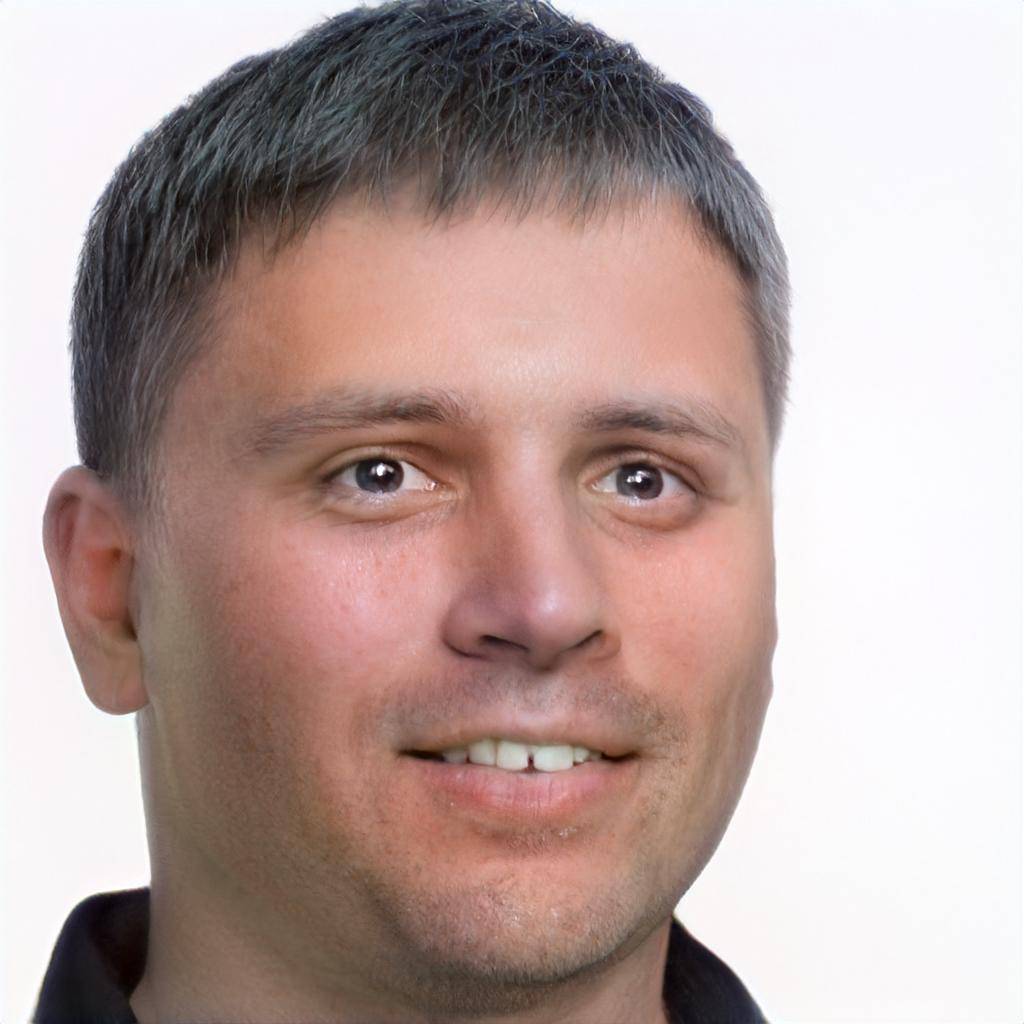}}}
    \subfigure[DST]{\framebox{\includegraphics[width=0.13\linewidth]{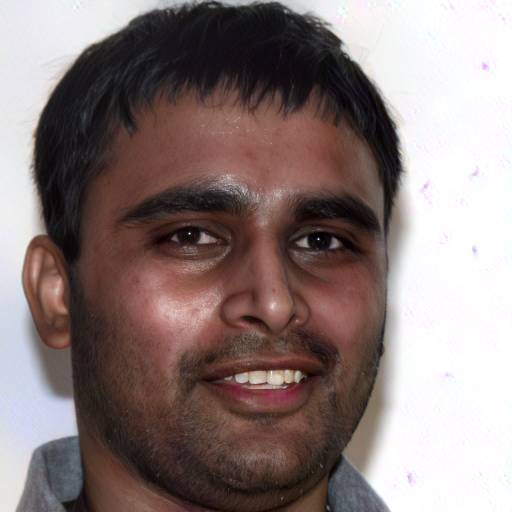}}}
    \subfigure[Ours]{\framebox{\includegraphics[width=0.13\linewidth]{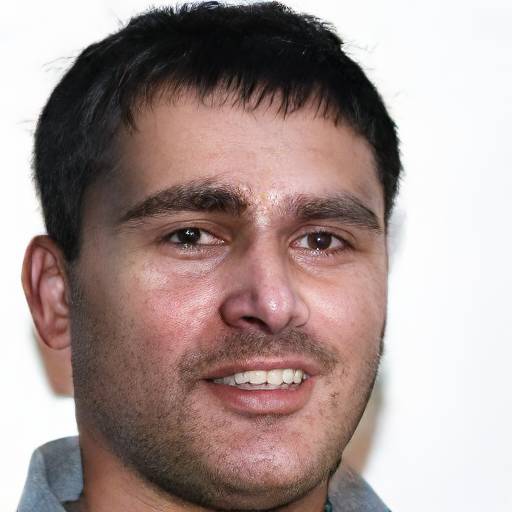}}} \\
    
    \end{tabular}
    \caption{Comparisons with state-of-the-art methods for face style transfer. In each row, (a) is a geometry reference image and (b) is an appearance reference image. (c)$\sim$(f) are the results by the existing methods and our results are shown in (g). WCT$^2$ could better preserve the geometry of (a), but cannot well reconstruct the appearance. As shown in the third example, DST cannot capture the appearance for the hair region. Our method achieves visually the best results. Original images courtesy of ImagineCup, Dan Milner, Willie Williams, Drama League, John Benson, ShashiBellamkonda and Sebastiaan ter Burg.
    }
    \vspace{-2mm}
    \label{fig:compare with style transfer}
\end{figure*}

\item[-] {Cyclic swapping loss:}
To disentangle the geometry and appearance features more thoroughly, we adopt swapping to generate face images from geometry and appearance features that are
from different sources, \i.e. $I_1 \neq I_2$.
The cyclic swapping loss $L_{Swap}$ contains two terms: $L_{Geo}$ and $L_{Cycle}$.
Next we introduce the $L_{Geo}$ and $L_{Cycle}$ in detail.

Our network aims at a complete disentanglement of geometry and appearance feature for a facial image. When generating $I^\prime$ after swapping the appearance of $I_1$ to that of $I_2$, the structure of $I^\prime$ should be maintained. So we introduce geometry loss ${L}_{Geo}$ to keep the geometry unchanged of the generated image by comparing it to input image:

\begin{equation}
        {L}_{Geo} = ||\mathcal{E}_{image}(I_1)- \mathcal{E}_{image}(I^\prime)||_1,
\end{equation}

To ensure the appearance of the swapped image $I^\prime$ is the same as that of $I_2$, we introduce a cycle consistency loss. 
With the geometry of $I_2$ and the appearance of the swapped image $I^\prime$, the generated image $I_2^\prime =\mathcal{G}_{syn}(\mathcal{E}_{image}(I_2), \mathcal{E}_{appearance}(I^\prime))$ should cyclically reconstruct image $I_2$. 
We use the above formulation of reconstruction loss ${L}_{Recon}$ to achieve the cycle consistency constraint:
\begin{equation}
    \begin{aligned}
        {L}_{Cycle} = \alpha_1{L}_{Lab}(I_2,I_2^\prime) + \alpha_2{L}_{FM}(I_2,I_2^\prime) + \alpha_3{L}_{VGG}(I_2,I_2^\prime),\\
    \end{aligned}
\end{equation}
where the hyper-parameters $\alpha_1,\alpha_2,\alpha_3$ are same as the above setting.
Finally, $L_{Swap}$ can be formulated as follows:
\begin{equation}
    L_{Swap}= \tau_1 L_{Geo} + \tau_2 L_{Cycle},
\end{equation}
where we empirically set $\tau_1,\tau_2=1,1$ in our experiments.

\item[-] {Adversarial loss:}
Also, we adopt the muti-scale discriminator $\mathcal{D}$ to encourage the distribution of generated images to match the distribution of  real images:
\begin{equation}
\centering
    \begin{split}
        {L}_{GAN} &= \gamma_1\mathbb{E}[\log\mathcal{D}(I_1)] + \gamma_2\mathbb{E}[\log\mathcal{D}(I_2)]\\ 
        ~&+\gamma_3\mathbb{E}[1-\log(\mathcal{D}(I_1^\prime))] +  \gamma_4\mathbb{E}[1-\log(\mathcal{D}(I_2^\prime))]\\
        ~&+\gamma_5\mathbb{E}[1-\log\mathcal{D}(I^\prime))], \\
    \end{split}
\end{equation}
where we balance the weights of generator $\mathcal{G}_{syn}$ and discriminator $\mathcal{D}$ with $\gamma_1 = \gamma_2 = 0.5,\, \gamma_3 = \gamma_4 = \gamma_5 = 0.33$ in all our experiments.
\end{itemize}

Our final objective $L_{LD}$ is simply the sum of the above three losses (as we have already weighted each term in these losses) and minimizing $L_{LD}$ will lead to the optimization of three networks: $\mathcal{G}_{syn}$, $\mathcal{E}_{appearance}$, and $\mathcal{D}$. $L_{LD}$ is formulated as follows:
\begin{equation}
    L_{LD}(\mathcal{G}_{syn}, \mathcal{E}_{appearance}, \mathcal{D})= L_{Recon} + L_{Swap} + L_{GAN}. 
\end{equation}

\paragraph{{Global Fusion Training}}
With pre-trained \moduleOneFull ~\ modules, we are able to train the \moduleTwoFull ~\ module which fuses the features encoded by \moduleOneFull ~\ modules together and generates the final results. Similar to the previous stage, we use the adversarial loss, feature matching loss, and perceptual loss for the \moduleTwoFull ~\ module. 
Note that we do not need the swapping strategy in this stage since it does not involve any disentanglement.

\section{Experiments}\label{sec:exp}

In this section, we show our experimental setup as well as discuss the results of our experiments. We have done extensive experiments from three aspects, namely comparison with state-of-the-art methods, ablation study and user study. The results show the effectiveness of our proposed method and its superiority to the existing and alternative approaches. 

First, we make extensive comparisons with the state-of-the-art methods.
Since the source of geometry feature for image generation can be either real facial images or sketches, we compare our proposed method with two main branches of image-to-image translation methods: 1) style-transfer-based methods like WCT$^2$~\cite{Yoo_2019_ICCV}, STROTSS~\cite{Kolkin_2019_CVPR},  DST~\cite{kim2020deformable} and StyleGAN2~\cite{Karras2019stylegan2}, where we take real images as inputs and evaluate the ability of face style transfer; 2) sketch-to-image translation methods like pSp~\cite{pSp}, Liu et al. \shortcite{liu2020selfsupervised} and Zhang et al.~\shortcite{zhang2020cross}, where we take sketches as inputs and evaluate the ability of sketch-based face image editing. 
We also compare with some image editing methods based on sketch inputs.
Then, we conduct ablation studies to justify the necessity of each component in our framework.
Finally, a perception study is conducted to test whether the images generated by our method look more appealing to users compared with alternative methods, indicating the visual quality of our controllable face generation results.
As shown in our demo video, our interactive editing interface achieves real-time performance.
All the above experiments are carried out on a PC with an Intel i7-7700CPU, 32GB RAM, and two Nvidia RTX 2080Ti GPUs. 
And DeepFaceEditing is implemented in Pytorch~\cite{NEURIPS2019_9015} and Jittor~\cite{hu2020jittor}.
We use ADAM~\cite{kingma2014adam} with 0.0002 learning rate, $\beta_1=0.5$ and $\beta_2=0.99$. We use the maximum batch size 2 that fits in memory on our GPUs for 512 $\times$ 512 resolution. We will release the code and data for facilitating future research. Please find the details of the network architectures in the supplemental material.

\begin{figure}[ht]
\centering
    \setlength{\fboxrule}{0.5pt}
    \setlength{\fboxsep}{-0.01cm}
    \begin{spacing}{1}
    \begin{tabular}{cc}
        \hspace{-3mm}
        \rotatebox{90}{\small{\hspace{1mm}(a) Geometry}}
        &
        \hspace{-3mm}
        \framebox{\includegraphics[width=0.22\linewidth]{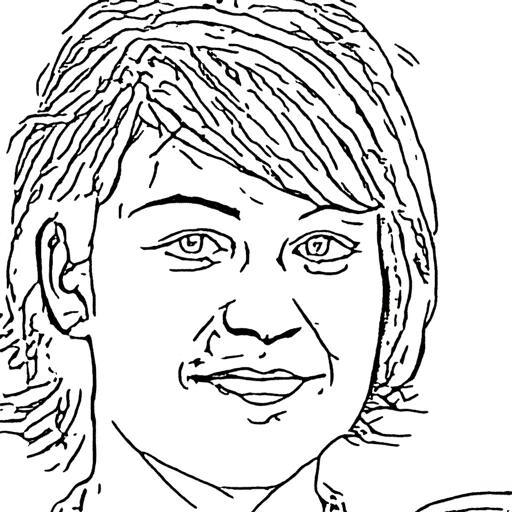}}
        \framebox{\includegraphics[width=0.22\linewidth]{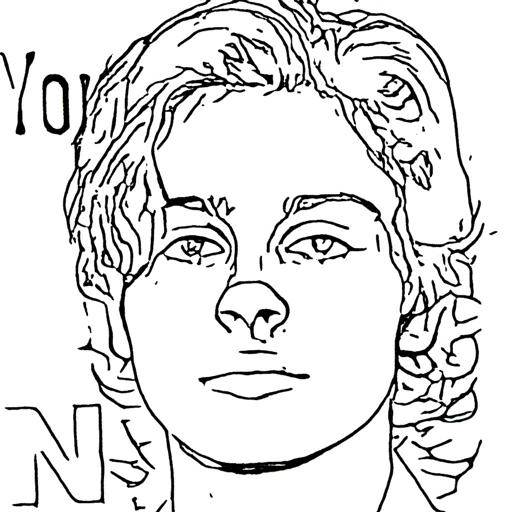}}
        \framebox{\includegraphics[width=0.22\linewidth]{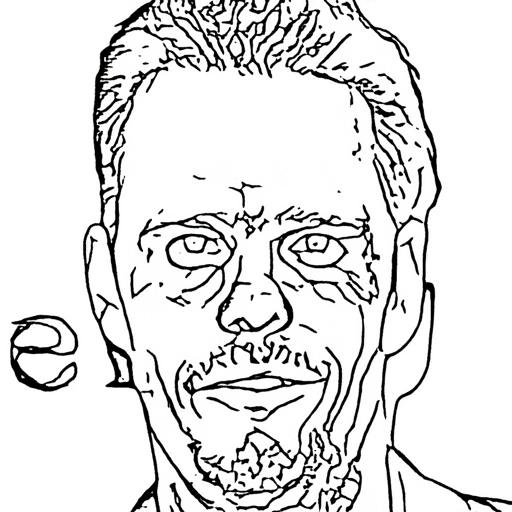}}
        \framebox{\includegraphics[width=0.22\linewidth]{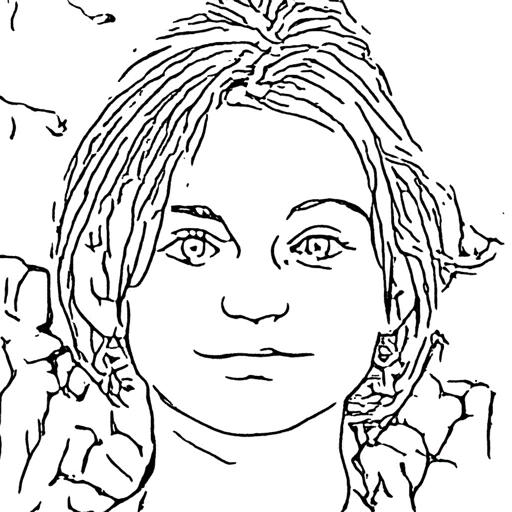}}\\
        
        \hspace{-3mm}
        \rotatebox{90}{\small{\hspace{1mm}(b) Appearance}}
        &
        \hspace{-3mm}
        \framebox{\includegraphics[width=0.22\linewidth]{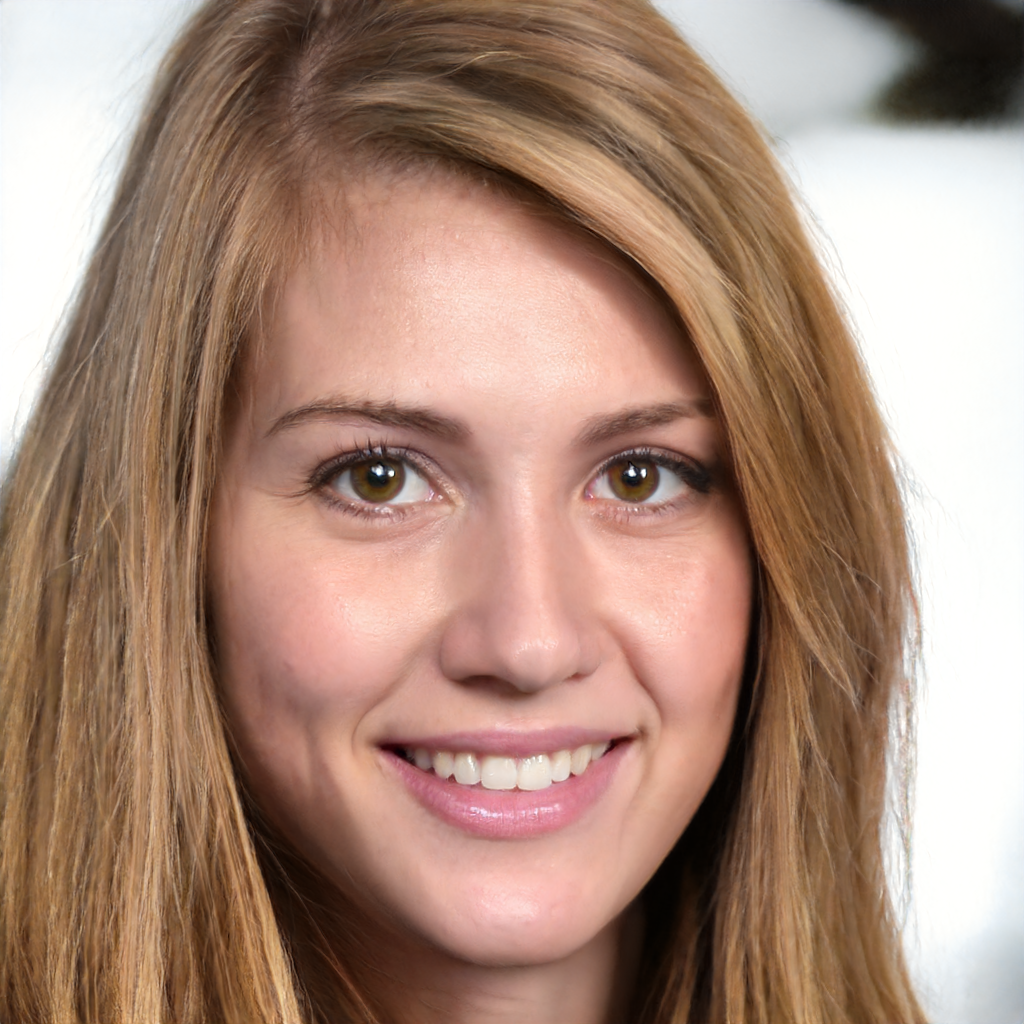}}
        \framebox{\includegraphics[width=0.22\linewidth]{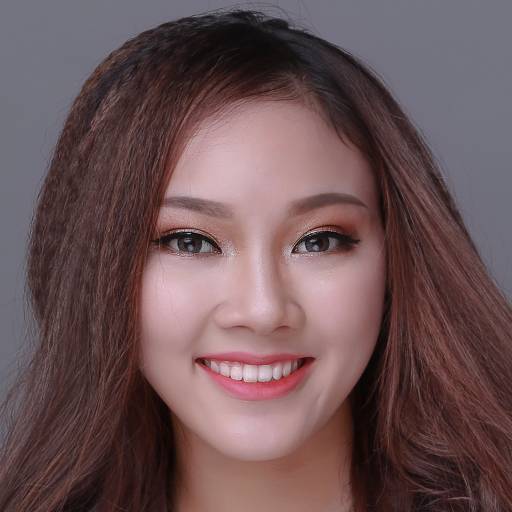}}
        \framebox{\includegraphics[width=0.22\linewidth]{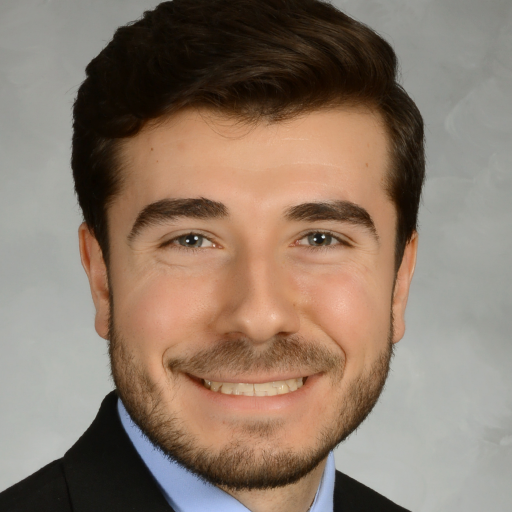}}
        \framebox{\includegraphics[width=0.22\linewidth]{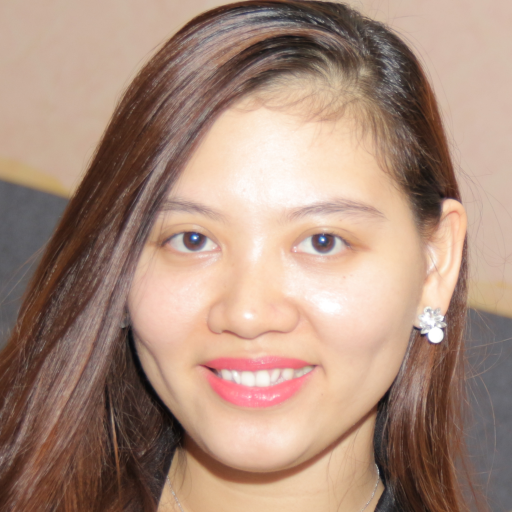}}\\
        
        \hline 
        \vspace{-3mm}
        \\
        
        \hspace{-3mm}
        \rotatebox{90}{\small{\hspace{3mm}(c) pSp.}}
        &
        \hspace{-3mm}
        \framebox{\includegraphics[width=0.22\linewidth]{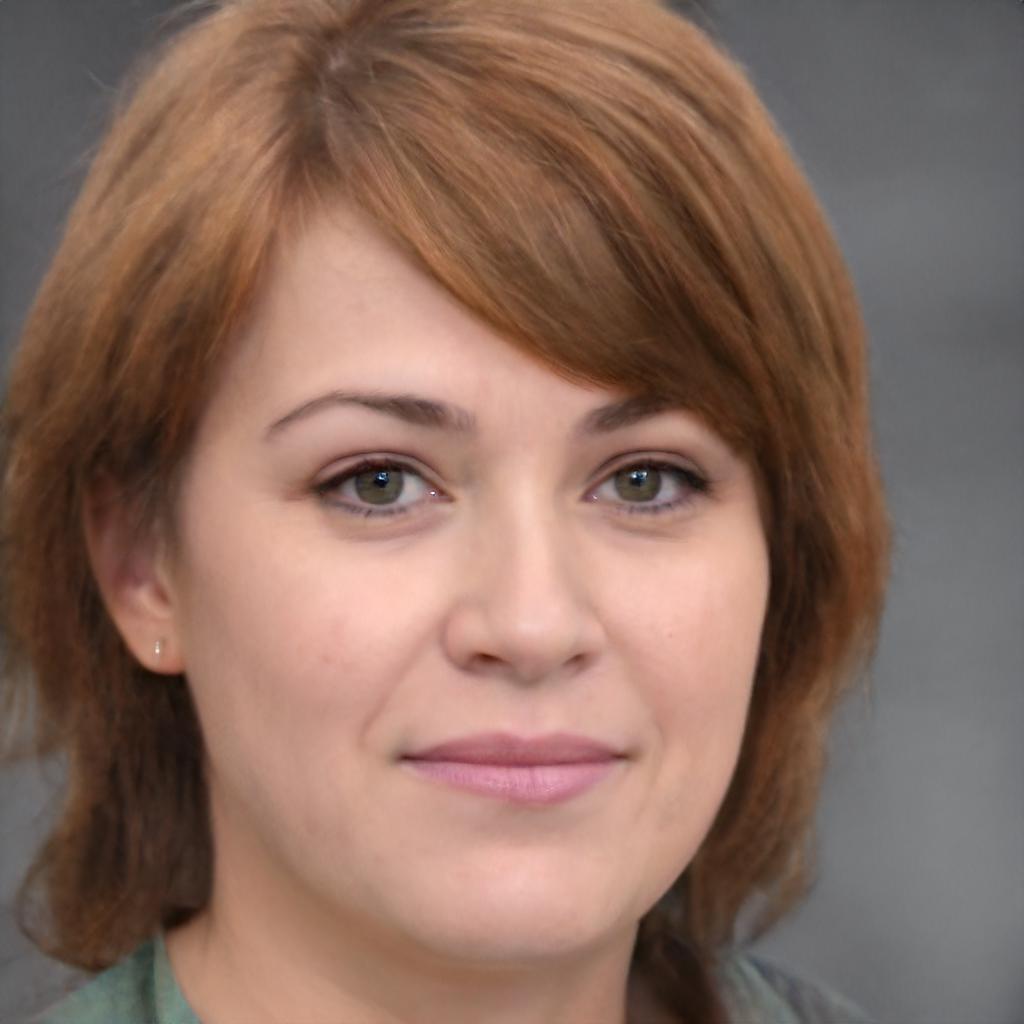}}
        \framebox{\includegraphics[width=0.22\linewidth]{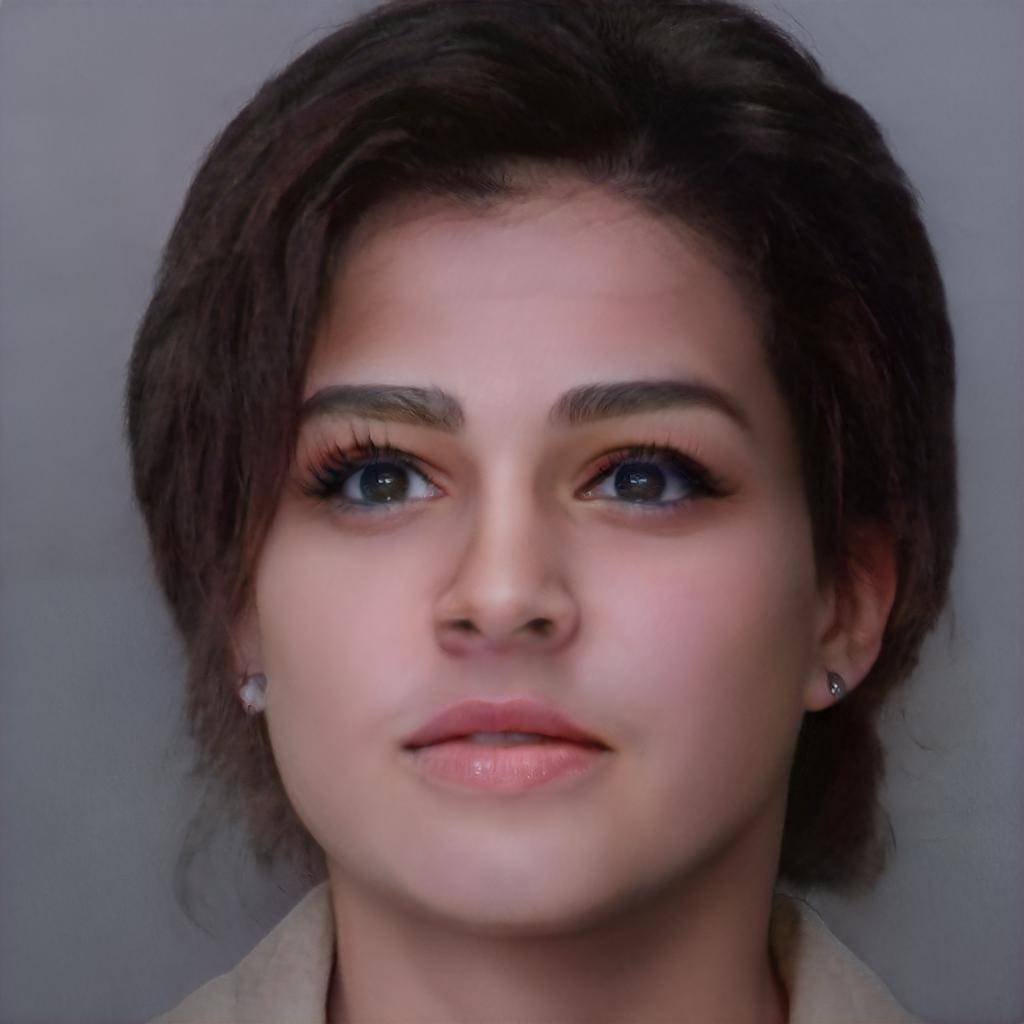}}
        \framebox{\includegraphics[width=0.22\linewidth]{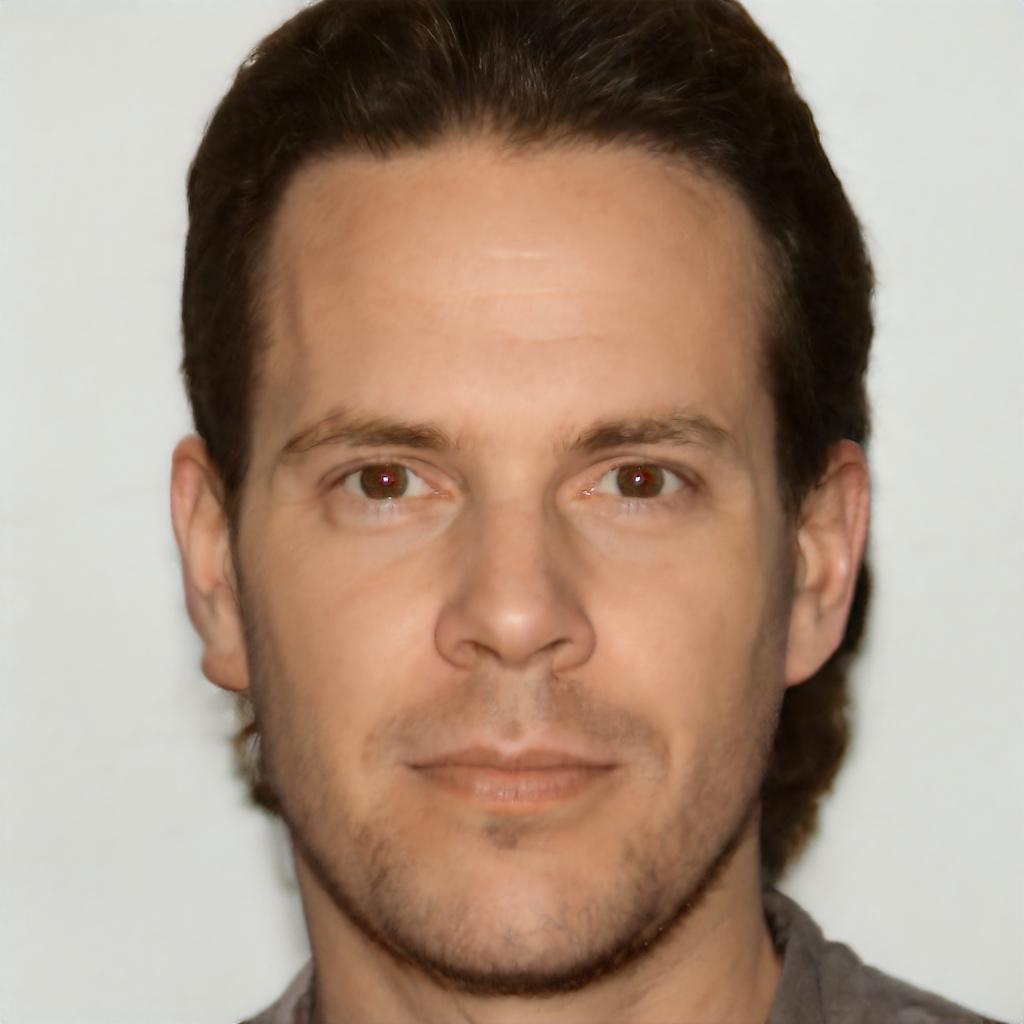}}
        \framebox{\includegraphics[width=0.22\linewidth]{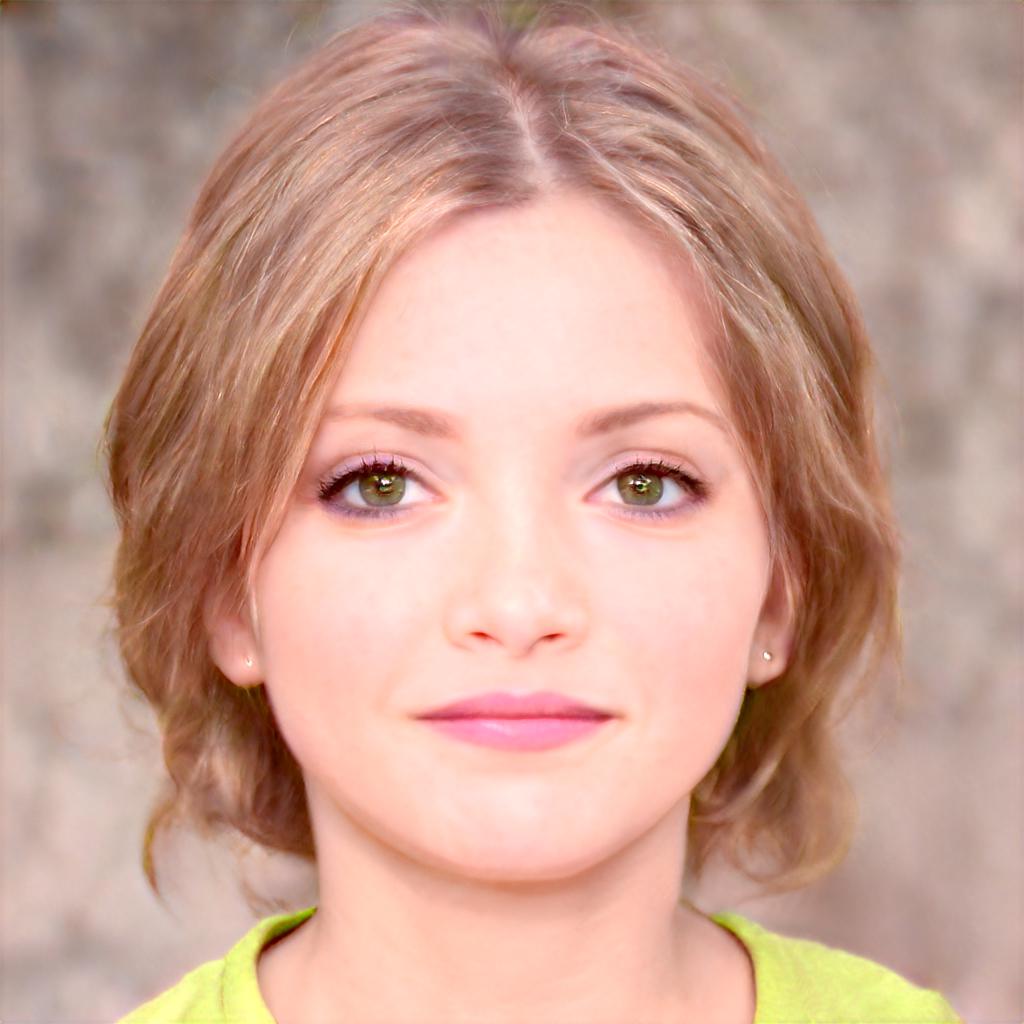}}\\
        
        \hspace{-3mm}
        \rotatebox{90}{\small{\hspace{3mm}(d) Liu et al.}}
        &
        \hspace{-3mm}
        \framebox{\includegraphics[width=0.22\linewidth]{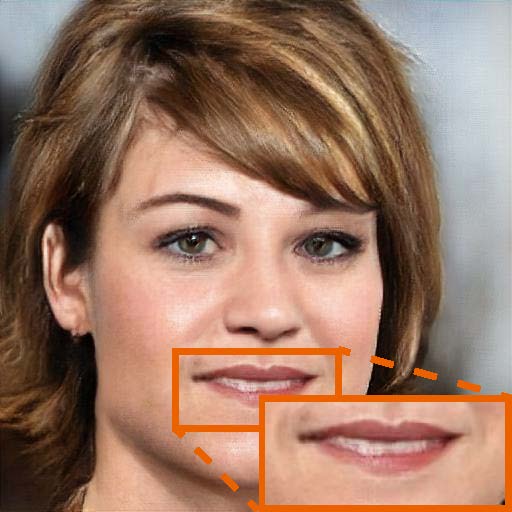}}
        \framebox{\includegraphics[width=0.22\linewidth]{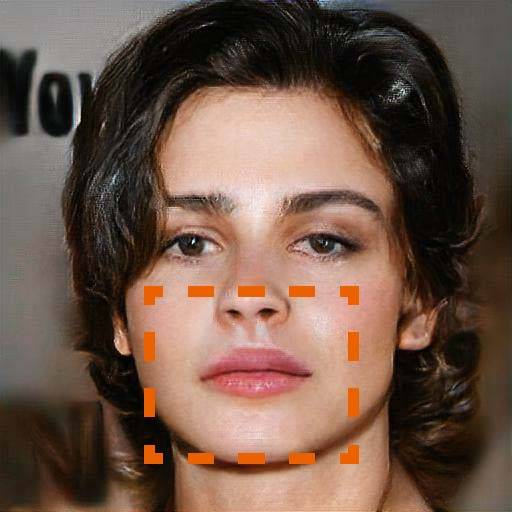}}
        \framebox{\includegraphics[width=0.22\linewidth]{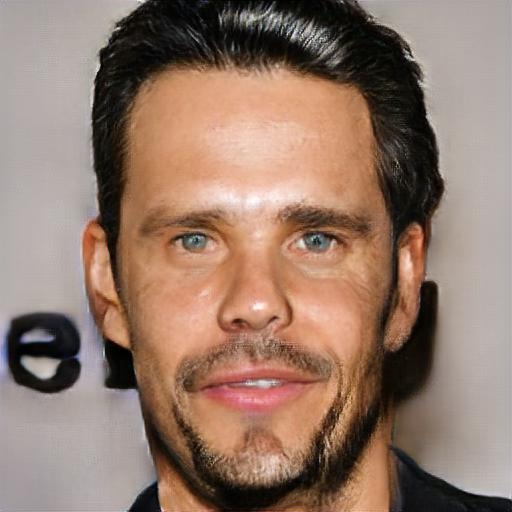}}
        \framebox{\includegraphics[width=0.22\linewidth]{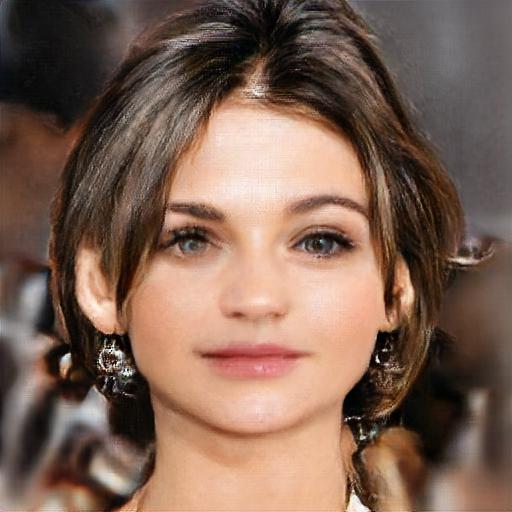}}\\
        
        \hspace{-3mm}
        \rotatebox{90}{\small{\hspace{0mm}(e) Zhang et al.}}
        &
        \hspace{-3mm}
        \framebox{\includegraphics[width=0.22\linewidth]{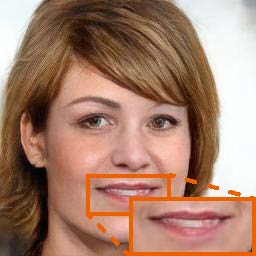}}
        \framebox{\includegraphics[width=0.22\linewidth]{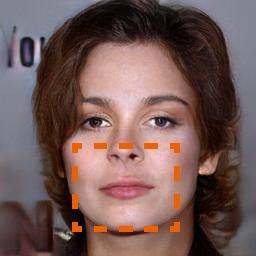}}
        \framebox{\includegraphics[width=0.22\linewidth]{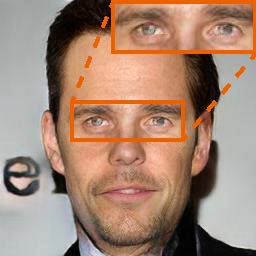}}
        \framebox{\includegraphics[width=0.22\linewidth]{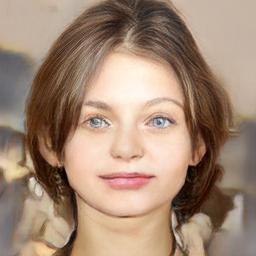}}\\

        \hspace{-3mm}
        \rotatebox{90}{\small{\hspace{4mm}(f) Ours}}
        &
        \hspace{-3mm}
        \framebox{\includegraphics[width=0.22\linewidth]{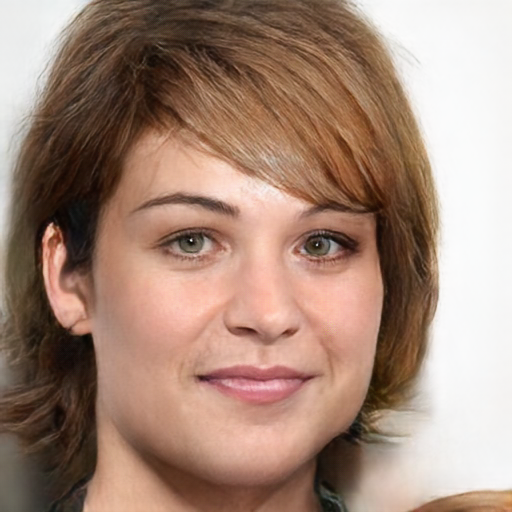}}
        \framebox{\includegraphics[width=0.22\linewidth]{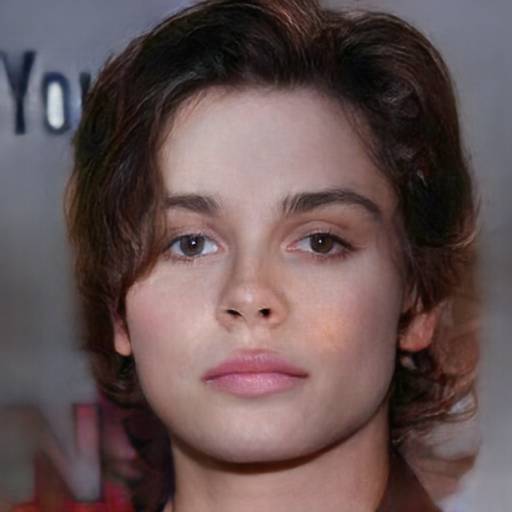}}
        \framebox{\includegraphics[width=0.22\linewidth]{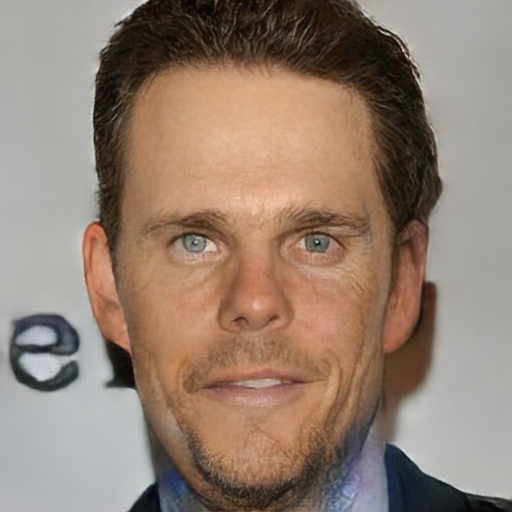}}
        \framebox{\includegraphics[width=0.22\linewidth]{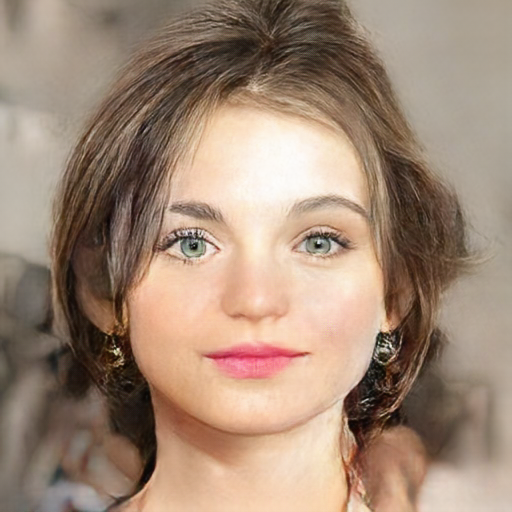}}\\
    \end{tabular}
    \end{spacing}

    \caption{Comparisons with sketch-to-image methods with the same input sketch (a) and appearance reference (b). The results of pSp (c) often {fail to} respect the input sketches, in particular the synthesized hair deviates from what the sketches depict. The results of Liu et al. \shortcite{liu2020selfsupervised} (d) {and Zhang et al.~\shortcite{zhang2020cross} (e) could well preserve} the geometry {indicated in the sketches}, but the former suffers from the changes of appearance, while the latter exhibits some blurry artifacts in local regions.
    Our method is able to generate results with detailed local regions and respect the geometry and appearance references well. Original images courtesy of Vi\d{\^{e}}t Anh Tru$^,$o$^,$ng, pdtghq and Ryan Nguyen.
    \vspace{-2mm}
    }
    \label{fig:compareSketch}
\end{figure}

\begin{figure}[htb]
    \centering
    \setlength{\fboxrule}{0.5pt}
    \setlength{\fboxsep}{-0.01cm}
    \framebox{\includegraphics[width=0.24\linewidth]{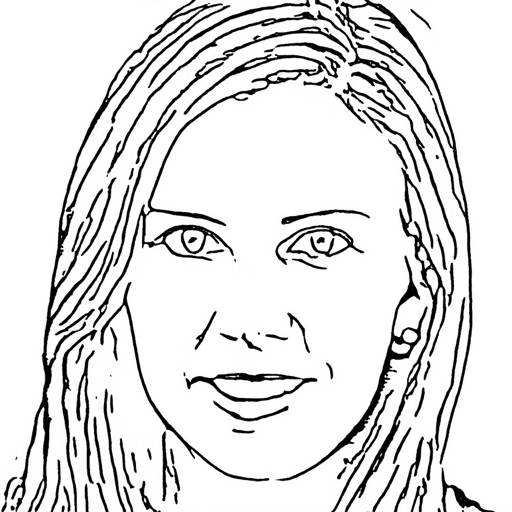}}
    \framebox{\includegraphics[width=0.24\linewidth]{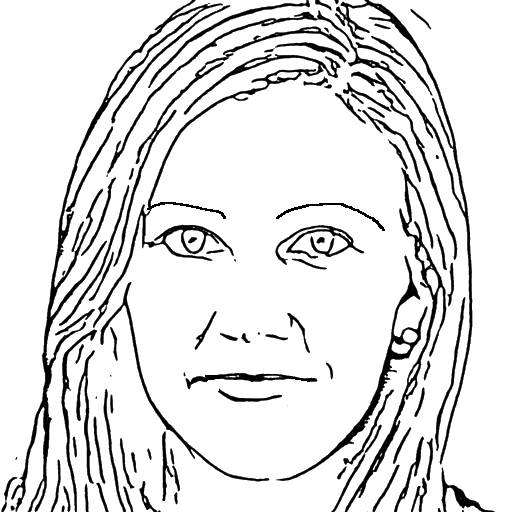}}
    \framebox{\includegraphics[width=0.24\linewidth]{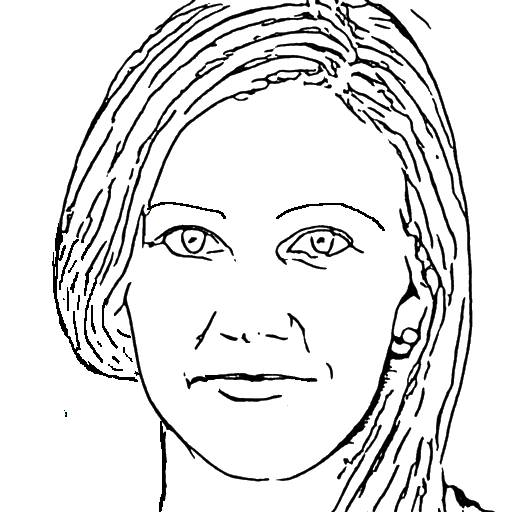}}
    \framebox{\includegraphics[width=0.24\linewidth]{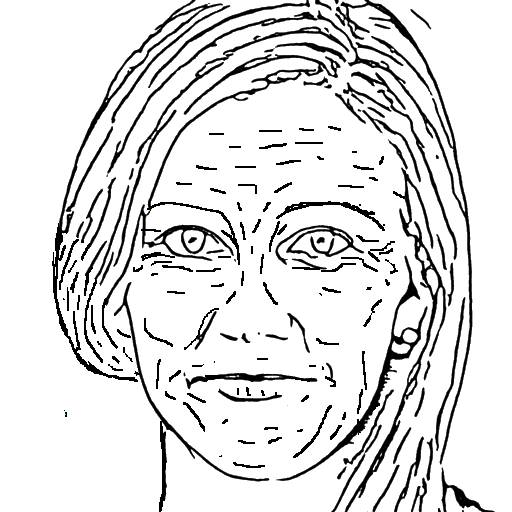}}\\
    
    \framebox{\includegraphics[width=0.24\linewidth]{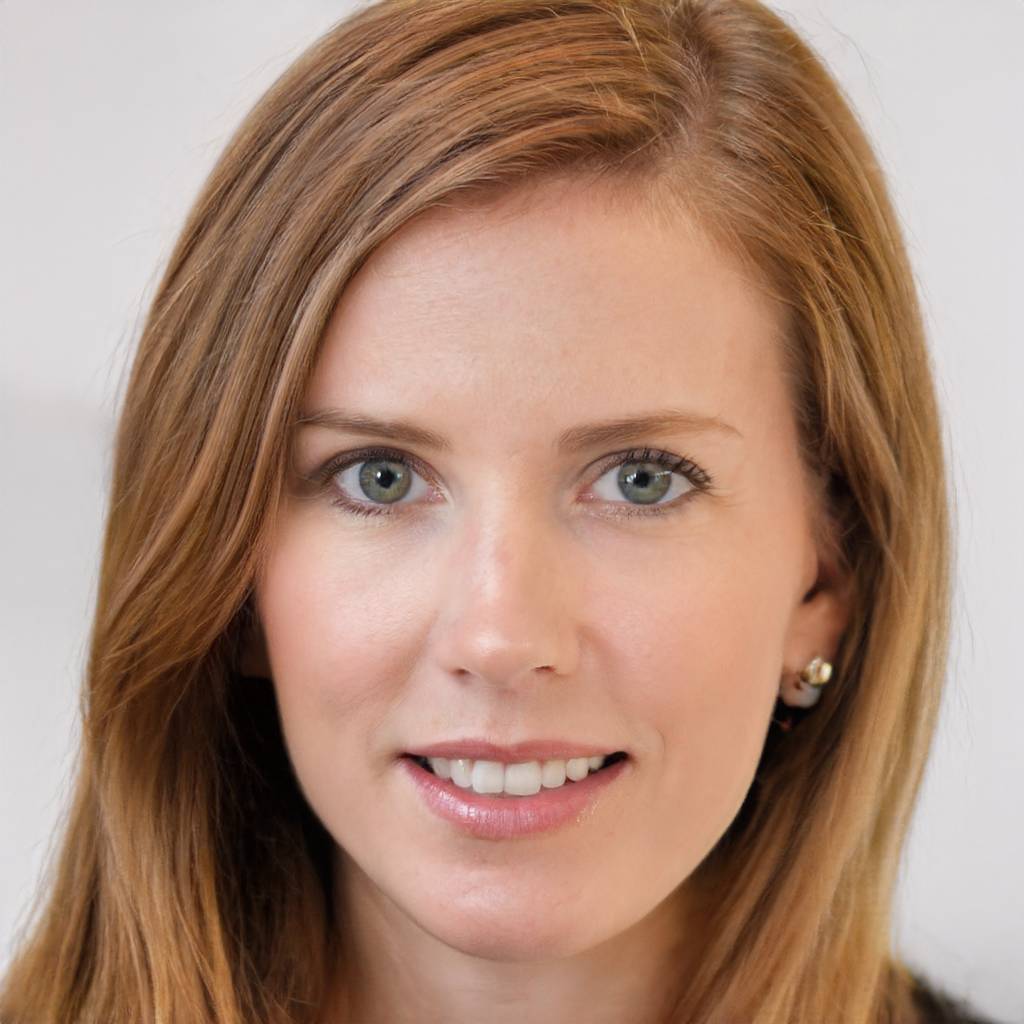}}
    \framebox{\includegraphics[width=0.24\linewidth]{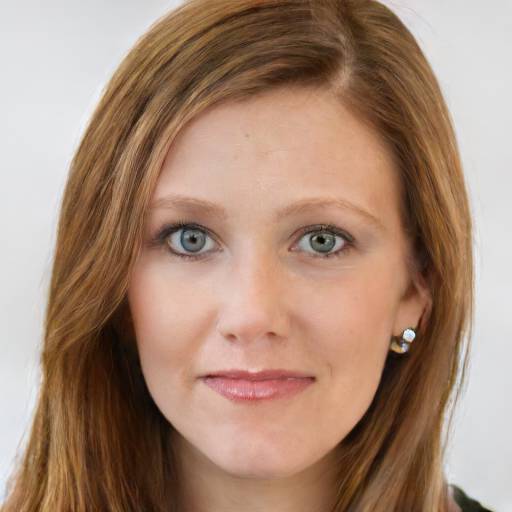}}
    \framebox{\includegraphics[width=0.24\linewidth]{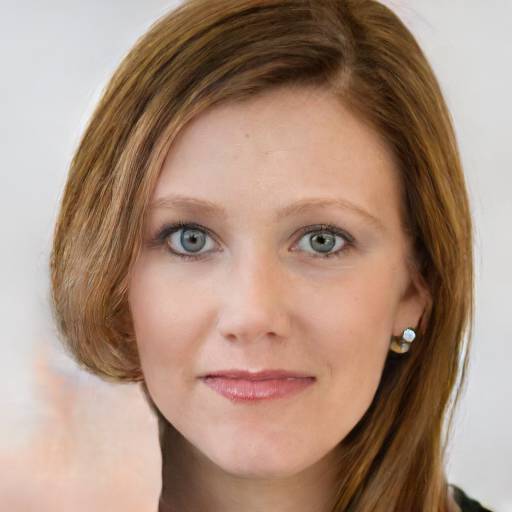}}
    \framebox{\includegraphics[width=0.24\linewidth]{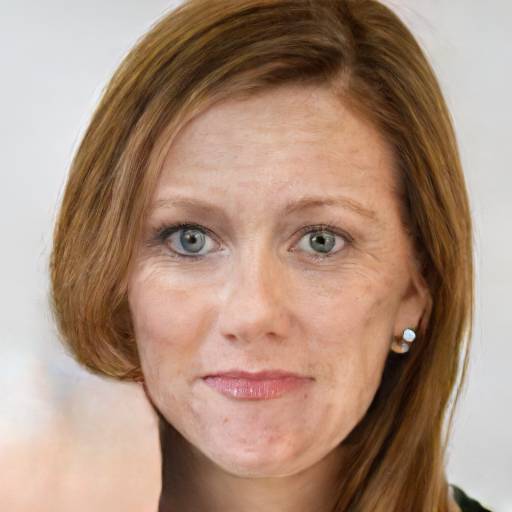}}\\

    \caption{
    Our model supports flexible face editing based on sketches, without requiring users to pre-define a region of interest. Thus our method provides users with more space of freedom and creativity. This figure shows a sequence of edited sketches and their corresponding generated face images.
    The first column shows the original image and its corresponding geometry.
    From left to right, we gradually add wrinkles on the sketch (through changing the geometry) while maintaining the appearance feature, achieving the effect of gradual aging.
    }
    \vspace{-2mm}
    \label{fig:edit squence}
\end{figure}

\begin{figure}[htb]
    \centering
    \setlength{\fboxrule}{0.5pt}
    \setlength{\fboxsep}{-0.01cm}
    
    \includegraphics[width=0.23\linewidth]{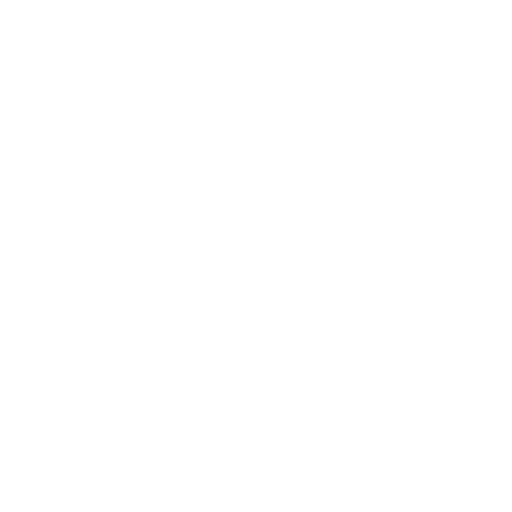}
    \framebox{\includegraphics[width=0.23\linewidth]{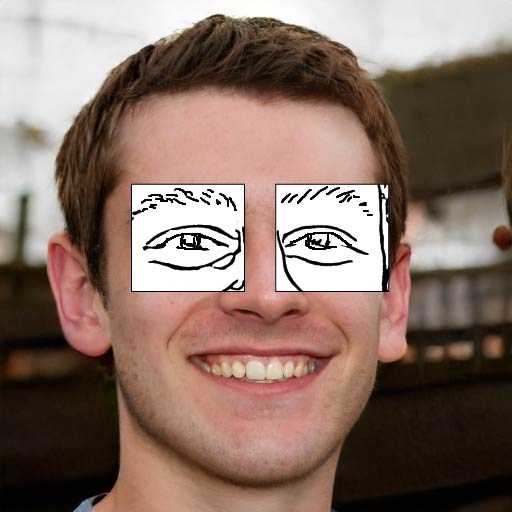}}
    \framebox{\includegraphics[width=0.23\linewidth]{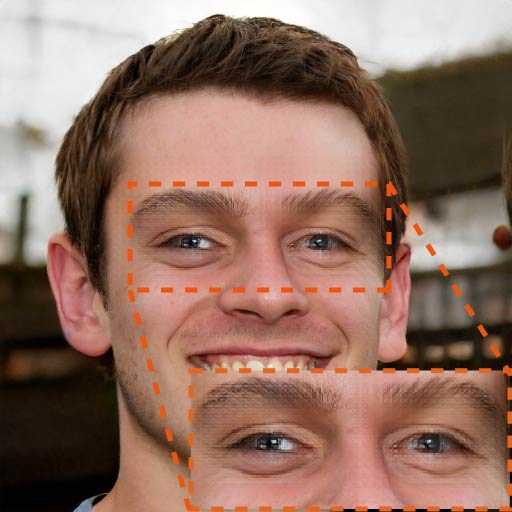}}
    \framebox{\includegraphics[width=0.23\linewidth]{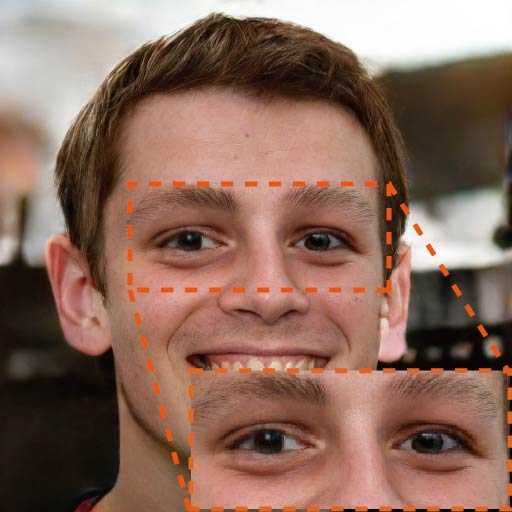}}
    \\
    
    \vspace{-1mm}
    \subfigure[Original  Image]{\framebox{\includegraphics[width=0.23\linewidth]{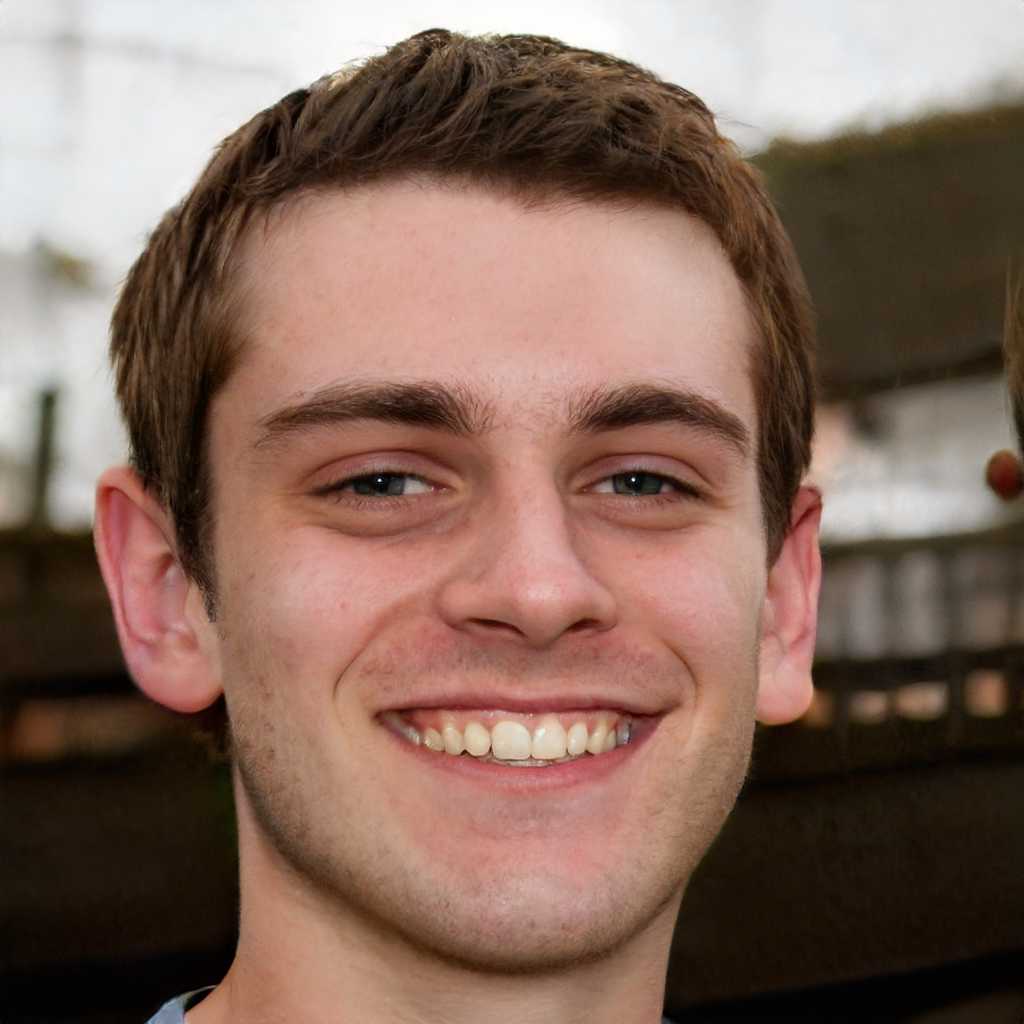}}}
    \framebox{\includegraphics[width=0.23\linewidth]{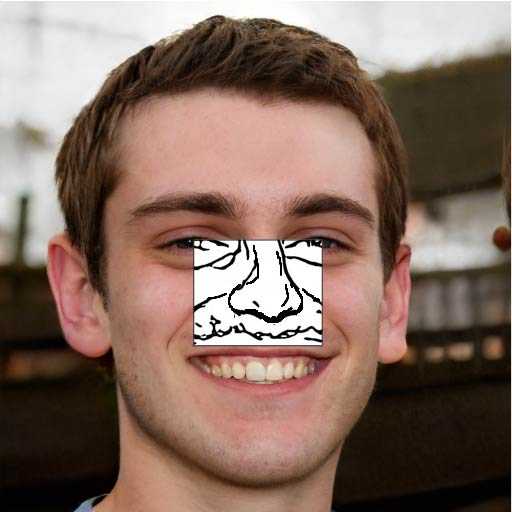}}
    \framebox{\includegraphics[width=0.23\linewidth]{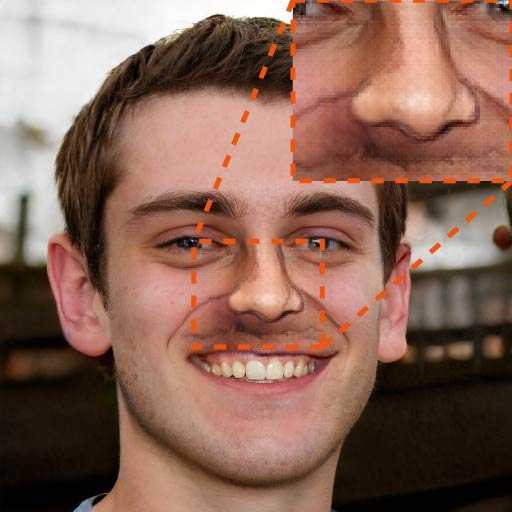}}
    \framebox{\includegraphics[width=0.23\linewidth]{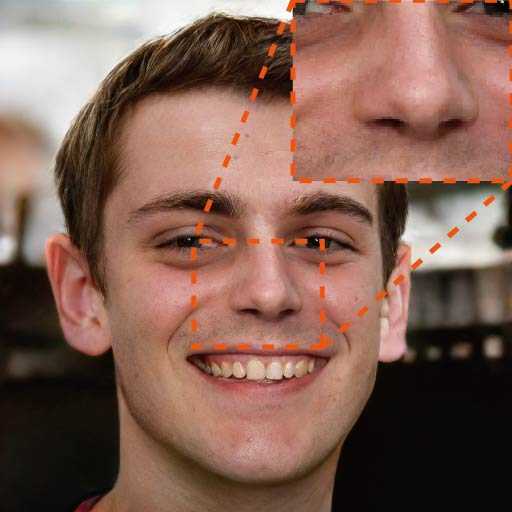}}
    \\
    
    \vspace{-7mm}
    \includegraphics[width=0.23\linewidth]{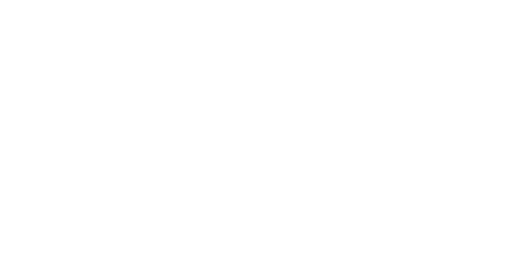}
    \subfigure[Edit  Image]{\framebox{\includegraphics[width=0.23\linewidth]{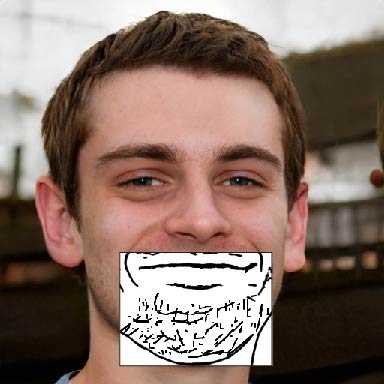}}}
    \subfigure[SC-FEGAN]{\framebox{\includegraphics[width=0.23\linewidth]{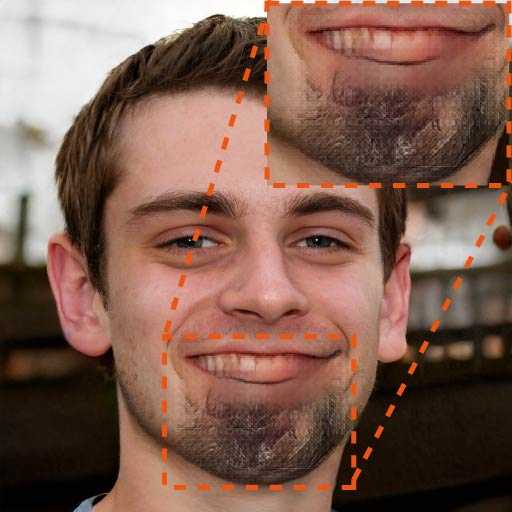}}}
    \subfigure[Ours]{\framebox{\includegraphics[width=0.23\linewidth]{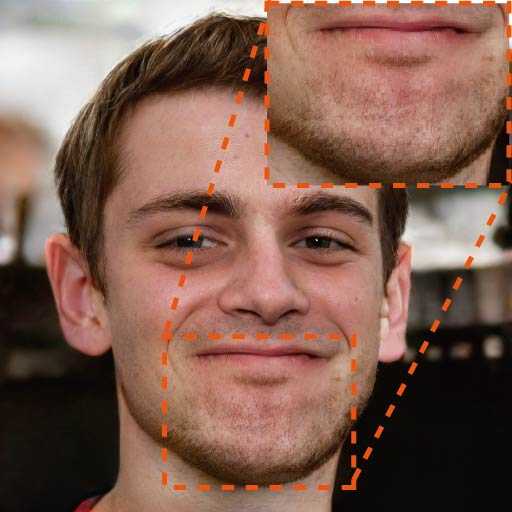}}}
    
    \caption{Sketch-based face editing results. Our method leads to more realistic results than SC-FEGAN, a state-of-the-art approach based on image completion.
    }
    \label{fig:sketch editing generation}
\end{figure}

\subsection{Results and Evaluations}

\paragraph{Face Style Transfer} 
In our framework, we can extract geometry features from either images or sketches.
Hence, we can disentangle the geometry and appearance of an image, and generate new images by swapping geometry and/or appearance.
This allows us to compare our method with state-of-the-art style transfer methods, including WCT$^2$~\cite{Yoo_2019_ICCV}, STROTSS~\cite{Kolkin_2019_CVPR},  DST~\cite{kim2020deformable}, and StyleGAN2~\cite{Karras2019stylegan2}.
We use the official implementations for all the above methods.
By default, $512 \times 512$ resolution is adopted, except that StyleGAN2 uses input and output images of $1024\times1024$.
The comparisons between all the methods are shown in Fig.~\ref{fig:compare with style transfer}.
Our method is capable of generating high-quality photo-realistic results and combining the geometry and appearance of different images without obvious artifacts, and outperforms state-of-the-art style-transfer methods.
STROTSS~\cite{Kolkin_2019_CVPR} overall can capture the style of appearance images, but their method does not produce spatially consistent transfer results, leading to salient artifacts. 
DST~\cite{kim2020deformable} generates high-quality results, but their results fail to capture the appearance of the reference images, possibly due to their strong geometry constraint. For StyleGAN2~\cite{Karras2019stylegan2}, the identities of the generated faces are changed, since their projected latent code could not reconstruct the images accurately and the mixing operation can affect the geometry of the generated images. 
STROTSS~\cite{Kolkin_2019_CVPR}, DST~\cite{kim2020deformable} and StyleGAN2~\cite{Karras2019stylegan2} all require an optimization process to generate style transfer images or projected latent code, thus cannot be real-time. 
Different from the above three algorithms, WCT$^2$~\cite{Yoo_2019_ICCV} is a fast feed-forward method without any optimization for style transfer.
Despite the realistic face images generated by WCT$^2$~\cite{Yoo_2019_ICCV}, the style transfer results of their method show a mixed appearance of two reference images without disentangled geometry and appearance.

\begin{figure}
    \centering
    \includegraphics[width=0.95\linewidth]{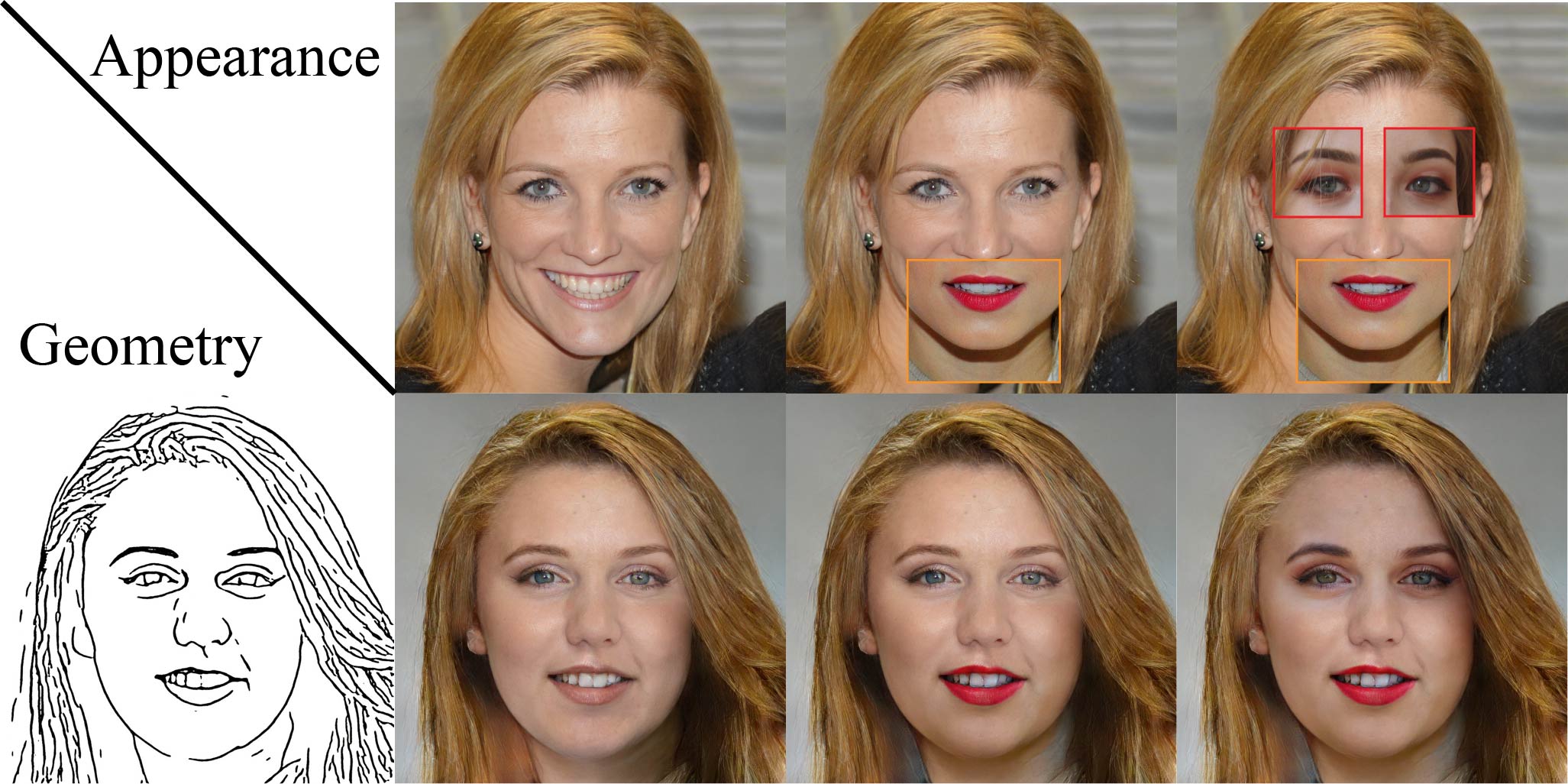}
    \caption{Our model enables face appearance editing, where we fix the sketch (therefore fixing the geometry feature) and replace target components of the face image. The model is able to robustly fuse the newly added component-level image patches with the original face component images and generate photo-realistic edited face images.
    }
    \label{fig:appearance_editting}
\end{figure}

\begin{figure}
    \centering
    \setlength{\fboxrule}{0.5pt}
    \setlength{\fboxsep}{-0.01cm}
    \begin{spacing}{1}
    \begin{tabular}{cc}
        \hspace{-3mm}
        \rotatebox{90}{\small{\hspace{1mm}Input Sketch}}
        &
        \hspace{-3mm}
        \framebox{\includegraphics[width=0.22\linewidth]{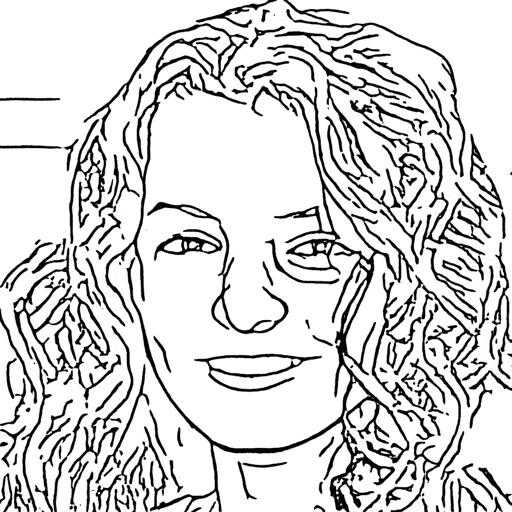}}
        \framebox{\includegraphics[width=0.22\linewidth]{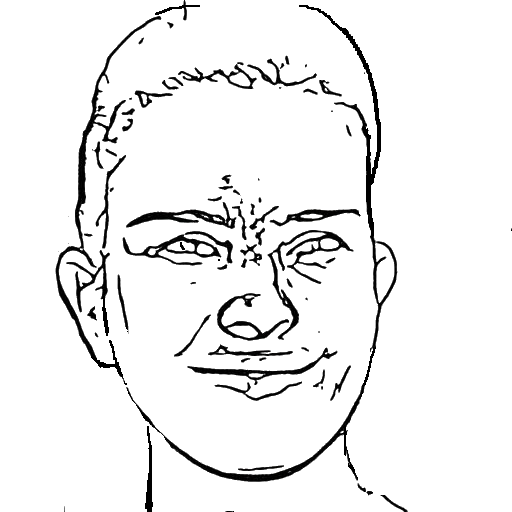}}
        \framebox{\includegraphics[width=0.22\linewidth]{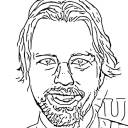}}
        \framebox{\includegraphics[width=0.22\linewidth]{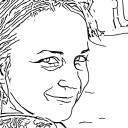}}\\
        
        \hspace{-3mm}
        \rotatebox{90}{\small{\hspace{2.5mm}Appearance}}
        &
        \hspace{-3mm}
        \framebox{\includegraphics[width=0.22\linewidth]{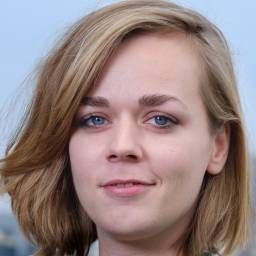}}
        \framebox{\includegraphics[width=0.22\linewidth]{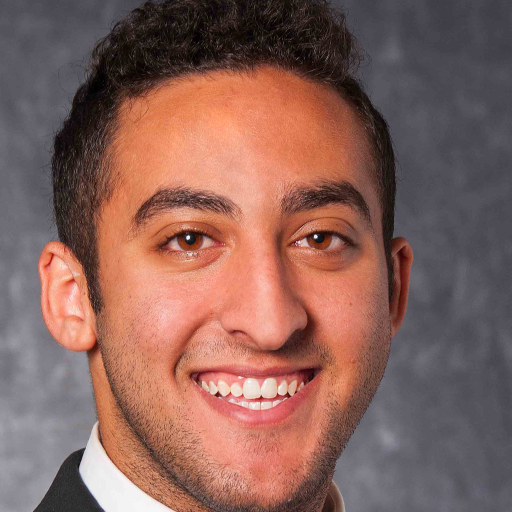}}
        \framebox{\includegraphics[width=0.22\linewidth]{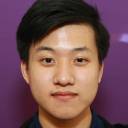}}
        \framebox{\includegraphics[width=0.22\linewidth]{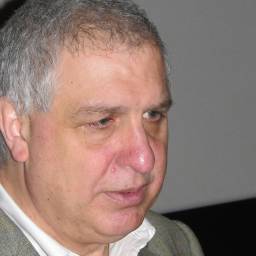}}\\
        
        \hspace{-3mm}
        \rotatebox{90}{\small{\hspace{3mm} w/o local}}
        &
        \hspace{-3mm}
        \framebox{\includegraphics[width=0.22\linewidth]{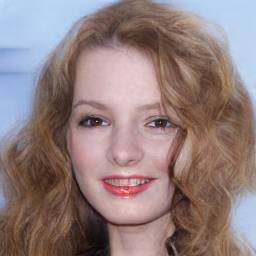}}
        \framebox{\includegraphics[width=0.22\linewidth]{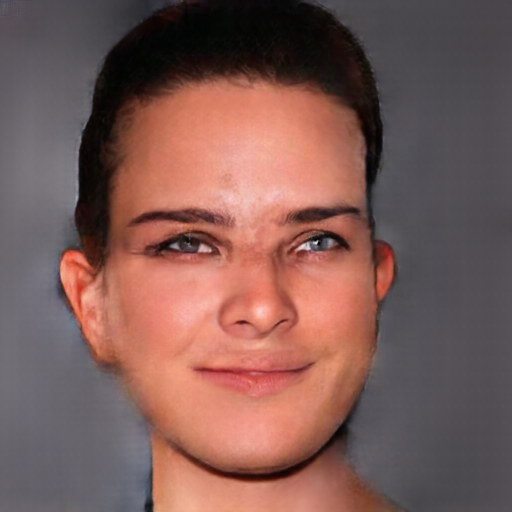}}
        \framebox{\includegraphics[width=0.22\linewidth]{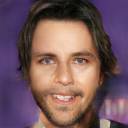}}
         \framebox{\includegraphics[width=0.22\linewidth]{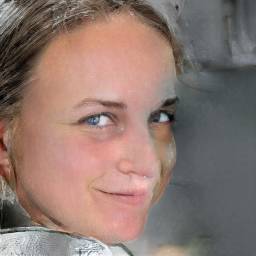}}\\
        
        \hspace{-3mm}
        \rotatebox{90}{\small{\hspace{-3mm} w/o swap \& $L_{cycle}$}}
        &
        \hspace{-3mm}
        \framebox{\includegraphics[width=0.22\linewidth]{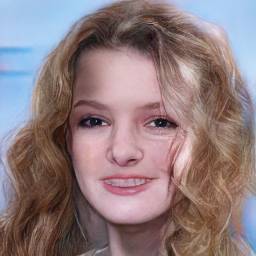}}
        \framebox{\includegraphics[width=0.22\linewidth]{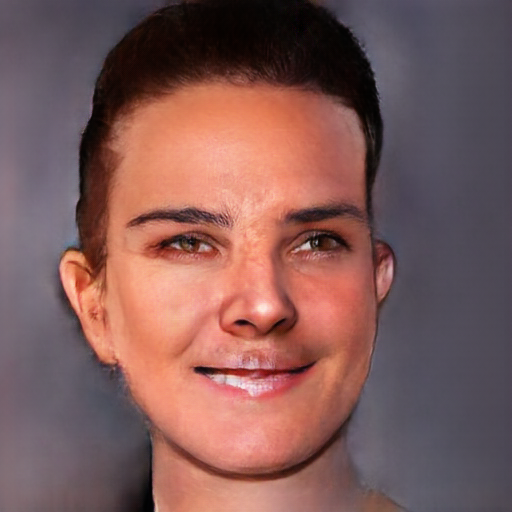}}
        \framebox{\includegraphics[width=0.22\linewidth]{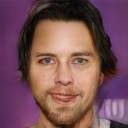}}
        \framebox{\includegraphics[width=0.22\linewidth]{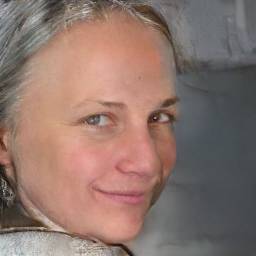}}\\

        \hspace{-3mm}
        \rotatebox{90}{\small{\hspace{2mm}w/o $L_{cycle}$}}
        &
        \hspace{-3mm}
        \framebox{\includegraphics[width=0.22\linewidth]{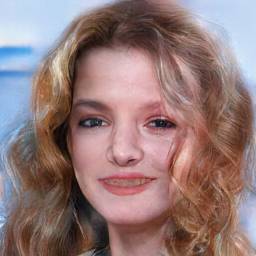}}
        \framebox{\includegraphics[width=0.22\linewidth]{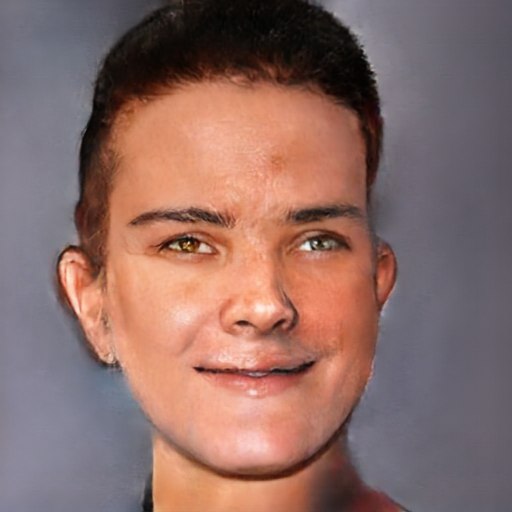}}
        \framebox{\includegraphics[width=0.22\linewidth]{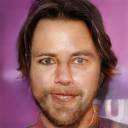}}
        \framebox{\includegraphics[width=0.22\linewidth]{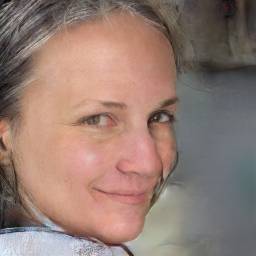}}\\

        \hspace{-3mm}
        \rotatebox{90}{\small{\hspace{6mm}Ours}}
        &
        \hspace{-3mm}
        \framebox{\includegraphics[width=0.22\linewidth]{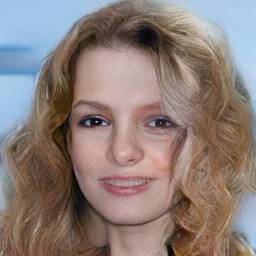}}
        \framebox{\includegraphics[width=0.22\linewidth]{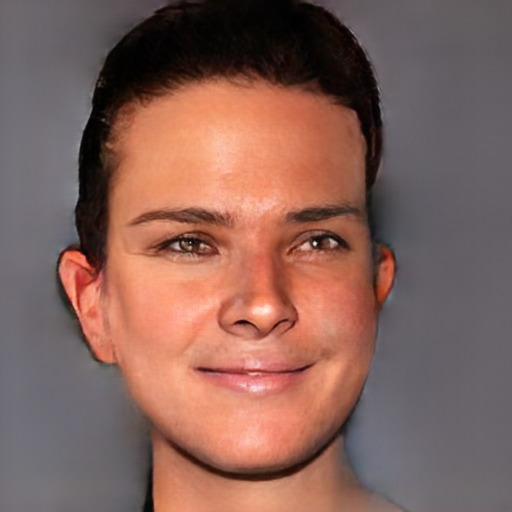}}
        \framebox{\includegraphics[width=0.22\linewidth]{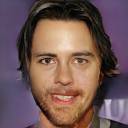}}
        \framebox{\includegraphics[width=0.22\linewidth]{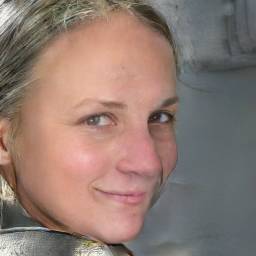}}\\

    \end{tabular}
    \end{spacing}
    \vspace{-4mm}
    \caption{The results of the ablation study.
    ``w/o local'', ``w/o swap \& $L_{Cycle}$'', and ``w/o $L_{Cycle}$'' are our method without the local-to-global strategy,  without swapping and cyclic loss, and without the cyclic loss, respectively.
    Original images courtesy of Sebastiaan ter Burg, Dan Milner, ImagineCup and Stuart Caie.
    }
    \label{fig:ablation}
\end{figure}

\paragraph{{Sketch-based Face Image Synthesis}}
As our method can directly extract the geometry feature from a sketch and combine it with the appearance feature extracted from a reference image, it can be used for sketch-to-image translation. As shown in Fig.~\ref{fig:compareSketch}, we thus compare our method with the existing solutions for this task, including
pSp~\cite{pSp}, \cite{liu2020selfsupervised}, and \cite{zhang2020cross}.
For a fair comparison, we use their official source code but re-train their networks on our training dataset. 
pSp combines an encoder with the StyleGAN decoder and can be applied to sketch-to-image translation. However, since the sketch geometry is encoded into the latent code, generated faces by pSp often do not faithfully respect the input sketches.
The style mixing operation adopted in pSp also affects the geometry of synthesized faces undesirably.
Liu et al. \shortcite{liu2020selfsupervised} design a general network to translate a sketch to an image given a reference image.
The geometry of their synthesized faces is maintained well while the appearance shows variations from the reference images (see the color of mouth). 
Zhang et al.~\shortcite{zhang2020cross} propose a cross-domain correspondence network by finding a dense mapping between an input image and a reference image, and then use a warped exemplar to generate results, which, however, may lack the control of details. 
There are obvious artifacts in the results of Zhang et al. \shortcite{zhang2020cross}, e.g., the fuzzy nose and the blurred eyes shown in the second and third columns of Fig.~\ref{fig:compareSketch} ({e}).
It is easy to see that our method combines the geometry and appearance features consistently and generates realistic and detailed results (e.g., the resulting eyebrows, mouth and gradually changed hair). More results generated by our method are shown in the supplementary material with diverse geometry and appearance.

\paragraph{{Face Editing}}
As for face image editing, we compare our method with the state-of-the-art technique SC-FEGAN~\shortcite{Jo_2019_ICCV}.
Taking sketches as inputs, our method is able to edit the face images based on modifications to sketches, both globally and locally. 
In contrast, SC-FEGAN~\shortcite{Jo_2019_ICCV} is not able to synthesize a face image entirely from an input sketch since their method is based on image completion. 
For a fair comparison, we thus only edit partial sketches and set paired rectangle masks for SC-FEGAN.
Without released training code from the authors, we use their official editing GUI with pre-trained model for comparison.
As shown in Fig.~\ref{fig:sketch editing generation}, our method produces more realistic and visually consistent results than SC-FEGAN (see the second row for an edited nose). 
In addition, their method requires a pre-defined mask for users to specify an area of interest, thus constraining the user experience.
Pre-defining such areas sometimes is not easy, especially for detailed lines.
For example, in Fig.~\ref{fig:edit squence}, we draw lines around the chin to generate wrinkles, resulting in lighting and shadow variations.
In such cases it is difficult for users to pre-define areas of variations.

With disentangled geometry and appearance, we can also edit the global and/or local appearance without changing the geometry of the face.
As shown in Fig.~\ref{fig:appearance_editting}, given a sketch and appearance image as inputs, our method can generate a face retaining both global sketch wrinkles and appearance colors.
With the fixed sketch, {i.e.} fixed geometry feature, we give two new reference images replacing the mouth and/or two eyes cropped from other images respectively.
Our model is able to robustly fuse the newly added component-level image patches with the original face component images and generate photo-realistic edited face images.

\subsection{Ablation Study}
We conduct an ablation study on the test set to show the impact of individual key components of our system. Since our method is based on a local-to-global framework, we first show the results without the fusion of local parts.
As our swapping strategy contains appearance swapping and cyclic reconstruction, we perform ablation study in two ways, one is to remove swapping and $L_{Cycle}$, the other is to remove $L_{Cycle}$ but keep $L_{Geo}$.
From Fig.~\ref{fig:ablation}, it is obvious that without the local-to-global strategy, the results exhibit artifacts in local details. As shown in the last column, even for non-frontal faces, the local-to-global strategy improves the quality of local details. Results in the fourth row to the last row illustrate that when swapping and cycle consistency loss are utilized, the quality of the synthesized results improves. On the other hand, our baseline model without swapping and $L_{Cycle}$ only successfully performs reconstruction during training, where the geometry reference and appearance reference are matched.
When the appearance or geometry features are changed during testing, the color translation seems to behave like a pixel-by-pixel copy, resulting in obvious inappropriate colorization of pixels. 
These results confirm that this network does poorly in separating the appearance and geometry. By adding the swapping strategy, the baseline without $L_{Cycle}$ avoids such wrong colors, but does not properly combine those two features. To further enhance the information flow, we add the cycle consistency loss $L_{Cycle}$ to decouple the features from the generated results. This constraint effectively improves the quality of image synthesis.

\subsection{Perception Study}

As for the tasks of style transfer and sketch-to-image translation, the face images generated by alternative methods tend to generate reasonably realistic face images (see Figs.~\ref{fig:compare with style transfer},~\ref{fig:compareSketch}), and the visual quality differences are subtle. It is therefore difficult for existing general image quality measures such as FID~\cite{heusel2017gans} and LPIPS~\cite{zhang2018perceptual} to distinguish between their qualities.
To evaluate the visual quality and the faithfulness (\ie, the similarity to the geometry and appearance images) of synthesized faces, we conducted a perception study.

\begin{figure}
    \centering
    \includegraphics[width=0.99\linewidth]{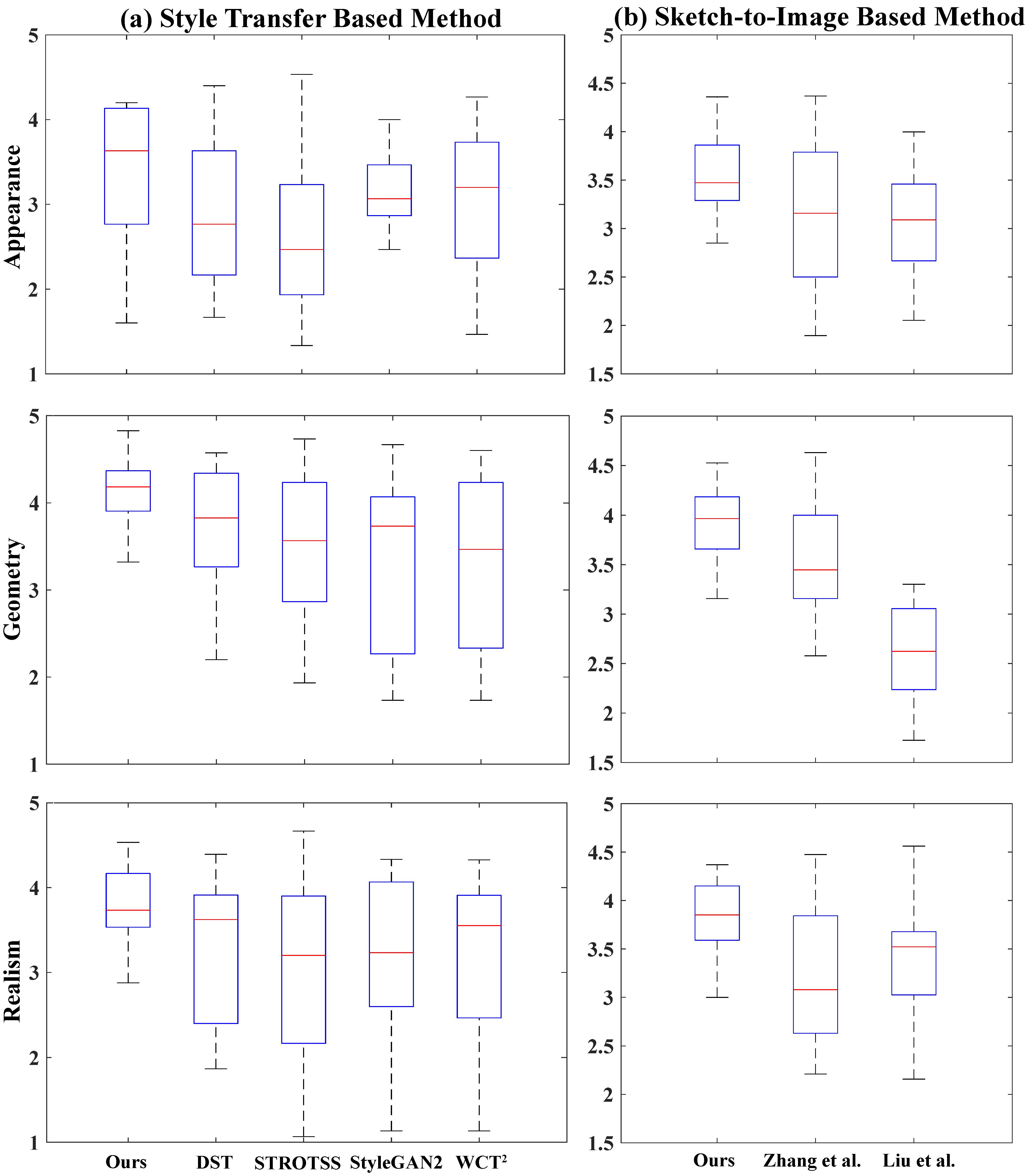}
    \caption{Box plots of the average quality and faithfulness perception scores 
    in terms of similarity between generated faces and the reference faces in appearance and geometry respectively, as well as visual realism of the generated faces, based on the participants in the two user studies.
    (a) The comparison of style transfer with five methods: WCT$^2$~\cite{Yoo_2019_ICCV}, STROTSS~\cite{Kolkin_2019_CVPR},  DST~\cite{kim2020deformable}, StyleGAN2~\cite{Karras2019stylegan2}, and ours.
    (b) the evaluation of sketch-to-image synthesis with Liu et al. \shortcite{liu2020selfsupervised}, Zhang et al. \shortcite{zhang2020cross}, and our method.
    }
    \label{fig:userstudy}
\end{figure}

\begin{figure}
    \centering
    \includegraphics[width=0.99 \linewidth]{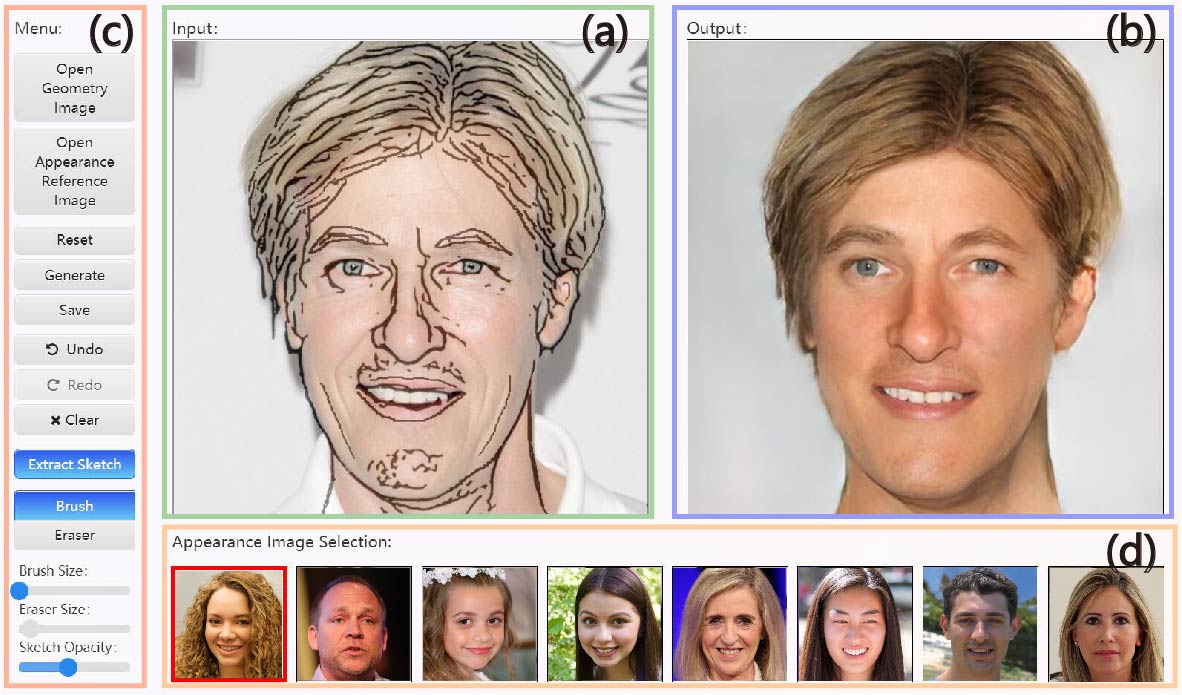}
    \caption{A screenshot of our sketch-based interface for face editing. 
    (a) and (b) are an input sketch overlaid on top of a geometry image and the corresponding generated face image, respectively. (c) and (d) are the control function panel and the appearance image list, respectively.
    Original images courtesy of WRI Ross Center for Sustainable Cities, Pawel Loj, seanccochran, Marcus Harzem, Malcolm Slaney, and TRE - RJ.
    }
    \label{fig:windows}
\end{figure}

\begin{figure*}[htb]
\centering
\includegraphics[width=0.98\linewidth]{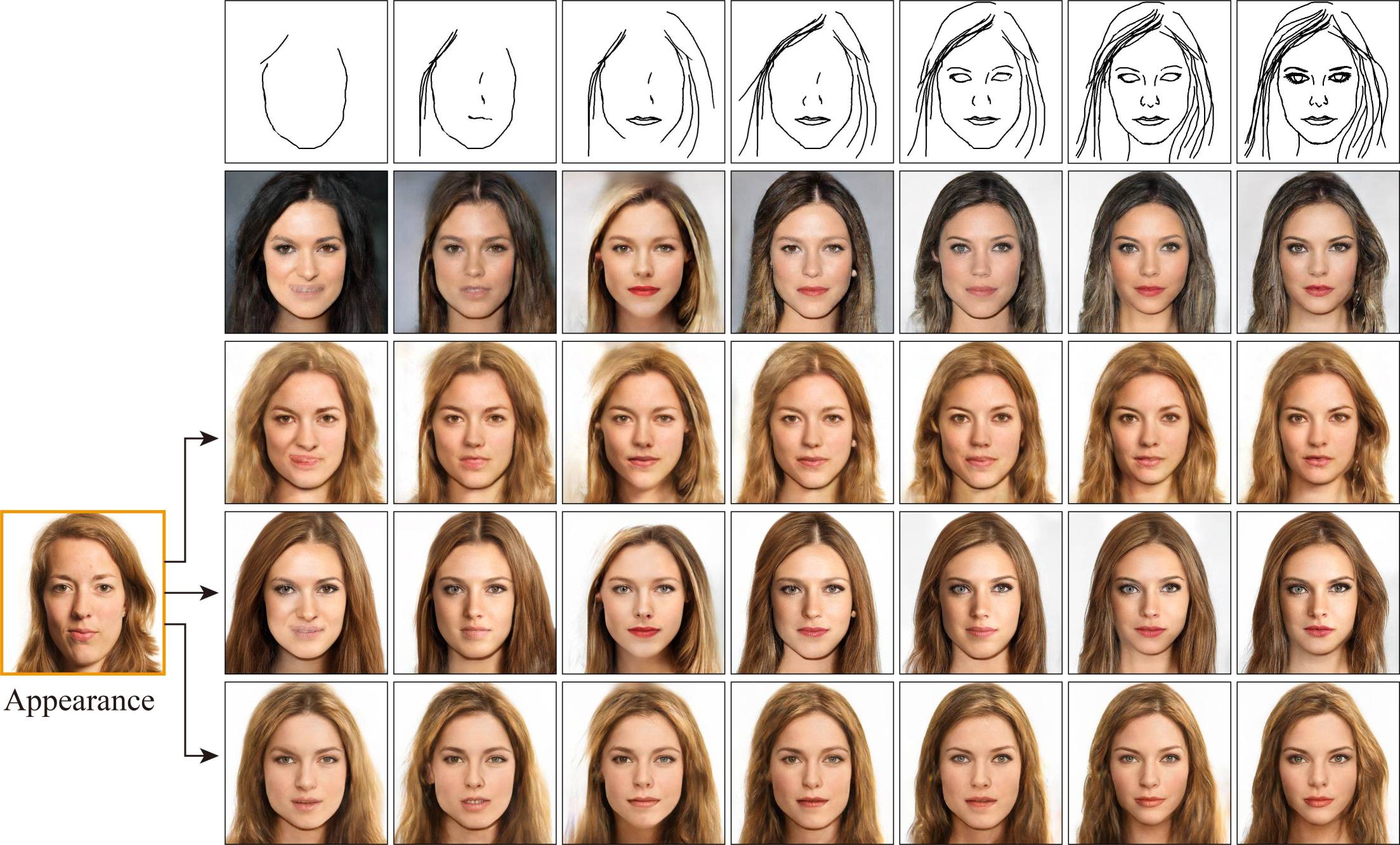}
\caption{
Comparisons of image generation given a sequence of hand-drawn sketches (1st row). The results by DeepFaceDrawing~\cite{10.1145/3386569.3392386} (2nd row) exhibit clear appearance changes during editing. The 3rd row shows the results of using a style transfer technique ~\cite{Kolkin_2019_CVPR} to transfer the style of a reference image (Left) to the images generated by DeepFaceDrawing. The 4th row shows the results of using swapping autoencoder ~\cite{park2020swapping} with the same input as the 3rd row. Taking the sketches (1st row) and the appearance reference (Left) as input, our method (bottom row) generates more stable and realistic results. 
Original images courtesy of Sebastiaan ter Burg.
}
\vspace{-2mm}
\label{fig:sketch-to-image-application}
\end{figure*}

The evaluation was done via two online questionnaires. The first perception study aimed to evaluate the effect of style transfer. We showed two input images, the geometry reference image and the appearance reference image, and five synthesized images (including  WCT$^2$~\cite{Yoo_2019_ICCV}, STROTSS~\cite{Kolkin_2019_CVPR},  DST~\cite{kim2020deformable}, StyleGAN2~\cite{Karras2019stylegan2} and ours) for each example, placed side by side in a random order to avoid bias.
Each participant was asked to evaluate 20 examples according to three criteria: the maintenance of the geometry reference, the similarity to the appearance reference and the realism of generated images, each in a five-point Likert scale (1 = strongly negative to 5 = strongly positive).
In total, 40 participants participated in this study and we got 40 (participants) $\times$ 20 (questions) = 800 subjective evaluations for each method.
We performed one-way ANOVA tests on five methods in aspects of `Geometry', `Appearance' and `Realism' corresponding to the three criteria respectively. 
As shown in Fig.~\ref{fig:userstudy}(a), the statistics of the evaluation results
were plotted. 
We found significant effects of our method for all three criteria: geometry ($F_{(4,95)} = 3.4164, p < 0.05$), appearance ($F_{(4,95)} = 2.6196, p < 0.05$) and realism ($F_{(4,95)} = 2.6012, p < 0.05$). 

The second perception study was conducted to evaluate the quality of sketch-to-image translation.
Similar to the former one, for each example we showed the user two input images, the sketch image for geometry and the appearance image, and three synthesized images by Liu et al.~\shortcite{liu2020selfsupervised}, Zhang et al.~\shortcite{zhang2020cross} and ours, placed side by side in a random order. 
Each participant was asked to evaluate 20 examples according to three criteria: the visual quality of synthesized images, the similarity to the input sketch and the faithfulness in appearance, each in a five-point Likert scale (1 = strongly negative to 5 = strongly positive). 
In total, we got 40 (participants) $\times$ 20 (questions) = 800 subjective evaluations for each method.
The second column of Fig.~\ref{fig:userstudy}(b) shows the statistics of these three methods. We also performed the ANOVA tests on the three aspects, and get the values for  geometry ($F_{(2, 57)} = 8.9514, p<
0.05$), appearance ($F_{(2, 57)} = 3.8033, p<
0.05$) and realism ($F_{(2, 57)} = 6.0887, p<
0.05$). It is clear that our method achieved a significant improvement over the other two methods.

\begin{figure}[htb]
    \centering
    \includegraphics[width=0.98\linewidth]{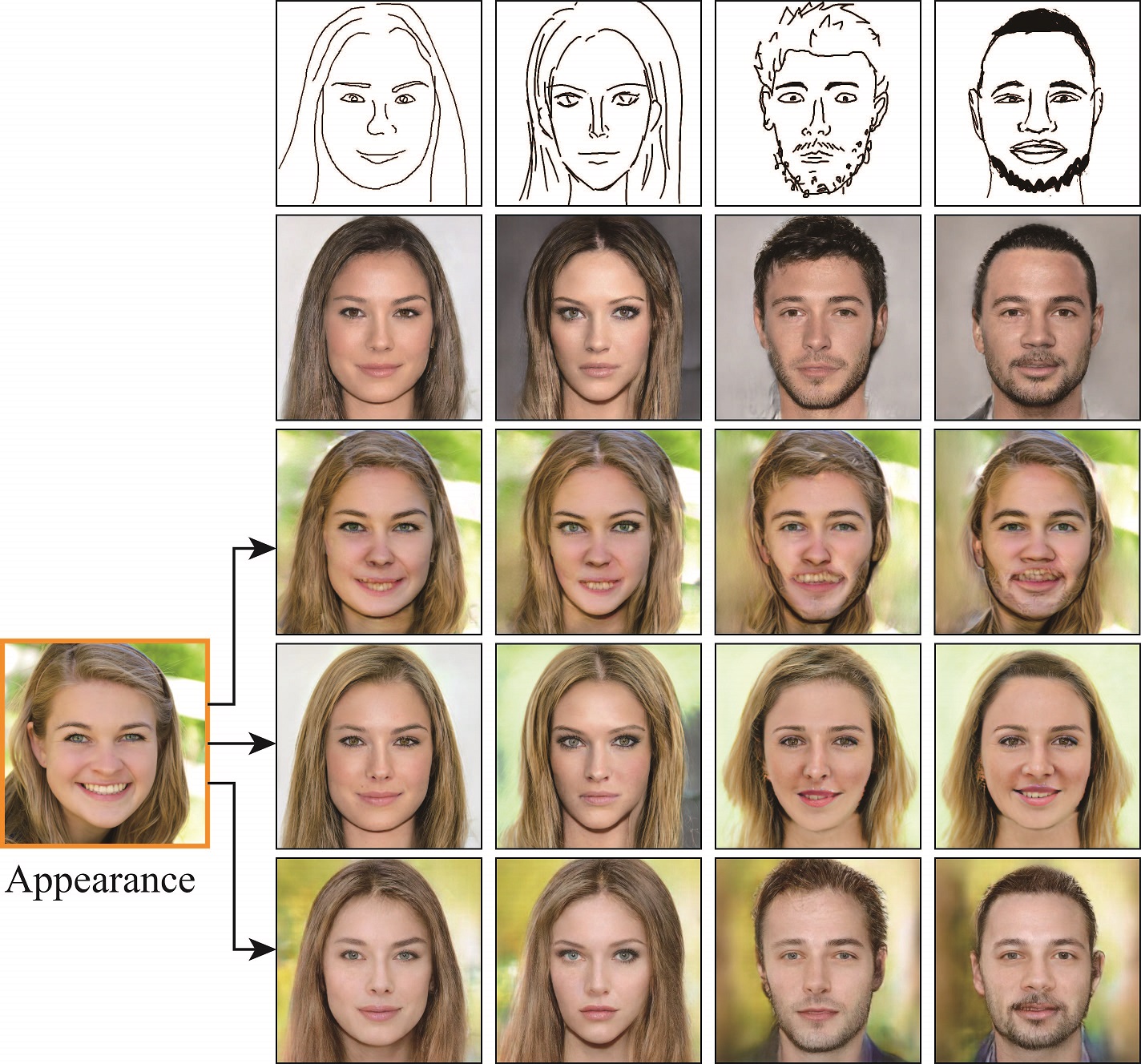}
        \caption{
        Comparisons of face generation given hand-drawn sketches (1st row). DeepFaceDrawing~\cite{10.1145/3386569.3392386} (2nd row) has no control of appearance. The results of using a style transfer technique ~\cite{Kolkin_2019_CVPR} (3rd row) have some artifacts on the mouth and hair. When using the swapping autoencoder~\cite{park2020swapping}, the results (the last two columns in the 4th row) do not retain the face geometry specified by the sketches well. Taking the sketches (1st row) and the appearance reference (Left) as input, our method (bottom row) generates stable results.
        Original images courtesy of Tom Kunz.
        }
        \vspace{-2mm}
    \label{fig:sketch-to-image-application2}
\end{figure}

\begin{figure}
    \centering
    \setlength{\fboxrule}{0.5pt}
    \setlength{\fboxsep}{-0.01cm}
    
    \framebox{\includegraphics[width=0.19\linewidth]{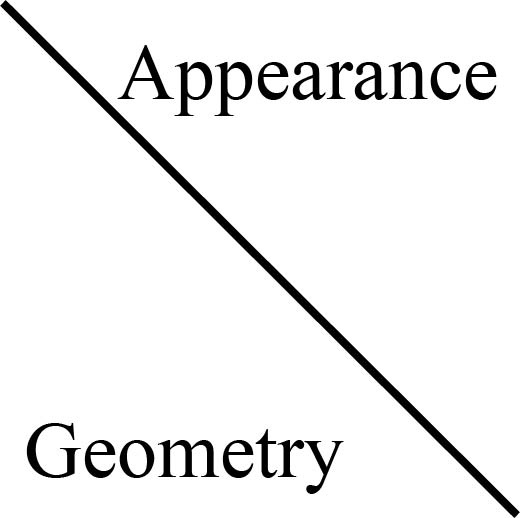}}
    \framebox{\includegraphics[width=0.19\linewidth]{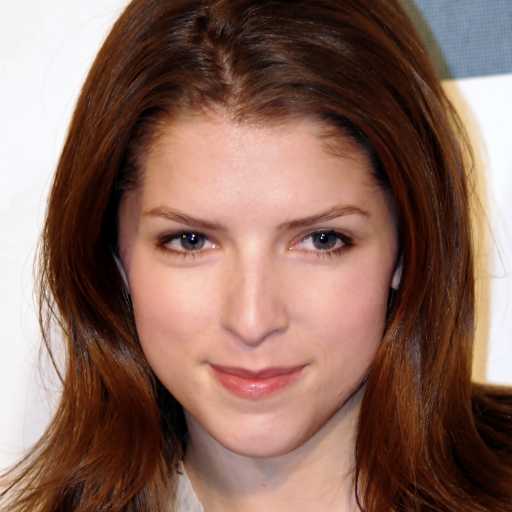}}
    \framebox{\includegraphics[width=0.19\linewidth]{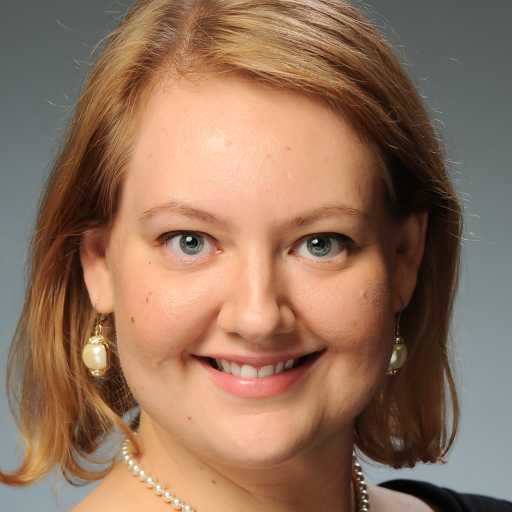}}
    \framebox{\includegraphics[width=0.19\linewidth]{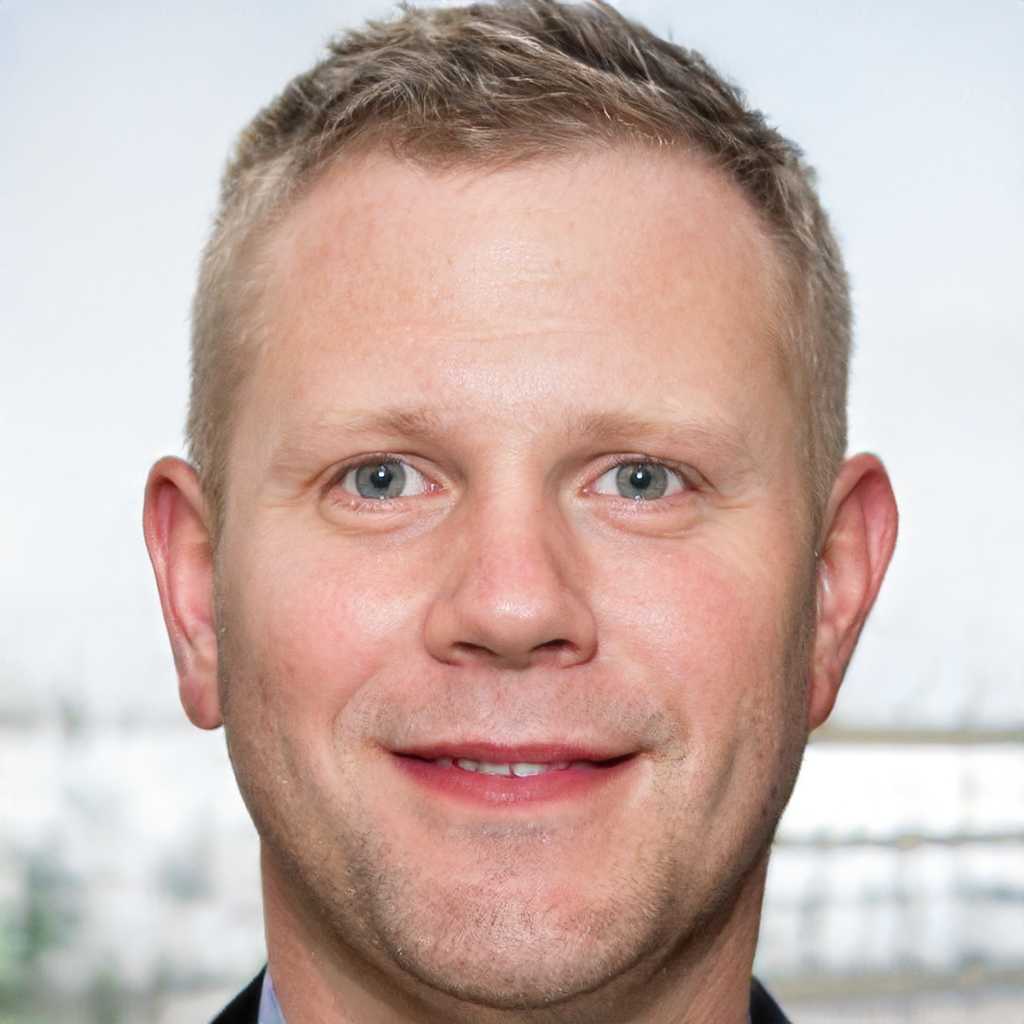}}
    \framebox{\includegraphics[width=0.19\linewidth]{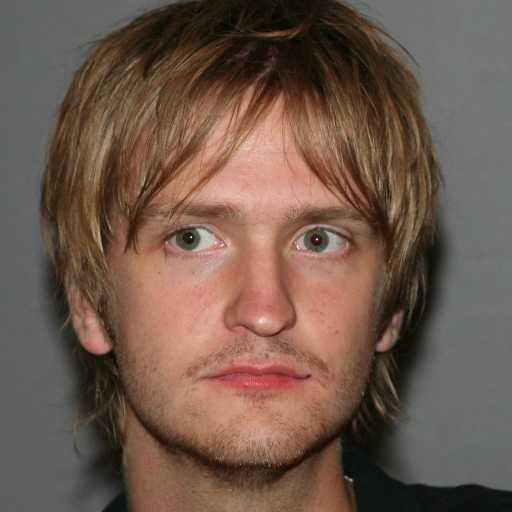}}
    \\
    
    \framebox{\includegraphics[width=0.19\linewidth]{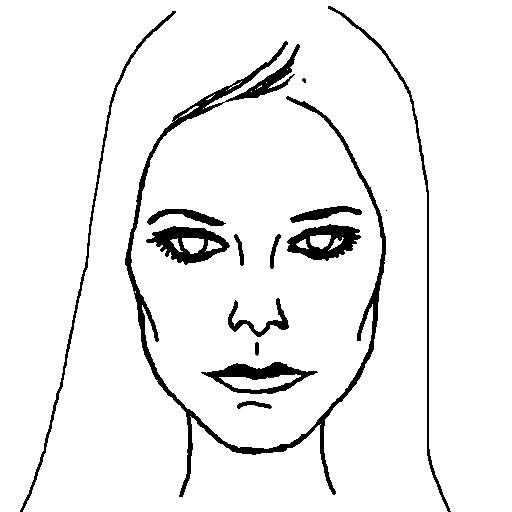}}
    \framebox{\includegraphics[width=0.19\linewidth]{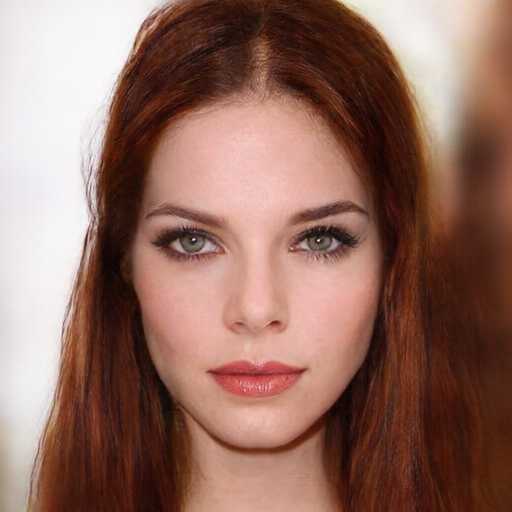}}
    \framebox{\includegraphics[width=0.19\linewidth]{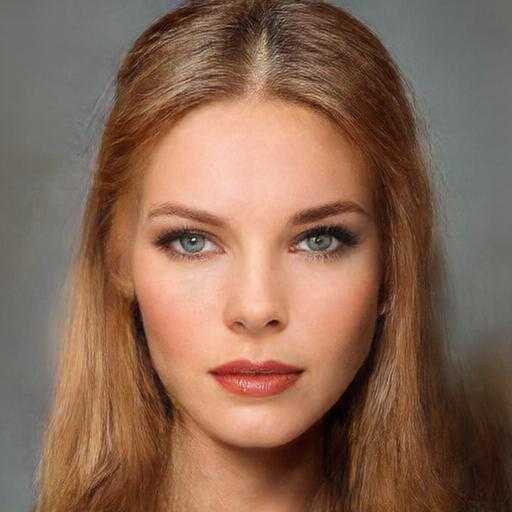}}
    \framebox{\includegraphics[width=0.19\linewidth]{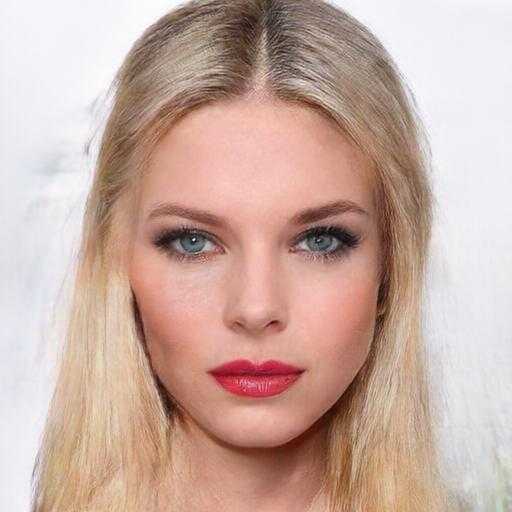}}
    \framebox{\includegraphics[width=0.19\linewidth]{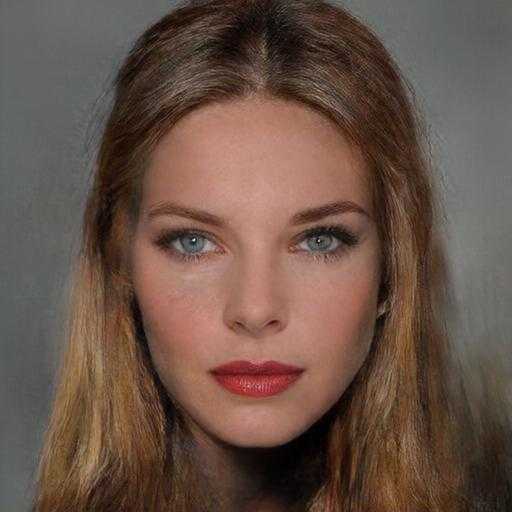}}
    \\
    
    \framebox{\includegraphics[width=0.19\linewidth]{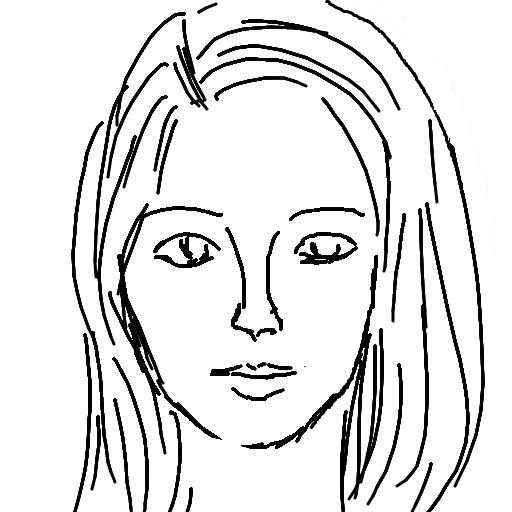}}
    \framebox{\includegraphics[width=0.19\linewidth]{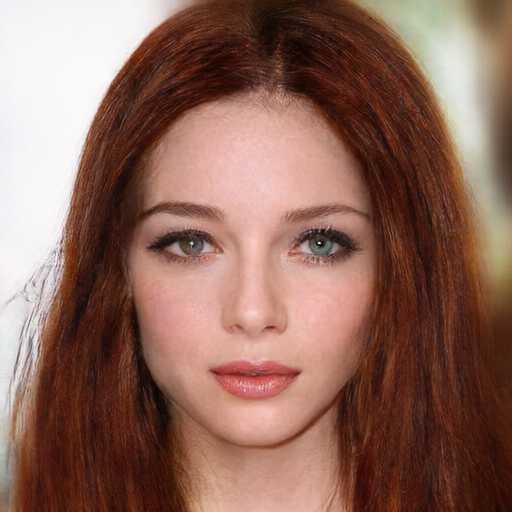}}
    \framebox{\includegraphics[width=0.19\linewidth]{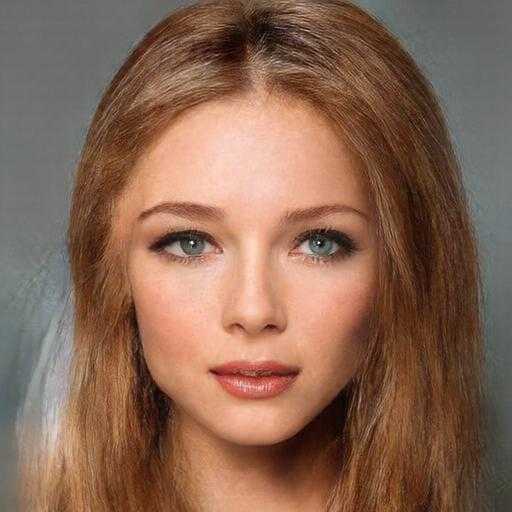}}
    \framebox{\includegraphics[width=0.19\linewidth]{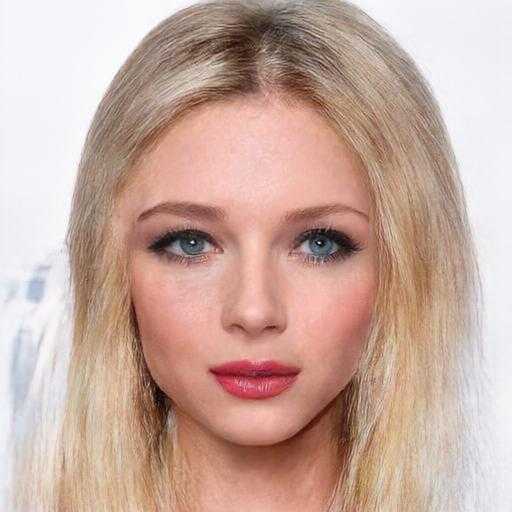}}
    \framebox{\includegraphics[width=0.19\linewidth]{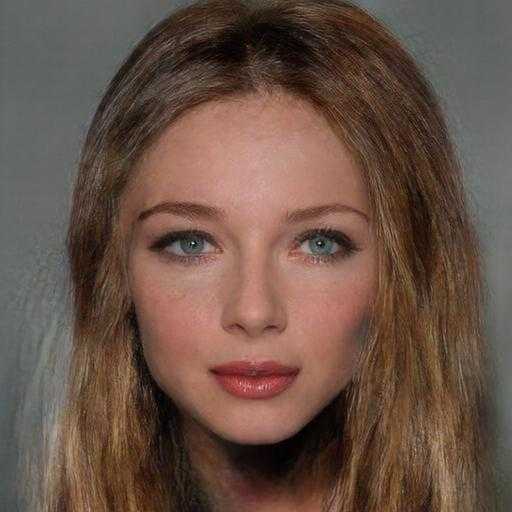}}
    \\

    \framebox{\includegraphics[width=0.19\linewidth]{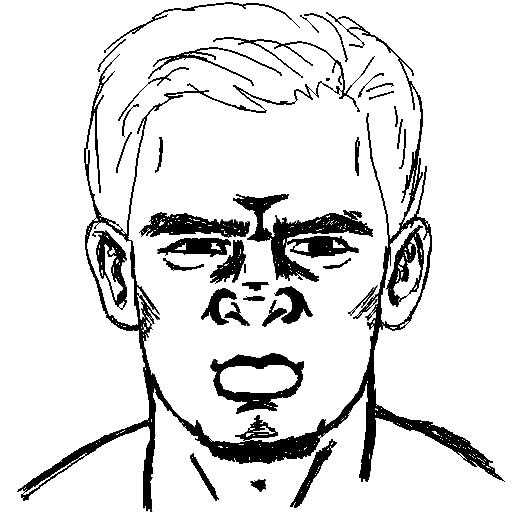}}
    \framebox{\includegraphics[width=0.19\linewidth]{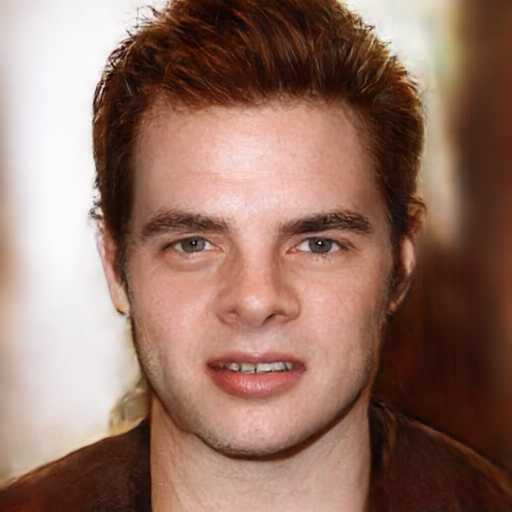}}
    \framebox{\includegraphics[width=0.19\linewidth]{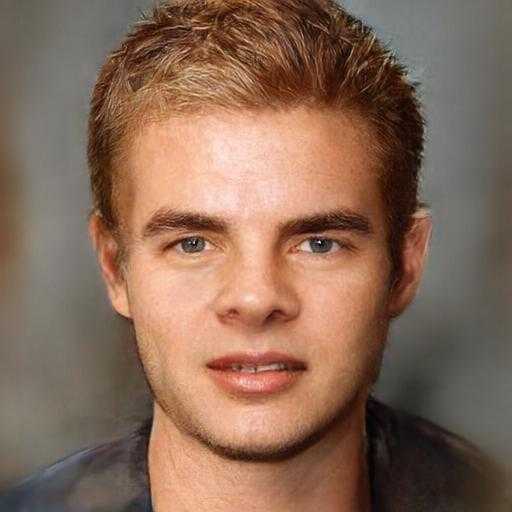}}
    \framebox{\includegraphics[width=0.19\linewidth]{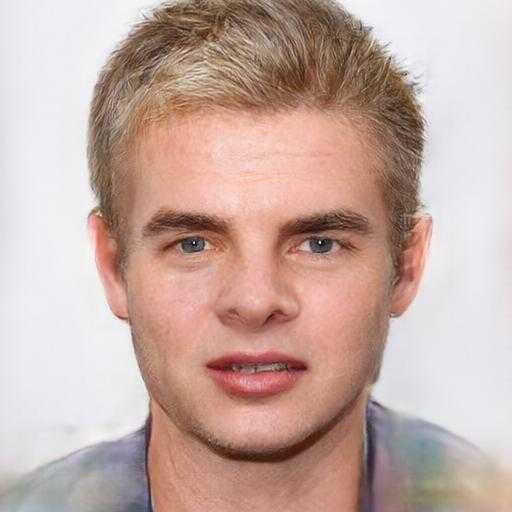}}
    \framebox{\includegraphics[width=0.19\linewidth]{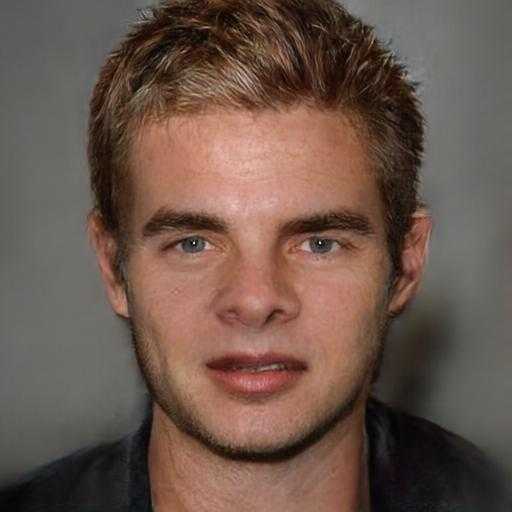}}
    \\
    
    \framebox{\includegraphics[width=0.19\linewidth]{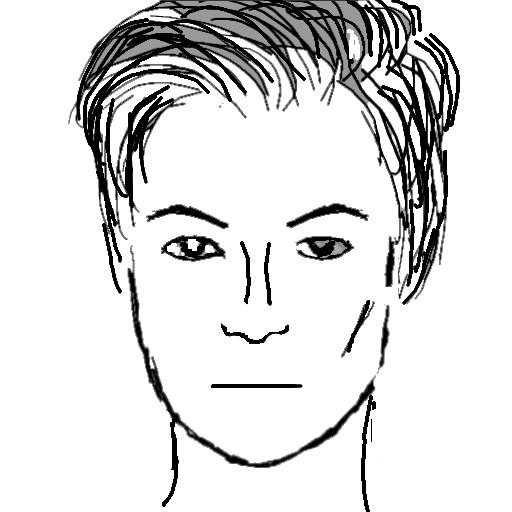}}
    \framebox{\includegraphics[width=0.19\linewidth]{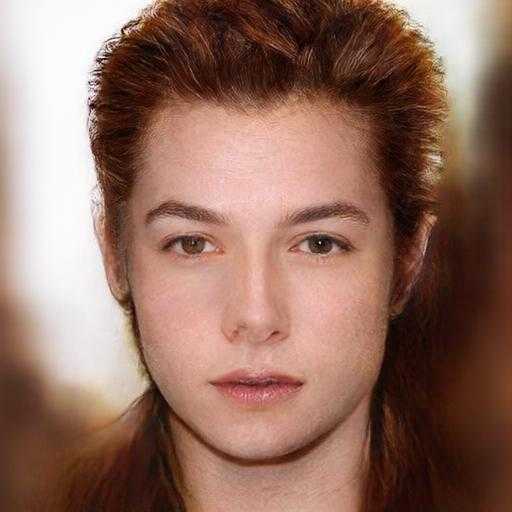}}
    \framebox{\includegraphics[width=0.19\linewidth]{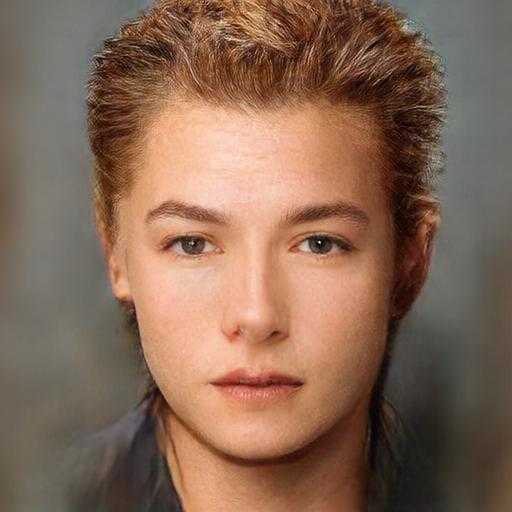}}
    \framebox{\includegraphics[width=0.19\linewidth]{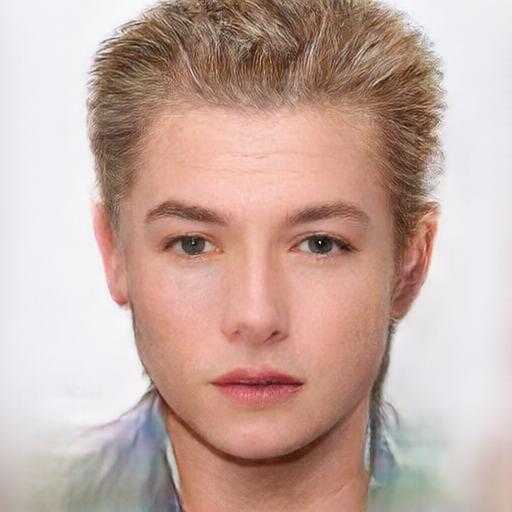}}
    \framebox{\includegraphics[width=0.19\linewidth]{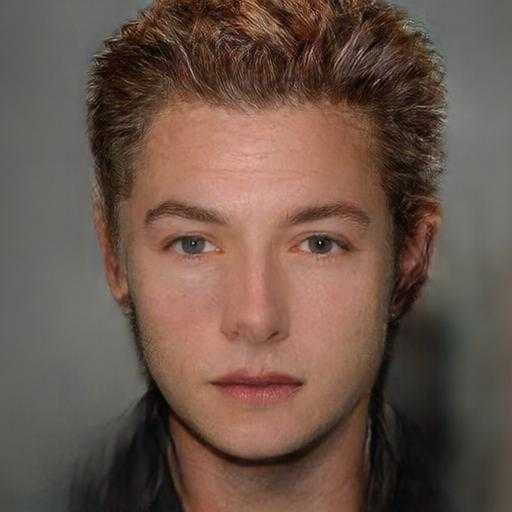}}
    \\
    
    \caption{Face synthesis results with hand-drawn sketches. With the control of appearance, our method can generate diverse photo-realistic faces from hand-drawn sketches.
    Original images courtesy of David Shankbone,Greg Mooney and danieljordahl.
    }
    \label{fig:hand-draw sketch generation}
\end{figure}

\section{Applications}
Thanks to the disentanglement of geometry and appearance, our method can be adapted to many applications. In this section, we describe three applications: 1) Sketch editing interface; 2) Hand-drawn sketch to face image generation; 3) Disentangled morphing.

\subsection{Sketch Editing Interface}

First, we design a real-time sketch-based face editing user interface (Fig.~\ref{fig:windows}), enabling detailed face editing via sketching.
The control panel (Fig.~\ref{fig:windows} (c)) consists of some necessary tools such as image opening tools, editing tools, etc. We also offer a sketch extraction function, facilitating users to translate a photo to a sketch for further editing. 
Our system generates synthesized results in real-time, based on modifications of sketches. 
A list of reference appearance images are placed at the bottom of interface (Fig.~\ref{fig:windows} (d)), and users can upload their own face images to the list. By selecting an appearance image, users can control the appearance of the generated results. 

\subsection{Hand-drawn Sketch to Image Generation}
The editing module in our system enables users to draw faces via coarse or fine sketches.
For users with little experience in drawing, they can produce realistic faces even from simple sketches (see the first sketch image only drawing an outline of a face in Fig.~\ref{fig:sketch-to-image-application}).
If the user is expert at drawing, he/she can edit the sketches with a brush or an eraser in our system to generate a more specific face closer to his/her conception. 
As shown in Fig.~\ref{fig:sketch-to-image-application}, a progressive sketch sequence is drawn by the user, and we compare the synthesized face sequence with DeepFaceDrawing~\cite{10.1145/3386569.3392386}, DeepFaceDrawing with style transfer and swapping autoencoder from a reference image, and our results based on the geometry of sketches and appearance of the same reference.
It is evident that the results of DeepFaceDrawing (the 2nd row) show faces with diverse appearances, e.g. varied hair and skin colors. Such sudden and uncontrollable changes are not desired by the user who is making incremental changes to the sketch.
Even combined with a style transfer method~\cite{Kolkin_2019_CVPR} using the left image as a reference, the results of DeepFaceDrawing show some artifacts in local areas like the mouth or eyes. Besides, our method is more efficient than \cite{Kolkin_2019_CVPR}, which takes around 1 minute to generate a result. Combining DeepFaceDrawing with the swapping autoencoder~\cite{park2020swapping}, the results show slight color difference with the reference image and are also affected by the generation of DeepFaceDrawing (4th column). Even with the appearance swapping, differences between editing frames also exist (especially on hair),
while our results are more robust and faithful to incremental changes. 
A similar comparison is shown in Fig.~\ref{fig:sketch-to-image-application2} where we use full hand-drawn sketches instead of an editing sequence. DeepFaceDrawing~\cite{10.1145/3386569.3392386} generates plausible face images (2nd row), but without appearance control. Applying a style transfer technique~\cite{Kolkin_2019_CVPR} and the swapping autoencoder~\cite{park2020swapping} to the results of DeepFaceDrawing can partially address this problem. However, applying these two steps in succession may accumulate errors. The results of~\cite{Kolkin_2019_CVPR} (3rd row) have some artifacts on the mouth and hair. The swapping autoencoder~\cite{park2020swapping} (4th row) can generate attractive results, but the resulting faces may not retain the geometry of the sketches (third and last columns).
As shown in Fig.~\ref{fig:sketch-to-image-application2} (bottom row) and Fig.~\ref{fig:hand-draw sketch generation}, in our system, users can control both the geometry with hand-drawn sketches and the appearance with diverse references.

\begin{figure}[htb]
    \centering
    \includegraphics[width=0.90
    \linewidth]{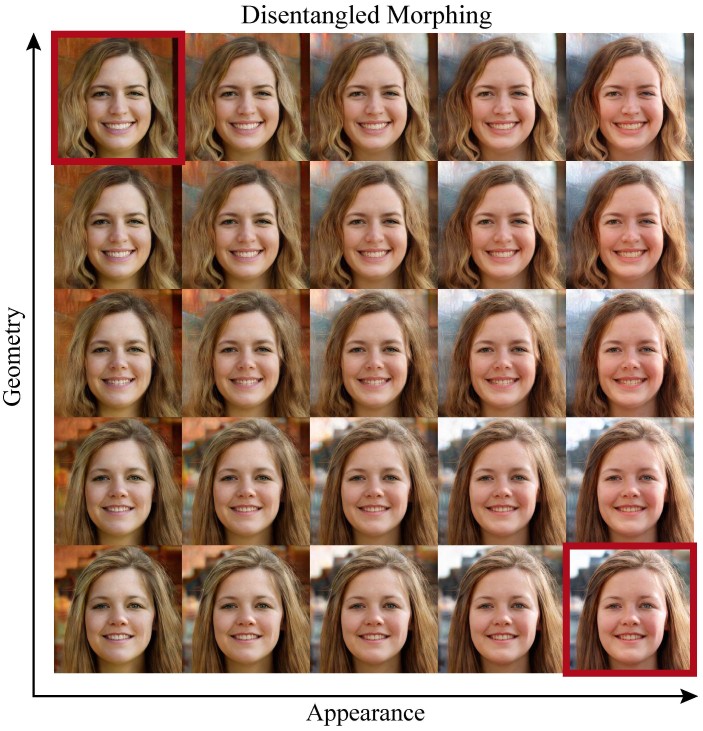}
    \caption{
    The results of disentangled morphing by interpolating the top-left and bottom-right faces (in red boxes).
    The remaining face images are generated automatically with our method. In each row, the appearance of the images is interpolated while keeping the geometry unchanged, while in each column, the geometry is interpolated while retaining the appearance. 
    Original images courtesy of Lydia Liu and SISTERKLAAS Lore. 
    }
    \vspace{-4mm}
    \label{fig:disentangled}
\end{figure}

\subsection{Disentangled Morphing}

Our method learns two disentangled latent spaces of the geometry and appearance respectively and enables a novel application of generating face images with reference of given geometry or appearance patterns. 
We demonstrate via disentangled face image morphing that a clear disentanglement between geometry and appearance has been achieved by our method.
As shown in Fig.~\ref{fig:disentangled}, we achieve controllable interpolation 
along two dimensions (Appearance and Geometry).
Given two input images $(I_1, I_2)$, our method can extract the geometry code $f_{G}^1, f_{G}^2$ and the appearance code $f_{A}^1,f_{A}^2$ from both images. 
Then, we conduct linear interpolation in the geometry and/or appearance latent spaces, using the corresponding latent codes from image $I_1$ and $I_2$.
The results demonstrate that our interpolation faces are controllable and every interpolated face enjoys high fidelity.

\begin{figure}
    \centering
    \subfigure[Geometry]{\includegraphics[width=0.30\linewidth]{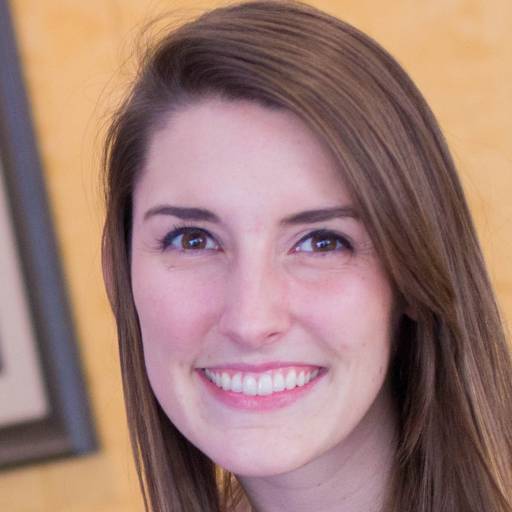}}
    \subfigure[Appearance]{\includegraphics[width=0.30\linewidth]{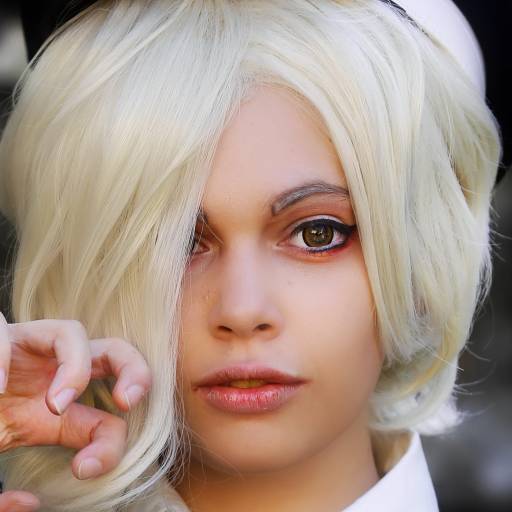}}
    \subfigure[result]{\includegraphics[width=0.30\linewidth]{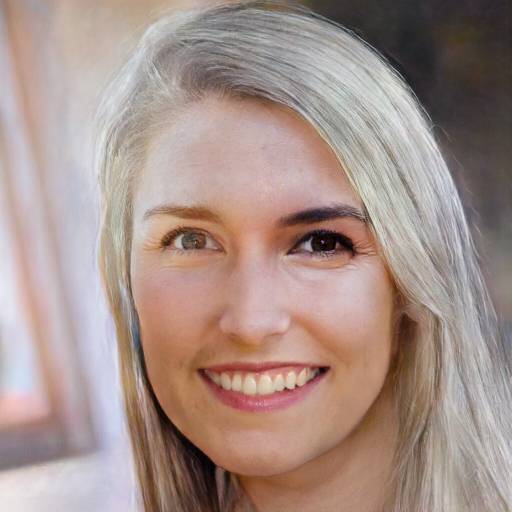}}
    \vspace{-2mm}
    \caption{A less successful example. When an appearance image contains a large occlusion, there may be some color bias in the generated result.
    Original images courtesy of UC Davis College of Engineering and taymtaym.
    \vspace{-4mm}
    }
    \label{fig:limit} 
\end{figure}

\begin{figure}
    \centering
    \setlength{\fboxrule}{0.5pt}
    \setlength{\fboxsep}{-0.01cm}
    \begin{tabular}{cc}
    
    \adjustbox{cfbox = teal 1pt 0pt}
    {\includegraphics[width=0.175\linewidth]{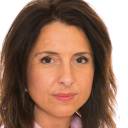}} & 
    \framebox{\includegraphics[width=0.175\linewidth]{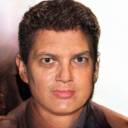}}
    \framebox{\includegraphics[width=0.175\linewidth]{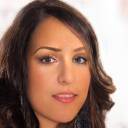}}
    \framebox{\includegraphics[width=0.175\linewidth]{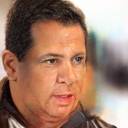}}
    \framebox{\includegraphics[width=0.175\linewidth]{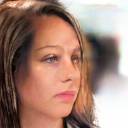}}
    \\
    
    \adjustbox{cfbox = teal 1pt 0pt}
    {\includegraphics[width=0.175\linewidth]{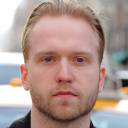}} & 
    \framebox{\includegraphics[width=0.175\linewidth]{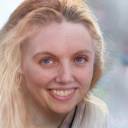}}
    \framebox{\includegraphics[width=0.175\linewidth]{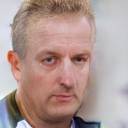}}
    \framebox{\includegraphics[width=0.175\linewidth]{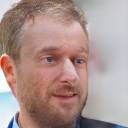}}
    \framebox{\includegraphics[width=0.175\linewidth]{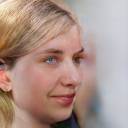}}
    \\
    \end{tabular}
    
    \caption{
    The results for non-frontal faces. Images in the first column provide appearance and the rest images are generated by our method. The inputs of sketches can be seen in the supplementary material. As the rotation angle increases, the quality of the results decreases. 
    Original images courtesy of Senterpartiet (Sp) and Phillip Nguyen. 
    }
    \vspace{-2mm}
    \label{fig:no_frontal_images}
\end{figure}

\section{Conclusion, Limitations and Future Work}
This work presented a structured disentanglement framework for face generation and editing, carried out in a local-to-global manner. Our key observation is that geometry and appearance features of face images can be effectively disentangled, and sketches serve as an ideal intermediate representation for geometry features. Therefore, sketches can impose a strong constraint during disentanglement. The component-level geometry and appearance disentanglement is achieved by \moduleOneFull ~\ modules which are trained using a swapping strategy, while the \moduleTwoFull ~\ module performs coherent local-to-global image generation from feature maps of facial image patches. Through extensive experiments, we prove that our approach can generate much more realistic results than existing methods. We also adapted our system for novel applications such as sketch editing, hand-drawn sketch based face image generation and disentangled face morphing.

One limitation of this work is that we only disentangle the geometry and appearance while other attributes such as head pose have not been considered in our current implementation. 
As shown in Figs.~\ref{fig:limit} and~\ref{fig:no_frontal_images}, when there is a large rotation of the subject's head or there is substantial occlusion, there may be some color bias in the generated results.
Lighting is another challenging problem for most existing face synthesis methods and is not explicitly disentangled in our framework, so it is difficult to finely control complicated lighting conditions by sketches or reference images. 
As future work, it would be useful to explore disentanglement of other attributes such as head pose and lighting to make the method more general.
Besides, sketches have semantic ambiguity and in some extreme cases, it is hard even for humans to distinguish the accurate boundary between neck, hair and background. This ambiguity may sometimes cause some artifacts on the outer boundary of the face foreground, leading to blurred hair.
As future work, semantic masks may be combined with sketches to generate more attractive results. Furthermore, while we allow detailed editing of geometry through sketching, our current approach uses an appearance reference image to control the appearance of the generated face image. 
This could be improved using other forms of input, such as strokes with color, which would be more flexible, such as~\cite{Sangkloy_2017_CVPR}. This is a promising research direction in the future but it remains challenging, because the face appearance not only contains the color but also the material attributes, and it is very difficult to stroke the complex materials of the face. The consistency between the color strokes should also be maintained. These will be explored in the future research work.

\begin{acks}
This work was supported by National Natural Science Foundation of China (No. 61872440 and No. 62061136007), Science and Technology Service Network Initiative of the Chinese Academy of Sciences (No. KFJ-STS-ZDTP-070, No. KFJ-STS-QYZD-129 and No. KFJ-STS-QYZD-2021-11-001), Royal Society Newton Advanced Fellowship (No. NAF$\backslash$R2$\backslash$192151),
Youth Innovation Promotion Association CAS, and Beijing Program for International S\&T Cooperation Project (No. Z191100001619003).
Hongbo Fu was supported by HKSAR RGC General Research Fund (No. 11212119) and City University of Hong Kong (SCM ACIM Collaborative Research Fellowship).

\end{acks}

\bibliographystyle{ACM-Reference-Format}
\bibliography{bibliography}

\end{document}